\documentclass[aps,twocolumn,prb,superscriptaddress,floatfix]{revtex4-2}
\setcitestyle{super}
\usepackage{amsmath,mathtools,graphicx,xcolor,bm,multirow,enumitem,soul,subfigure,setspace,xr,float,mhchem,overpic,array}
\def\fontfig{\normalsize\bf }
\def\fonttext{\normalsize }
\usepackage{ragged2e}
\allowdisplaybreaks

\def\br{{\bf r}}

\def\bq{{\bf q}}
\def\bu{{\bf u}}

\def\utoden{Oden Institute for Computational Engineering and Sciences, The University of Texas at Austin, Austin, Texas 78712, USA}
\def\utphysics{Department of Physics, The University of Texas at Austin, Austin, Texas 78712, USA}
\def\ubilbao{Department of Physics, University of the Basque Country UPV/EHU, 48940 Leioa, Basque Country, Spain}

\begin{document}

\title{Symmetry-protected topological polarons}

\author{Kaifa Luo}
\affiliation{\utphysics}
\affiliation{\utoden}

\author{Jon Lafuente-Bartolome}
\affiliation{\ubilbao}

\author{Feliciano Giustino}
\email{fgiustino@oden.utexas.edu}
\affiliation{\utoden}
\affiliation{\utphysics}

\date{\today}

\maketitle

\onecolumngrid
\vspace{-10pt}
\noindent\textbf{%
Emergent quasiparticles in solids often exhibit unique topological properties as a result of the complex interplay between charge, orbital, spin and lattice degrees of freedom. Among these quasiparticles, the polaron occupies a special place as the first known manifestation of the interaction between a fermion and a boson field. While polarons have been investigated for almost a century, whether these quasiparticles exhibit topological properties and why remain open questions. Here, we establish the universal symmetry principles governing the topology of polar textures in large polarons. Using a group-theoretic analysis, we identify four distinct classes of polar textures in time-reversal-invariant systems, and we show that they carry integer topological charges. We validate this classification by performing state-of-the-art first-principles calculations of materials representative of each class. For these materials, we compute the fingerprints of polaron topology in Huang diffuse scattering, and propose ultrafast electron and X-ray scattering experiments to detect these quasiparticles.
}
\vspace{10pt}
\twocolumngrid


Topological invariants of emergent quasiparticles, as encoded in the real-space spin textures of skyrmions \cite{nagaosa2013topological,fert2017magnetic} and the momentum-space Berry curvature of Dirac, Weyl, and Majorana fermions \cite{RMP-2010,RMP-2011,zhang2009topological,weng2015weyl}, often give rise to exotic physical phenomena, and constitute promising candidates for information carriers in next-generation quantum devices. 
Despite extensive research on topological invariants, their potential role in the physics of one of the most fundamental quasiparticles, the polaron, is only beginning to emerge \cite{grusdt2016interferometric,grusdt2019topological,JLB_topological_2024}. Polarons are electronic excitations dressed by a phonon field \cite{landau1948effective,feynman1955slow}; they are ubiquitous in materials \cite{emin2013polarons,wang2016tailoring,franchini2021polarons} and play key roles in artificial photosynthesis \cite{Ren2024photocat,Sokolovic2024RealSpace,Dai2024Identification}, neuromorphic computing \cite{Yao2020Protonic,Onen2023Nanosecond}, and emerging photovoltaics \cite{Zhu2016,Miyata2018,Puppin2020,Guzelturk2021,Wu2021}. A significant body of knowledge exists on polaron energetics and dynamics \cite{Alexandrov2012,Devreese2020,franchini2021polarons}, spanning model Hamiltonian approaches \cite{Millis-1996,Mishchenko-2000s,Berciu2006,Devreese-2009j,kloss2019,milli-2023} and first-principles calculations \cite{Wiktor2018,Falletta2022,birschitzky2022machine,vasilchenko_polarons_2024,Chang2025}; however, very little is known about the real-space polar textures emerging from electron-phonon couplings, whether and under which conditions they exhibit nontrivial topology, and which physical observables may be associated with such topological properties. 

Very recently, topologically non-trivial polar textures accompanying polarons have been identified in the halide double perovskite \ce{Cs2AgBiBr6} \cite{JLB_topological_2024} and in the quantum paraelectric \ce{SrTiO3} \cite{xu_ultrasmall_2023} via first-principles calculations. In addition, several experimental observations of nontrivial polar textures have been reported in engineered ferroelectric oxides like \ce{Pb(Zr,Ti)O3} and \ce{BaTiO3} \cite{junquera_topological_2023,hong_stability_2017}, and their moir\'es \cite{sanchez-santolino_2d_2024,Tsang-2024g}, including polar analogs of magnetic skyrmions \cite{das_observation_2019,shao_emergent_2023}, merons \cite{yadav_observation_2016,damodaran_phase_2017}, and hopfions \cite{govinden_spherical_2023}. These topological objects attracted considerable interest as they could offer novel pathways for storing and manipulating quantum information via all-electrical means. For example, single-polaron write/move/erase operations using STM tips have recently been demonstrated \cite{Liu2022,Cai2024}.

Here, we set out to systematically identify and analyze topologically nontrivial textures of polarons. For other quasiparticles, topological invariants are fundamentally dictated by symmetry, such as time-reversal, particle-hole, and chiral symmetries, as well as crystalline point-group and space-group symmetries \cite{chiu_classification_2016,bradlyn_topological_2017,zhang_catalogue_2019,vergniory_complete_2019,xu_high-throughput_2020}. By analogy, we hypothesize that symmetry principles offer the most natural starting point for investigating polaron topology.

The symmetry of polarons in coupled electron-phonon systems remains an open question. For non-interacting electrons and phonons, the relevant symmetry is the little group of operations that preserves their wavevectors. 
However, when electrons and phonons couple to form small polarons, the resulting electron wavefunction and polar distortion are coherent superpositions of states from the entire Brillouin zone \cite{Sio2019a,sio_polarons_2023,JLB_topological_2024,Dai-2024a}, incorporating low-symmetry wavevectors with trivial little groups. This complexity makes a symmetry-based analysis impractical. 
In contrast, large polarons primarily involve low-energy valley electrons and long-wavelength phonons, which imposes symmetry constraints on the effective interaction Hamiltonian. 
These constraints enable a general symmetry analysis of large polarons, as demonstrated for cubic crystals in Ref.~\citenum{vasilchenko_polarons_2024}, and provide a framework to investigate their topological properties.

\section*{Results}

\subsection*{Symmetry-based classification of polarons}

In the presence of large polarons, the host crystal can be considered as a continuous medium, and the symmetry of the undistorted crystal is preserved on average, since atomic displacements are small. Under these approximations, the response of the lattice to an excess electron induces atomic displacement patterns that can be described via a cell-averaged, slowly-varying vector field $\bu$. 
This vector field
describes the change of the atomic positions from the pristine crystal structure without polarons to the distorted configuration in the presence of polarons, in analogy with the definition of macroscopic polarization in the modern theory of polarization \cite{resta1992theory,king1993theory,resta1994macroscopic}. 
This field is constrained by the crystal point-group symmetry:
    \begin{equation}\label{eq:disp-sym}
        \bu(\hat S \hspace{0.5pt}\br)=\hat{S} \hspace{0.5pt}\bu(\br)\,,
    \end{equation}
where $\hat{S}$ represents the point operation of any symmetry belonging to the crystal space group \cite{nye1985physical,dresselhaus2007group}, and $\br$ is the position relative to the polaron center. The above equation is proven in Supplemental Note~1, starting from the \textit{ab initio} polaron equations of Ref.~\citenum{Sio2019b} and including the leading terms in the electron-phonon couplings at long wavelength, namely Fr\"ohlich \cite{Frohlich1954,Vogl1976,Verdi2015,sjakste2015wannier} and piezoelectric couplings \cite{mahan1964piezoelectric,Martin1972,Royo2019,Brunin2020,Park2020}.

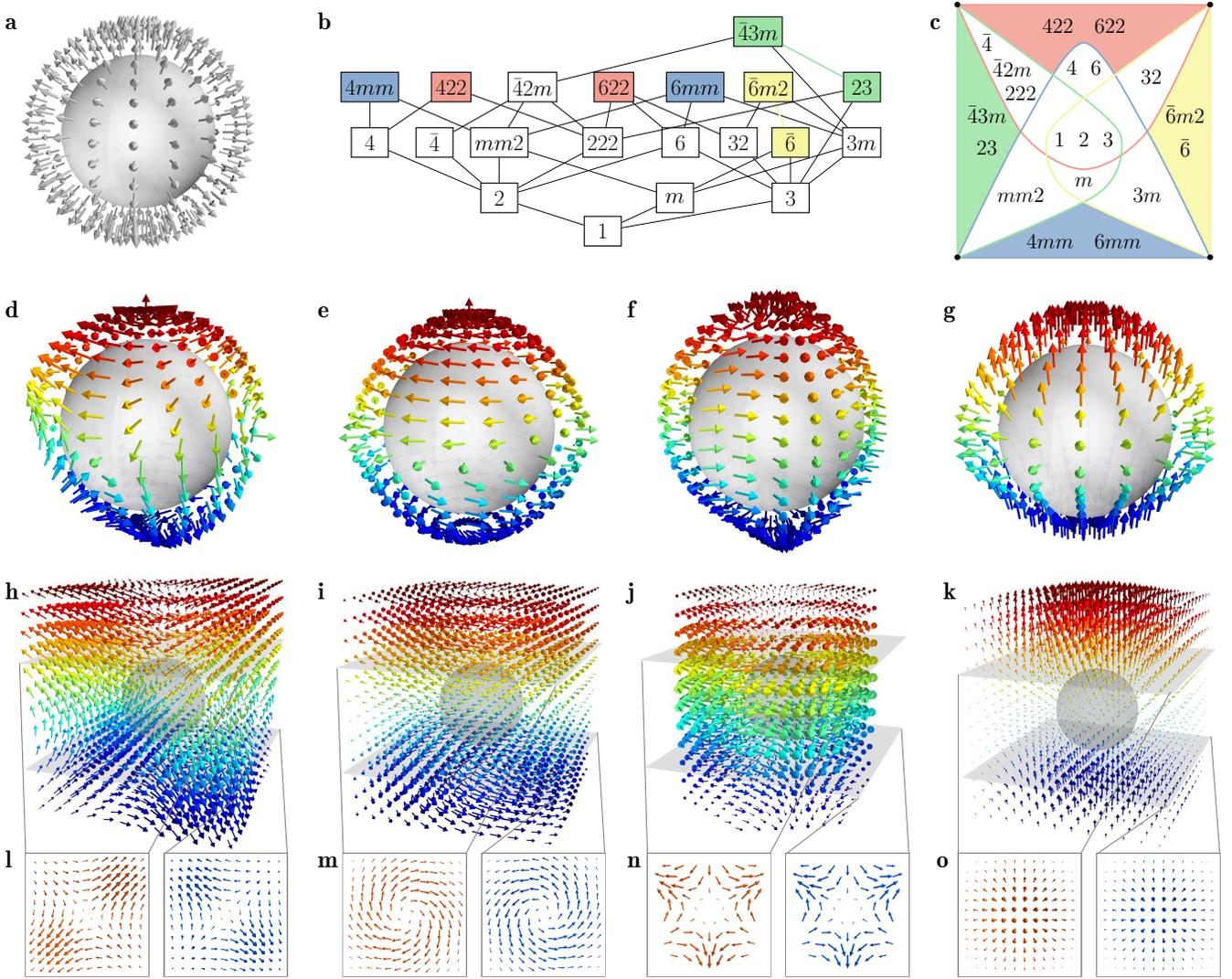
\begin{figure*}[tbh!]
    \centering
    \begin{overpic}[width=1.0\textwidth]{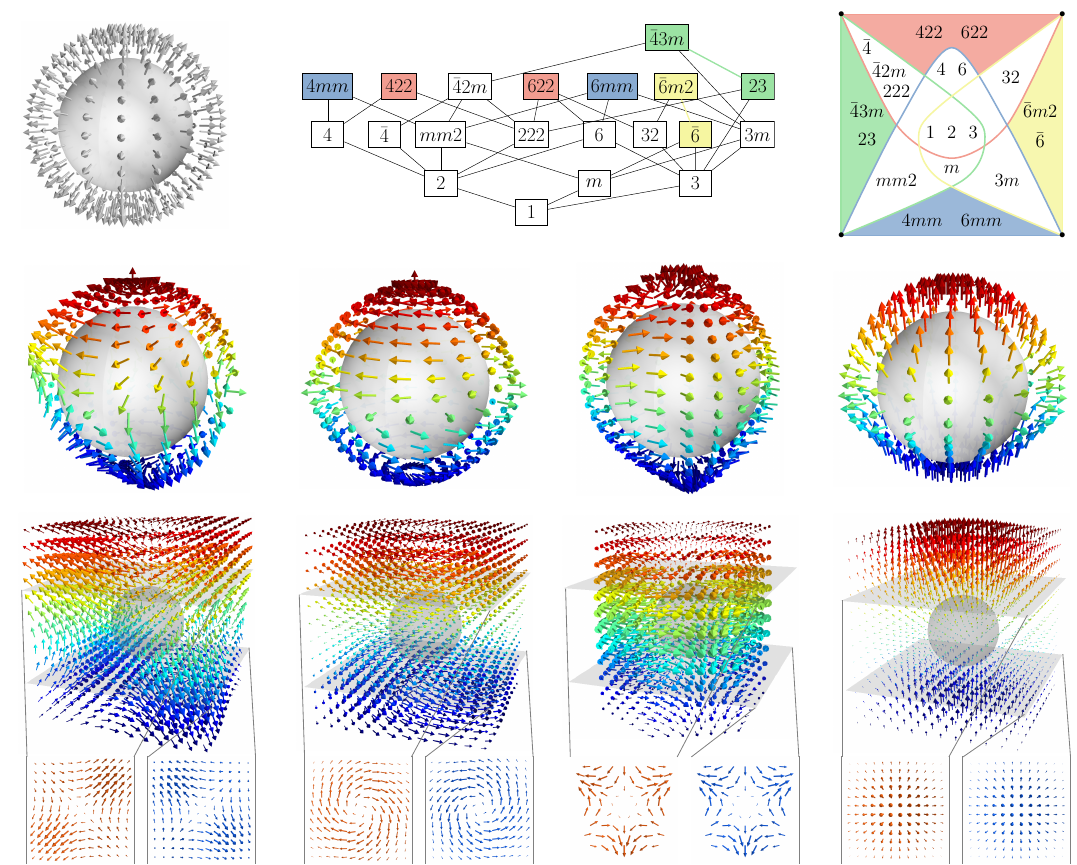}
        \put (0.15cm,13.8cm) {\fontfig a}
        \put (4.7cm, 13.8cm) {\fontfig b}
        \put (13.6cm,13.8cm) {\fontfig c}
        \put (0.15cm,9.6cm) {\fontfig d}
        \put (4.7cm, 9.6cm) {\fontfig e}
        \put (9.2cm, 9.6cm) {\fontfig f}
        \put (13.8cm,9.6cm) {\fontfig g}
        \put (0.15cm,5.5cm) {\fontfig h}
        \put (4.7cm, 5.5cm) {\fontfig i}
        \put (9.2cm, 5.5cm) {\fontfig j}
        \put (13.8cm,5.5cm) {\fontfig k}
        \put (0.15cm,1.6cm) {\fontfig l}
        \put (4.7cm, 1.6cm) {\fontfig m}
        \put (9.2cm, 1.6cm) {\fontfig n}
        \put (13.7cm,1.6cm) {\fontfig o}
    \end{overpic}
    \caption{
    \textbf{Symmetry-based classification of topological polarons}. 
    \textbf{a} Monopole-like hedgehog polaron with topological charge $|Q|=1$, vorticity $v=1$, and helicity $\gamma=0$. 
    The arrows show the atomic displacement field on a sphere enclosing the polaron center.
    \textbf{b} Pruned graph of group-subgroup relations between crystal point groups without inversion symmetry, extracted
    from Supplemental Fig.~1. Color-coded groups host symmetry-protected topological polarons 
    (green: antivortex; red: vortex; yellow: double-antivortex; blue: vertical flow).
    \textbf{c} The point groups in the colored regions admit only one type of polaron texture, while all other groups 
    admit combinations of two, three, or four textures that are not protected by symmetry.
    \textbf{d}~Antivortex polaron on a sphere, with topological charge $|Q|=3$, vorticity $v=-1$, and helicity $|\gamma|=90^\circ$. 
    The color code is a visual aid. 
    \textbf{e} Vortex polaron, with topological charge $|Q|=1$, vorticity $v=1$, and helicity $|\gamma|=90^\circ$. 
    \textbf{f} Double antivortex polaron, with topological charge $|Q|=2$, vorticity $v=-2$, 
      and helicity $|\gamma|=90^\circ$. 
    \textbf{g} Vertical flow polaron, with topological charge $|Q|=0$ (vorticity and helicity are not defined in this case). 
    \textbf{h}-\textbf{k} and \textbf{l}-\textbf{o}: Volumetric plots and planar cuts of the polaron textures shown in \textbf{d}-\textbf{g}, respectively. 
     The spheres are the same as those shown in \textbf{d}-\textbf{g}.
     A detailed analysis of each of these textures and their topological invariants
     is provided in Supplemental Fig.~2 and Supplemental Note 3. 
     }\label{fig-symmetry}
\end{figure*}

Equation~\eqref{eq:disp-sym} dictates the possible forms that $\bu$ can take in a given crystal. To illustrate this point, we consider two examples, leaving detailed derivations to Supplemental Note~2. 
(i) In the simplest case of cubic rock-salt crystals, which belong to the $m\bar{3}m$ point group, symmetry requires the displacement field $(u_x,u_y,u_z)$ near the polaron center to transform like $(x,y,z)$ plus terms of third order and higher in $x,y,z$. As a result, atomic displacements form a monopole-like field with a characteristic hedgehog pattern, as shown in Fig.~\ref{fig-symmetry}(a). This is precisely the behavior observed in recent \textit{ab initio} calculations of large polarons in LiF \cite{Sio2019a,vasilchenko_polarons_2024}.
(ii) If inversion symmetry is removed from the $m\bar{3}m$ group, the resulting subgroup is $\bar{4}3m$, which describes, for example, cubic zinc-blende crystals (see Supplemental Fig.~1 for group-subgroup relations). In this case, symmetry dictates a displacement field of the type $a\,(x,y,z) + b\,(yz,zx,xy)$ plus terms of third order and higher, with $a$ and $b$ being material-specific constants. Therefore, in addition to the hedgehog texture, in this case we also have a three-dimensional antivortex field, as shown in Figs.~\ref{fig-symmetry}(d),(h).  Here, the lack of inversion symmetry is precisely what enables nontrivial textures, because vortex-like fields involving $xy$, $xz$, $yz$ are parity-forbidden. This is analogous to the case of magnetic skyrmions driven by the Dzyaloshinskii-Moriya interaction, which is allowed only in non-centrosymmetric crystals \cite{fert2017magnetic,heinze2011spontaneous}.

Building on this insight, we perform a classification of nontrivial polaron textures by inspecting the group-subgroup relations of crystal point groups. From the full graph shown in Supplemental Fig.~1, we obtain the pruned graph in Fig.~\ref{fig-symmetry}(b) by eliminating all groups that possess inversion symmetry.
By applying the symmetry operations of each group to \eqref{eq:disp-sym}, we obtain four distinct nontrivial vector fields $\bu(\br)$: an antivortex with texture given by $(yz,zx,xy)$; a vortex with texture $(-yz,xz,0)$; a double antivortex $(2xy,x^2\hspace{-0.7pt}-\hspace{-0.7pt}y^2,0)$; and a vertical flow $(0,0,z^2)$. The rationale for this nomenclature will become apparent shortly. These vector fields are shown on a sphere in Fig.~\ref{fig-symmetry}(d)-(g), respectively, as well as 3D plots and 2D cuts in panels (h)-(k) and (l)-(o) of the same figure, respectively. Groups admitting these vector fields are  highlighted in color in Fig.~\ref{fig-symmetry}(b) and include $\bar{4}3m$, $4mm$, $422$, $622$, $6mm$, $\bar{6}m2$, as well as two subgroups, $\bar 6$ and $23$, which inherit the textures of their respective parent groups (see Supplemental Table~1 for a summary). Figure~\ref{fig-symmetry}(c) shows how all other groups admit combinations of two, three, or four of the above elementary vector fields; for materials in these other groups, polaron textures are not symmetry-protected, therefore we do not consider them further.

\subsection*{Topological invariants and topological protection}

A complete classification of polaron textures requires identifying their associated topological invariants. To this end, we focus on the skyrmion number or topological charge $Q$ of a vector field, which is also referred as the ``degree of mapping'' in Ref.~\citenum{belavin1975metastable}. This quantity is defined as the flux of the topological density $\bm{\Omega}$ through a unit sphere enclosing its center: $Q=(1/4\pi)\int\bm{\Omega}\cdot d{\bf S}$, with $\bm{\Omega}_\alpha = \epsilon_{\alpha\beta\gamma} \hat{\bf u} \cdot ( \partial_\beta \hat{\bf u} \times \partial_\gamma \hat{\bf u} )/2$ \cite{JLB_topological_2024,nagaosa2013topological}. Here, $\hat{\bf u}({\br})$ is the normalized atomic displacement field, Greek subscripts denote Cartesian directions, $\epsilon_{\alpha\beta\gamma}$ is the Levi-Civita symbol, $\partial_\alpha$ is the spatial derivative with respect to the direction $\alpha$, $d{\bf S}$ is the surface element, and summation over repeated indices is implied. The density $\bm{\Omega}$ measures the local winding of the field, and the charge $Q$ counts how many times the displacement field wraps a closed surface enclosing the polaron center. 

In Supplemental Note~3, we evaluate the topological charge for the hedgehog, antivortex, vortex, double antivortex, and vertical flow fields identified above. We find integer charges $Q=0, \pm 1, \pm 2, \pm 3$ across these vector fields, with each texture carrying a uniquely defined charge. The sign of the topological charge is determined by the constant $a$ and is a material-specific property. Supplemental Fig.~2 and Supplemental Table~2 report the complete assignment of topological charges for these textures. 

Different textures can uniquely be identified by their topological charge, except for the hedgehog and the vortex which share the same charge $Q=\pm 1$. To further distinguish these patterns and motivate our nomenclature, we also consider the vorticity $v$ and the helicity $\gamma$ of the vector field, which are well defined for any 2D slice at constant elevation $z$, cf.\ Fig.~\ref{fig-symmetry}(l)-(o) \cite{nagaosa2013topological}. The vorticity is the winding of the in-plane component of the vector field along a closed loop, $v = (1/2\pi)\oint d\varphi$ with $\varphi = \tan^{-1}(u_y/u_x)$; the helicity is the average angle between the field and the in-plane radial direction (see Supplemental Fig.~2). The polaron textures identified here correspond to $v=1$ for the hedgehog (no helicity, $\gamma=0$) and the vortex ($|\gamma|=90^\circ$); $v=-1$ for the antivortex; and $v=-2$ for the double antivortex. The latter texture winds twice around the polaron center, hence its name.

Beyond the topological charge, point group symmetries also dictate the vectorial character of these textures. Indeed, following Ref.~\citenum{hlinka_eight_2014}, all time-reversal-even vectors can be classified into four categories: neutral, polar, chiral, and axial. Among the textures identified here, the hedgehog field is invariant under inversion symmetry, and thus belongs to the neutral category. The antivortex, double antivortex, and vertical flow fields reverse sign under inversion, therefore they are polar vectors. The vortex field lacks both inversion and mirror symmetries, and is classified as a truly chiral field, similar to the chiral phonons in $\alpha$-HgS \cite{Zhang2015ChiralPhonons,ishito_truly_2023,zhang2025weyl,zhang2025measurement}. The only missing element in our list is the axial texture, which is associated with ferrotoroidic order \cite{spaldin2008toroidal,hlinka_symmetry_2016}; this polaron class has recently been identified in halide perovskites \cite{JLB_topological_2024}, but is absent here because we are focusing on noncentrosymmetric point groups. For completeness, we discuss the polar textures of Ref.~\citenum{JLB_topological_2024} in Supplemental Note~4. The neutral, polar, chiral, and axial classes exhaust all possible symmetry-protected topological polaron textures in time-reversal-even crystals. 
In Supplemental Table~3, we also discuss the relation between the dimensionality and the topological stability of these polarons.

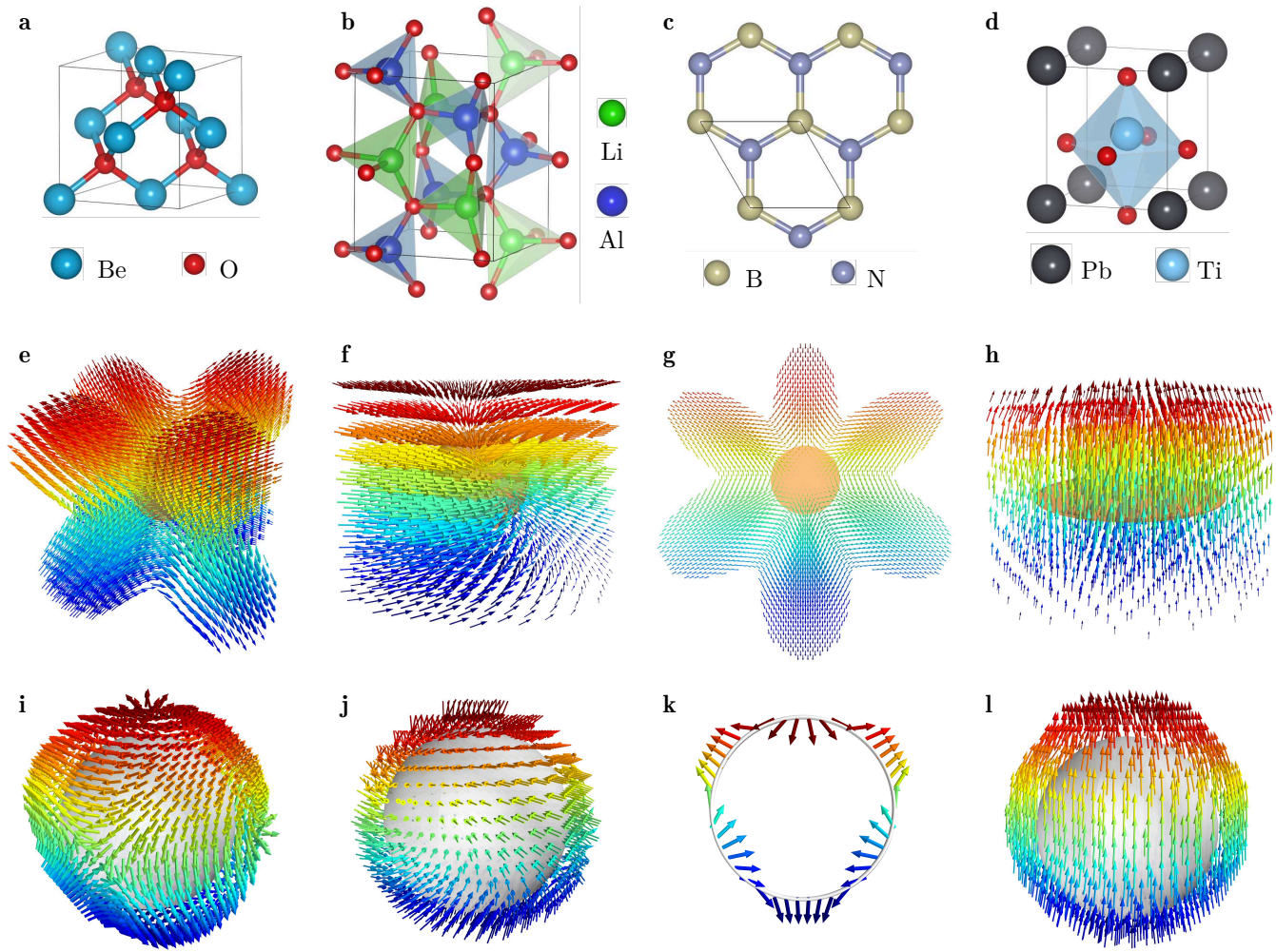
\begin{figure*}[tbh!]
    \centering
    \begin{overpic}[width=1.0\textwidth]{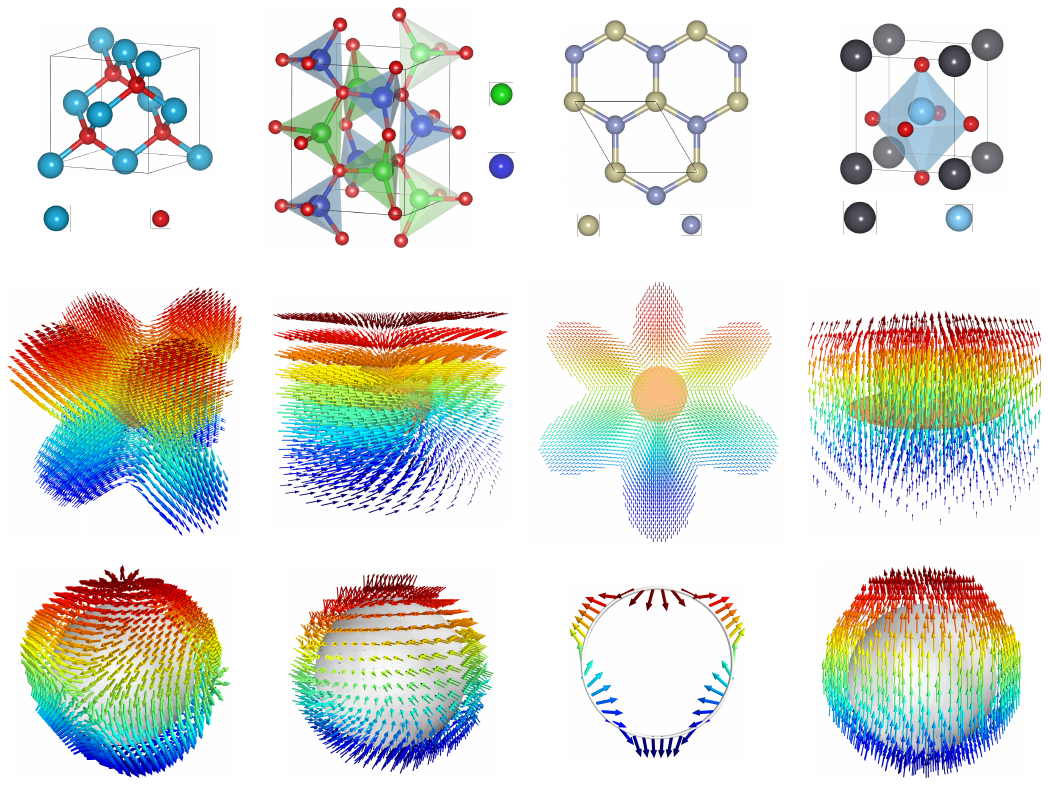}
        \put(0.3cm,13cm){\fontfig a}
        \put(4.8cm,13cm){\fontfig b}
        \put(9.3cm,13cm){\fontfig c}
        \put(13.8cm,13cm){\fontfig d}
        \put(0.3cm,8.35cm){\fontfig e}
        \put(4.8cm,8.35cm){\fontfig f}
        \put(9.3cm,8.35cm){\fontfig g}
        \put(13.8cm,8.35cm){\fontfig h} 
        \put(0.3cm,3.45cm){\fontfig i}
        \put(4.8cm,3.45cm){\fontfig j}
        \put(9.3cm,3.45cm){\fontfig k}
        \put(13.8cm,3.45cm){\fontfig l}
        \put(1.4cm,9.53cm){\fonttext Be}\put(3.1cm,9.53cm){\fonttext O}
        \put(8.43cm,11.18cm){\fonttext Li}\put(8.4cm,9.95cm){\fonttext Al}
        \put(10.45cm,9.4cm){\fonttext B}\put(12.15cm,9.4cm){\fonttext N}
        \put(15.14cm,9.52cm){\fonttext Pb}\put(16.76cm,9.52cm){\fonttext Ti}
    \end{overpic}
    \caption{
    \textbf{First-principles calculations of symmetry-protected topological polarons}. 
    \textbf{a}-\textbf{d} Ball-stick models of the conventional unit cells of 
    zb-BeO ($F\bar{4}3m$ space group),
    $\gamma$-\ce{LiAlO2} ($P4_{1}2_{1}2$),
    2D h-BN ($P3m1$),
    and \ce{PbTiO3} ($P4mm$), respectively. 
    \textbf{e}-\textbf{h} The polaron displacement field computed from first principles for each of the
    systems in the first row, in the same order. 
    The color code is a visual aid. 
    The orange ellipsoids visible in the center represent the envelope function of the electron wavefunction for 
    zb-BeO, $\gamma$-\ce{LiAlO2}, and \ce{PbTiO3}, and of the hole wavefunction for 2D h-BN.
    \textbf{i}-\textbf{l} Displacement field on a sphere centered at the polaron center with radius
    $2\,\sigma_{\rm p}$, where $\sigma_{\rm p}$ is the standard deviation obtained from the \textit{ab initio} polaron wavefunctions in Supplemental Fig.~6: 
    $\sigma_{\rm p}=$10.4\,\AA, 
    14.4\,\AA, and  
    10.8\,\AA\ for 
    zb-BeO,
    $\gamma$-\ce{LiAlO2},
    and 2D h-BN, respectively;
    $\sigma_{xy}=8.2$\,\AA\ and $\sigma_{z}=1.8$\,\AA\ 
    for \ce{PbTiO3}.
    Here we recognize the antivortex, vortex, double antivortex,
    and vertical flow fields, respectively, in the same order as in Fig.~1.
    In Supplemental Fig.~8, we perform a detailed comparison between these \textit{ab initio} results 
    and the symmetry-based polaron textures shown in Fig.~1.
    For ease of visualization, in each panel the displacement field is rendered for a single atomic species:
    O for BeO and \ce{PbTiO3}, Li for $\gamma$-\ce{LiAlO2}, and B for 2D h-BN. The displacements of the other species follow the same patterns. 
    Note that, for h-BN, we use the 2D monolayer for ease of visualization; the group-subgroup relations for 2D rosette groups are shown in Supplemental Fig.~1. 
    }\label{fig-abinitio}
\end{figure*}

\subsection*{Validation via first-principles calculations}

To validate our topological classification of polarons, we proceed to direct \textit{ab initio} calculations using the method of Ref.~\citenum{Sio2019b} (see Computational Methods). Based on the map in Fig.~\ref{fig-symmetry}(c), we consider one representative compound per class: 
(i) zinc-blende BeO [Fig.~\ref{fig-abinitio}(a)], a theoretically predicted ultra-wide-band-gap semiconductor \cite{chae2025extreme, park1999theoretical,duman2009structural} with point group $\bar{4}3m$, for which our theory predicts a 3D antivortex pattern; 
(ii) $\gamma$-\ce{LiAlO2} [Fig.~\ref{fig-abinitio}(b)], a representative oxide for nuclear fusion applications and battery cathodes with point group $422$ \cite{Liu_LiAlO2}, for which we predict a vortex-type polaron;
(iii) hexagonal BN \cite{zhang2016bandgap}, a common insulator in 2D electronics, with point group $\bar{6}m2$, for which we predict a double antivortex polaron; for ease of visualization, in Fig.~\ref{fig-abinitio}(c) we  consider instead an h-BN monolayer \cite{zhang2017hBN}, which shares the same in-plane symmetry; 
(iv) the tetragonal \ce{PbTiO3} perovskite [Fig.~\ref{fig-abinitio}(d)], a prototypical displacive ferroelectric with point group $4mm$ \cite{rabe2007physics}; for this compound, our theory predicts a vertical flow polaron texture.

For each of these crystals, we solved the \textit{ab initio} polaron equations \cite{Sio2019b} and we found large polarons with formation energies in the range 10-50\,meV (Supplemental Fig.~3). In all cases, the formation of these polarons is driven by long-wavelength phonons (Supplemental Fig.~4), and for these phonons we verified that the electron-phonon coupling matrix elements correctly satisfy crystal point-group symmetries, see Supplemental Fig.~5. In all cases, the electron or hole charge density resembles a Gaussian envelope with a standard deviation ranging between 0.8\,nm (\ce{PbTiO3}) and 1.4\,nm ($\gamma$-\ce{LiAlO2}); these solutions are shown in Supplemental Fig.~6. 

Figures~\ref{fig-abinitio}(e)-(h) show how, in each of these representative compounds, polarons exhibit precisely the displacement texture predicted by our theory: a 3D antivortex for BeO, a vortex for $\gamma$-\ce{LiAlO2}, a double antivortex for 2D h-BN, and a vertical flow for \ce{PbTiO3}. 
We note that the vertical flow pattern in \ce{PbTiO3} is aligned with the ferroelectric polarization in this compound; a detailed analysis is presented in Supplemental Fig.~7. 
Panels (i)-(l) in the same figure show the calculated displacement fields on a sphere for each of these textures. A side-by-side comparison of these patterns with the symmetry-protected fields of Fig.~\ref{fig-symmetry} is shown in Supplemental Fig.~8. The close agreement between our theoretical predictions and our explicit \textit{ab initio} calculations demonstrates the reliability of our symmetry-based approach.

\begin{figure*}[tbh!]
    \centering
    \begin{overpic}[width=0.8\textwidth]{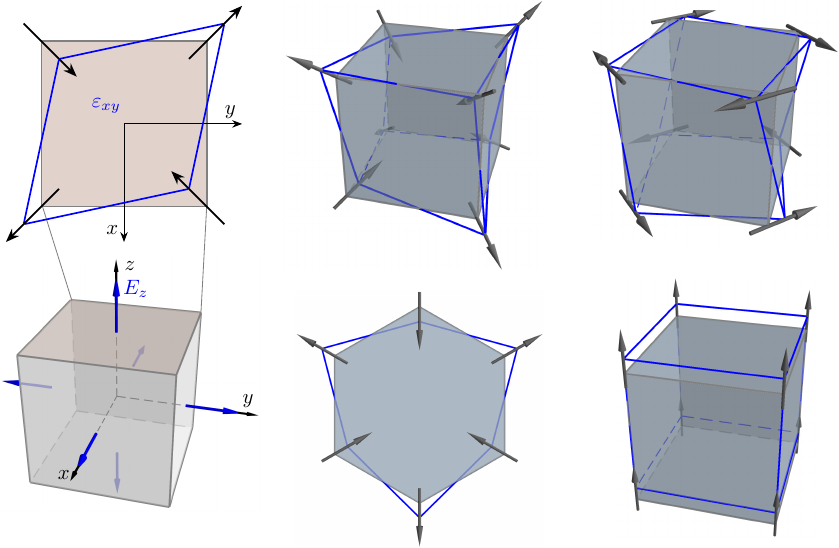}
        \put(0cm, 9.3cm){\fontfig a}
        \put(5cm, 9.3cm){\fontfig b}
        \put(10cm,9.3cm){\fontfig c}
        \put(5cm, 4.1cm){\fontfig d}
        \put(10cm,4.1cm){\fontfig e}
        \put(6.4cm, 5.0cm){\fonttext antivortex}
        \put(11.5cm,5.0cm){\fonttext vortex}
        \put(6.0cm, -0.2cm){\fonttext double antivortex}
        \put(11.1cm,-0.2cm){\fonttext vertical flow}
    \end{overpic}
    \caption{
    \textbf{Rationalizing polaron textures in terms of local strain}. 
    \textbf{a} Schematic of a cube enclosing the polaron excess charge. In a first approximation, this charge induces electric fields perpendicular to the cube faces, which generate local strains through the converse piezoelectric tensor (see Supplemental Table~4).
    With reference to the top face, and considering point group $\bar{4}m3$, the only nonvanishing component of the converse piezoelectric tensor for $E_z<0$ is $d_{xyz}>0$,
    leading to the strain $\varepsilon_{xy}<0$. This strain corresponds to a rhombohedral distortion. The cumulative distortions of all cube faces produce the pattern shown in \textbf{b}, which matches the antivortex polaron of zb-BeO in Fig.~2(i). In this panel, the unstrained cube is shown in dark blue, the strained cube is in blue, and arrows denote the displacements of each vertex.    
    \textbf{c} By repeating the same reasoning for point group $422$, the combination of the distortions of each face cause a counter-rotation of the top and bottom faces about the $z$-axis. This chiral distortion pattern matches the vortex polaron of $\gamma$-\ce{LiAlO2} in Fig.~2(j). 
    \textbf{d}, \textbf{e} Strain-induced distortion patterns allowed within the $\bar{6}m2$ and $4mm$ point groups, respectively. These patterns match the double antivortex polaron of 2D h-BN  in Fig.~2(k) and the vertical flow polaron of \ce{PbTiO3} in Fig.~2(l), respectively. 
    }\label{fig-model}
\end{figure*}

To further validate the theory, we computed the topological charge $Q$ for each of these polarons (see Computational Methods), and we found that each charge is indeed quantized and matches exactly what we predicted from symmetry analysis. We also verified that these polar textures fulfill the Poincar\'e-Hopf theorem \cite{Milnor1988g}, whereby the sum of the topological charges of a smooth vector field on a compact manifold equals its Euler characteristic, $Q_{\rm tot}\! =\! \chi$. Since the Born-von K\'arm\'an supercells used in our calculations are topologically equivalent to  hypertori ($\chi\!=\!0$), the theorem requires the total charge in the supercell to vanish. Focusing on 2D h-BN for illustration purposes, close examination of the displacement field in Supplemental Fig.~9 reveals two topological sources located in the interstitial regions between periodic images of the polaron. Each of these topological defects carries a charge $Q=+1$; furthermore, the polaron carries a charge $Q\!=\!-2$. Together, these contributions yield a total topological charge $Q_{\rm tot} \!=\! +1\! +\!1\!-\!2=0$, in agreement with the Poincar\'e-Hopf theorem. Similar considerations apply to the other systems considered here. The present findings confirm the existence of symmetry-protected, topologically nontrivial polaron textures in materials, at least at the level of atomistic first-principles calculations.

\subsection*{Analytical model of topological polaron textures}

The nontrivial polaron textures discovered here can be rationalized with a remarkably simple analytical model. Since the calculated polaron wavefunctions are relatively featureless (Supplemental Fig.~6), in a first approximation they can be described via Gaussian envelopes. The simplest polaron model that yields Gaussian wavefunctions is the Landau-Pekar model \cite{Alexandrov2012}. In this model, the Gaussian charge density of the excess electron or hole generates an electric field, which polarizes in turn the ionic lattice, establishing the atomic displacement texture. The simplest coupling term between electric field and atomic displacements, the Fr\"ohlich coupling, only allows for hedgehog-type polar textures (Supplemental Note~5). To enable more complex patterns, we must consider the next-to-leading order term, which is provided by piezoelectric couplings. 

To investigate the effect of piezoelectric couplings, we make the following observations: (i) the electric field ${\bf E}$ generates a local strain field $\varepsilon_{\alpha\beta}$ via the converse piezoelectric strain coefficients $d_{\alpha\beta\gamma}$, $\varepsilon_{\alpha\beta} = d_{\alpha\beta\gamma} E_\gamma$ \cite{nye1985physical}; (ii) the strain tensor is related to the displacement field via the standard relation $\varepsilon_{\alpha\beta}= (\partial u_\alpha/\partial r_\beta + \partial u_\beta/\partial r_\alpha)/2$; (iii) the electric field generated by a Gaussian charge density centered at $\br\!=\!0$ is linear in $\br$ near the origin; since topological properties are insensitive to continuous deformations, this field can be taken to be isotropic. By combining these three relations, the displacement field can be expressed in terms of the converse piezoelectric tensor, as we show in Supplemental Note~5:
  \begin{equation}\label{eq.disp}
      {\bf u} = (3\, d_{\alpha\beta\gamma} \!-\! d_{\beta\gamma\alpha} \!-\! d_{\gamma\alpha\beta}) \,\hat{\bf r}_\alpha r_\beta r_\gamma~.
  \end{equation}
Here, $\hat{\bf r}_\alpha$ denotes the unit vector along the direction $\alpha$. For instance, in point groups 422 and 622, the only nonvanishing tensor components are $d_{312}=-d_{321}$ \cite{Aroyo2006BilbaoCS,Aroyo:xo5013}; by using these components in \eqref{eq.disp}, we obtain the displacement field $(yz,-zx,0)$. The corresponding displacement patterns for all other groups are reported in Supplemental Table~3 and are illustrated schematically in Fig.~\ref{fig-model}; a comparison between the displacement field and the strain field is shown in Supplemental Fig.~10 for 2D h-BN. These patterns are fully consistent with our \textit{ab initio} calculations, indicating that the piezoelectric effect plays a key role in shaping these textures.

\begin{figure*}[hbt!]
    \centering
    \begin{overpic}[width=15cm]{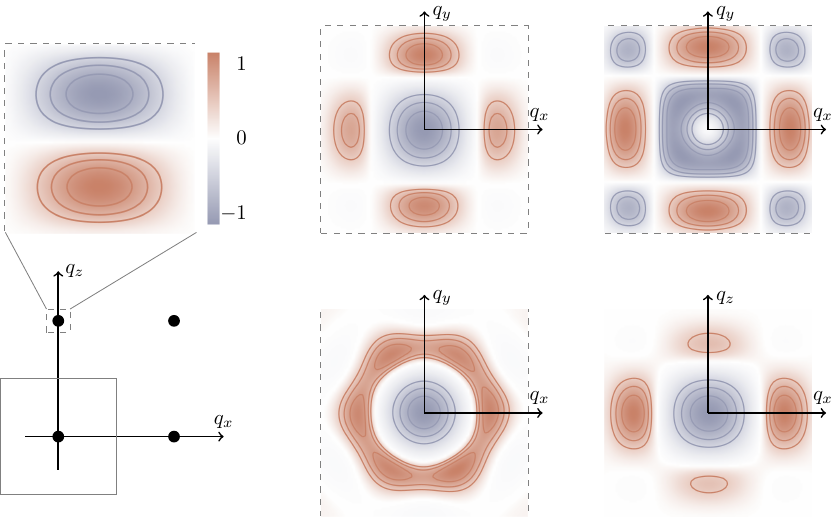}
        \put(0,8.9cm){\fontfig a}
        \put(0.15cm,8.1cm){\fonttext [001]}
        \put(2.15cm,8.65cm){\normalsize 0.1~\AA$^{-1}$}
        \linethickness{0.5mm}
        \put(1.75cm,8.55cm){\color{black}\line(1,0){1.7cm}}
        \put(5.4cm,8.9cm){\fontfig b}
        \put(5.8cm,8.4cm){\fonttext [001]}
        \put(6.8cm,4.75cm){\fonttext antivortex}
        \put(10.4cm,8.9cm){\fontfig c}
        \put(10.85cm,8.4cm){\fonttext [001]}
        \put(12.1cm,4.75cm){\fonttext vortex}
        \put(5.4cm,3.8cm){\fontfig d}
        \put(5.8cm,3.35cm){\fonttext [010]}
        \put(6.25cm,-0.3cm){\fonttext double antivortex}
        \put(10.4cm,3.8cm){\fontfig e}
        \put(10.85cm,3.35cm){\fonttext [001]}
        \put(11.75cm,-0.3cm){\fonttext vertical flow}
    \end{overpic}
    \vspace{5pt}\caption{
    \textbf{Huang diffuse scattering of topological polarons.} 
    \textbf{a} Huang diffuse scattering intensity calculated for a model hedgehog-type polaron in a simple cubic lattice, for the displacement pattern $(x,y,z)$ modulated 
    by the Gaussian profile $\exp(-r^2/2\sigma^2)$; $\sigma$ = 6~\AA\ from the electron polaron in LiF \cite{Sio2019b}. The double-drop structure is aligned with the Bragg vector [001], as shown at the bottom. Orange/blue 
    indicate normalized positive/negative intensity. 
    The solid square represents the first Brillouin zone, while the dashed square represents the region around the Bragg peaks shown in the other panels, extending from $-$0.1~\AA$^{-1}$ to 0.1~\AA$^{-1}$ along each direction. The contours are guides to the eye.
    \textbf{b}-\textbf{e} Huang diffuse scattering intensities calculated for the model antivortex [$(yz, zx, xy)$], vortex [$(yz, −xz, 0)$], 
    double antivortex [$(2xy, x^2 -y^2 , 0)$], and vertical flow polaron [$(0,0,z^2)$], respectively. The patterns are modulated by Gaussian profiles with $\sigma$ = 10~\AA~comparable to the \textit{ab initio} calculations in Fig.~2. 
    Contour lines correspond to 40\%, 60\%, and 80\% of the maximum value in each panel.   
    These idealized scattering intensities do not include thermal disorder, and correspond to the displacements of the acoustic phonons, which dominate at long timescales.
    The effect of thermal disorder and phonon contributions at short timescales are analyzed in Supplemental Fig.~11.
    }\label{fig-huang}
\end{figure*}

\subsection*{Huang diffuse scattering of topological polarons}

Experimental observation of the topological textures predicted here should be possible via Huang diffuse scattering \cite{huang1947}. When a crystal is illuminated by X-ray or electron beams, diffraction by the periodic arrangement of atoms leads to the standard Bragg peaks. Thermal fluctuations broaden these peaks by inducing dynamic distortions of the crystal lattice \cite{zacharias_multiphonon_2021,zacharias_efficient_2021}. In the presence of non-thermal distortions, such as the strain fields arising from point defects \cite{you_diffuse_1997}, extended defects \cite{larson_huang_1974}, alloying \cite{metzger_huang_1978}, quasicrystalline order \cite{de_boissieu_diffuse_1995}, and small polarons \cite{campbell_polaronic_2003}, diffraction lineshapes exhibit additional anisotropic components known as Huang scattering. These features carry the fingerprints of the underlying non-periodic crystal structures (see Computational Methods). Huang diffuse scattering has recently been employed in combination with ultrafast X-ray diffraction (UXRD) \cite{Guzelturk2021} and ultrafast electron diffraction (UED) \cite{rene_de_cotret_direct_2022,britt_momentum-resolved_2024,shi_formation_2024} to investigate the structure and dynamics of polarons in several classes of materials.

For an isolated polaron, the amplitude of Huang diffuse scattering is governed by the Fourier transform of its displacement field. As shown in Fig.~\ref{fig-huang}(a), a hedgehog-type polaron gives rise to a dipole-like pattern. This characteristic shape is well established in the crystallography of point defects \cite{defect1973}, and is commonly known as the ``double-drop'' pattern. Such a pattern was recently observed with high resolution in UED experiments on photoexcited GeS \cite{luo_ultrafast_2023}.

Figures~\ref{fig-huang}(b)-(e) show how different polarons imprint distinct diffuse scattering signatures, each carrying a unique nodal structure that is qualitatively different from the classic double-drop associated with point defects. Thermal broadening tends to smear out the fine structure of these fingerprints, but the essential features remain, as seen in Supplemental Fig.~11. These unique diffuse scattering patterns should be detectable with state-of-the-art time-resolved UXRD or UED at long timescales \cite{Puppin2020,Guzelturk2021,filippetto2022ultrafast,seiler_direct_2023}, and thus offer a potential pathway to directly observe topological polarons in materials.

\section*{Discussion}

The present discovery of several classes of symmetry-protected topological polarons establishes connections between electron-phonon physics, crystal symmetry, and topology \cite{Yu2024}. These topologically nontrivial real-space textures emerge clearly in first-principles calculations, and exhibit characteristic fingerprints that should be accessible via ultrafast Huang diffuse scattering. A potential consequence of topological polarons is that the accompanying strain fields should act as sources of pseudo-magnetic fields and Berry curvature, in analogy with the quantum Hall physics of strained graphene and related Dirac materials \cite{guinea_energy_2010,kim2015observation}. Some of these possibilities are investigated in Supplemental Note~6. 
Extension of our theory from a single polaron to multi-polaron scenarios may result in topological polar lattices in the form of polaron Wigner crystals \cite{wigner1934interaction,tanatar1989ground}. This possibility is investigated in Supplemental Fig. 12.

More generally, the present classification captures all four vectorial types of polaron textures allowed in time-reversal-invariant systems, including the recently identified ferrotoroidic fields \cite{JLB_topological_2024}. Further extending this framework to systems with broken time-reversal symmetry could reveal an even richer taxonomy, potentially bridging electron-phonon physics with ongoing efforts on electric Dzyaloshinskii-Moriya interactions \cite{zhao_dzyaloshinskiimoriya-like_2021}, magnetic skyrmions \cite{fert2017magnetic,heinze2011spontaneous}, and other magnetoelectric phenomena \cite{spaldin2019advances}. In this work we have not addressed cases where the electronic bands and phonon dispersions underlying the polaron also exhibit nontrivial topology. These cases may appear, for example, in the study of polarons in topological insulators \cite{RMP-2010,RMP-2011,zhang2009topological}, and could offer opportunities to investigate the interplay between the reciprocal-space topology of bands, phonons, or electron-phonon matrix elements and that of the real-space atomic displacement patterns. A promising class of materials where such effects may be investigated is that of correlated topological insulators \cite{Raghu2008,Lu2013}; in these materials, relatively narrow bands and heavy effective masses could favor polaron formation, while at the same time harboring quantum Hall phases where edge states interact with topological displacements patterns. 
Another interesting direction of future investigation could be to promote the classical displacement fields considered in this work to quantum fields, so as to investigate the interplay between fermionic and bosonic degrees of freedom and topology within a field-theoretic framework. This could be achieved, for example, by mapping the present displacement fields into the amplitude of coherent phonons \cite{PhysRevB.106.075119}. 
Similarly, it would be interesting to investigate whether the topology of the displacement field carries any observable imprints on the electronic wavefunctions; answering this question will require going beyond the adiabatic Born-Oppenheimer approximation, and consider coupled electron-nuclear dynamics, e.g., via exact factorization \cite{gross2010}. 
The present framework could also prove fruitful to investigate phonon vorticity around shallow impurities in semiconductors, as recently observed via atomic-resolution vibrational energy-loss spectroscopy \cite{Bao2024}.
Looking ahead, these far-reaching connections could present opportunities for engineering emergent gauge fields and electronic Berry phases through electron-phonon couplings, and thus open pathways for controlling charge, spin, and lattice degrees of freedom in quantum materials.

\section*{Materials and Methods}
All \textit{ab initio} calculations are performed using the Quantum ESPRESSO package \cite{giannozzi2017} (electronic structure and lattice vibrational properties), the Wannier90 code \cite{pizzi2020} (maximally-localized Wannier functions), the EPW code \cite{lee2023} (interpolation of electron-phonon matrix elements and polaron calculations), and the ABINIT package \cite{gonze2009abinit} (quadrupole tensors). We describe all the materials using the local density approximation (LDA) \cite{PhysRev.140.A1133}, optimized norm-conserving Vanderbilt (ONCV) pseudopotentials \cite{Hamann2013,van2018pseudodojo}, and a plane-wave basis with a kinetic energy cutoff of 100\,Ry. 
Phonon frequencies and electron-phonon matrix elements are computed within density functional perturbation theory \cite{baroni2001phonons}. Identical, unshifted uniform coarse grids of {\bf k}- and \bq-points are employed for these calculations, and are interpolated to dense grids by means of Wannier–Fourier interpolation \cite{souza-disentanglement,giustino_2007}. The methods presented in Refs.~\citenum{Verdi2015,sjakste2015wannier,Brunin2020,Park2020,Sio2022,Ponce2023} are used to handle the long-range contributions of the electron-phonon vertex, namely dipolar and quadrupolar effects. Complete computational details are available in SI Appendix.

\vspace{10pt}
\noindent\textbf{Acknowledgments}.
The authors are grateful to Jie-Cheng Chen and Tae Yun Kim for discussions. 
This research was supported by the Computational Materials Sciences Program funded by the US Department of Energy, Office of Science, Basic Energy Sciences, under award no. DE-SC0020129 (calculations and analysis). 
J.L.-B. also acknowledges Grant No. IT-1527-22, funded by the Department of Education, Universities and Research of the Basque Government, and Grant no. PID2022-137685NB-I00, funded by MCIN/AEI/10.13039/501100011033/ and by “ERDF A way of making Europe”.
This research used resources of the National Energy Research Scientific Computing Center and the Argonne Leadership Computing Facility, which are Department of Energy Office of Science User Facilities supported by the Office of Science of the US Department of Energy, under Contracts Nos. DE-AC02-05CH11231 and DE-AC02-06CH11357, respectively. We also acknowledge the Texas Advanced Computing Center at The University of Texas at Austin for providing access to Frontera, Stampede3, and Lonestar6 (http://www.tacc.utexas.edu).

\vspace{10pt}
\noindent\textbf{Author contributions}.
K.L. developed the theory, and performed calculations and data analysis. J.L.-B. contributed to calculations and data analysis. F.G. designed and supervised the project. All authors contributed to the writing of the manuscript.

\vspace{10pt}
\noindent\textbf{Competing Interests}.
The authors declare no competing interests.

\vspace{10pt}
\noindent\textbf{Code availability}.
All calculationes were performed using Quantum-ESPRESSO v7.4,\cite{giannozzi2017} EPW v5.9,\cite{lee2023} Wannier90 v3.1, \cite{pizzi2020} and ABINIT v9.10.\cite{gonze2009abinit} These codes are open-source software distributed via their corresponding websites. Raw data files and postprocessing scripts are available in the Materials Cloud Archive:  https://doi.org/10.24435/materialscloud:y1-js. 

\bibliography{refs}

\end{document}


\title{Symmetry-protected topological polarons:\\[3pt] Supplementary Materials}

\author{Kaifa Luo}
\affiliation{\utphysics}
\affiliation{\utoden}

\author{Jon Lafuente-Bartolome}
\affiliation{\ubilbao}

\author{Feliciano Giustino}
\email{fgiustino@oden.utexas.edu}
\affiliation{\utoden}
\affiliation{\utphysics}


\maketitle
\onecolumngrid

\setlength{\parindent}{0pt}
\setlength{\parskip}{2pt}

\def\br{{\bf r}}
\def\be{{\bf e}}
\def\bR{{\bf R}}
\def\bk{{\bf k}}
\def\bq{{\bf q}}
\def\bu{{\bf u}}
\def\bZ{{\bf Z}}
\def\hx{\hat{x}}
\def\hy{\hat{y}}
\def\hz{\hat{z}}
\def\hS{\hat{S}}
\def\a{{\alpha}}
\def\hbu{\hat{{\bf u}}}
\def\eps{{\epsilon}}
\def\ka{{\kappa}}
\def\sgn{\text{sgn}}

\noindent \textbf{Computational Methods}\\[6pt]
\textbf{Supplemental Notes}\\[-7pt]

\leftskip10pt
\noindent%
Supplemental Note 1: Transformation of polaron displacements under symmetry operations\\[3pt]
Supplemental Note 2: Symmetry-allowed polaron displacements\\[3pt]
Supplemental Note 3: Topological invariants and classification of polarons\\[3pt]
Supplemental Note 4: Analysis of ferroaxial polarons in inversion-symmetric groups\\[3pt]
Supplemental Note 5: Displacement fields from converse piezoelectric effect\\[3pt]
Supplemental Note 6: Pseudo-magnetic field of topological polarons \\

\leftskip0pt
\noindent%
\textbf{Supplemental Tables}\\[-7pt]

\leftskip10pt
\noindent%
Supplemental Table 1: Symmetry-allowed displacement fields by point group\\[3pt]
Supplemental Table 2: Topological invariants of displacement fields by polaron type\\[3pt]
Supplemental Table 3: Dimensionality and topological stability of piezoelectric displacement fields by polaron type\\[3pt]
Supplemental Table 4: Displacements induced by the converse piezoelectric effect by polaron type\\

\leftskip0pt
\noindent%
\textbf{Supplemental Figures}\\[-7pt]

\leftskip10pt
\noindent%
Supplemental Figure 1: Group-subgroup relations\\[3pt]
Supplemental Figure 2: Topological invariants\\[3pt]
Supplemental Figure 3: Convergence tests of polaron formation energies\\[3pt]
Supplemental Figure 4: Band and mode decomposition of \textit{ab initio} polarons\\[3pt]
Supplemental Figure 5: Symmetry of \textit{ab initio} electron-phonon matrix elements\\[3pt]
Supplemental Figure 6: Electron and hole wavefunctions of \textit{ab initio} polarons\\[3pt]
Supplemental Figure 7: Position-dependent macroscopic polarization induced by an electron polaron in \ce{PbTiO3}\\[3pt]
Supplemental Figure 8: Comparison between model polaron textures and \textit{ab initio} polarons\\[3pt]
Supplemental Figure 9: Numerical test of Poincar\'e-Hopf theorem\\[3pt]
Supplemental Figure 10: Comparison between polaron displacements and strain field in 2D h-BN\\[3pt]
Supplemental Figure 11: \textit{Ab initio} calculations of Huang diffuse scattering including thermal disorder\\[3pt]
Supplemental Figure 12: Polaron Wigner crystal in 2D h-BN\\[3pt]
Supplemental Figure 13: Electronic degeneracy and symmetry breaking\\

\clearpage
\section*{COMPUTATIONAL METHODS}

\subsection*{Computational setup for first-principles calculations}

First-principles density-functional theory calculations are performed using the Quantum ESPRESSO package \cite{giannozzi2017} (electronic structure and lattice vibrational properties), the Wannier90 code \cite{pizzi2020} (maximally-localized Wannier functions), the EPW code \cite{lee2023} (interpolation of electron-phonon matrix elements and polaron calculations), and the ABINIT package \cite{gonze2009abinit} (quadrupole tensors).

We describe zb-BeO, \ce{LiAlO2}, 2D h-BN, and \ce{PbTiO3} using the local density approximation (LDA) \cite{PhysRev.140.A1133}, optimized norm-conserving Vanderbilt (ONCV) pseudopotentials \cite{Hamann2013,van2018pseudodojo}, 
and a plane-wave basis with a kinetic energy cutoff of 100\,Ry. 
Our optimized lattice parameters are: $a=3.753$~\AA\ for zb-BeO; $a=5.093$~\AA\ and $c/a=1.209$ for $\gamma$-\ce{LiAlO2}; $a=2.488$~\AA\ for 2D h-BN; $a=3.858$~\AA\ and $c/a=1.043$ for \ce{PbTiO3}. These values are in good agreement with experimental data or prior calculations, namely
$a=3.72$~\AA\ (theory) \cite{park1999theoretical,duman2009structural,wang_phonon_2018}; $a=5.159$~\AA\ and $c/a=1.215$ (experiment) \cite{marezio1965crystal,wiedemann2016unravelling}; $a=2.50$~\AA\ (experiment) \cite{zhang2017hBN}; $a=3.904$~\AA\ and $c/a=1.063$ (experiment) \cite{shirane1956}, respectively.

Phonon frequencies and electron-phonon matrix elements are computed within density functional perturbation theory \cite{baroni2001phonons}. Identical, unshifted uniform coarse grids of {\bf k}- and \bq-points are employed for these calculations: 10$\times$10$\times$10 grids for both zb-BeO and \ce{PbTiO3}, a 10$\times$10$\times$8 grid for $\gamma$-\ce{LiAlO2}, and a 24$\times$24$\times$1 grid for 2D h-BN. Spin-orbit coupling is not included since it introduces only minor modifications to the low-energy band structures. 

Electron energies, phonon frequencies, and electron-phonon matrix elements are interpolated to denser grids by means of Wannier-Fourier interpolation \cite{giustino_2007}. 
The methods presented in Refs.~\citenum{Verdi2015,sjakste2015wannier,Brunin2020,Park2020,Sio2022,Ponce2023} are used to handle the long-range contributions of the electron-phonon vertex, namely dipolar and quadrupolar effects.
In polaron calculations, we employ 1, 8, 3, and 8 maximally-localized Wannier functions \cite{souza-disentanglement,marzari2012maximally} for zb-BeO, $\gamma$-\ce{LiAlO2}, 2D h-BN, and \ce{PbTiO3}, respectively. For the fine Brillouin zone grids in these calculations, we perform extensive convergence tests, as shown in Supplemental Fig.~\ref{figS:extrapolation}.

\subsection*{\textit{Ab initio} polaron equations}

Within the \textit{ab initio} theory of polarons developed in Refs.~\citenum{Sio2019a,Sio2019b}, the formation energy of the polaron $\Delta E_{\text{f}}$ is expressed as a self-interaction-free functional of the polaron wavefunction $\psi(\br)$ and the associated atomic displacements $\Delta\tau_{\ka\a p}$, where $\br$ is the electron or hole position and $\kappa,p,\alpha$ denote the atom, unit cell, and Cartesian direction, respectively:
\begin{equation}
    \Delta E_{\text{f}}
    =\langle\psi(\br)|\hat{H}_{\text{KS}}^{0}+\sum_{\ka\a p}\frac{\partial V_{\text{KS}}^{0}}{\partial\tau_{\ka\a p}}\Delta\tau_{\ka\a p}|\psi(\br)\rangle
    +\frac{1}{2}\sum_{\ka\a p,\ka'\a' p'}C_{\ka\a p,\ka'\a' p'}^{0}\Delta\tau_{\ka\a p}\Delta\tau_{\ka'\a' p'}\,.
\end{equation}
In this expression, $\hat{H}_{\text{KS}}^{0}$ and $V_{\text{KS}}^{0}$ are the Kohn-Sham Hamiltonian and potential, respectively, and $C_{\ka\a p,\ka'\a' p'}^{0}$ is the matrix of interatomic force constants. 
Upon variational minimization with respect to the polaron wavefunction $\psi(\br)$ and displacements $\Delta\tau_{\ka\a p}$, one finds the coupled system of equations:
    \begin{equation}\begin{aligned}\label{eq:polaron-eq1}
        &\hat{H}_{\text{KS}}^{0}\psi(\br)+\sum_{\ka\a p}\frac{\partial V_{\text{KS}}^{0}}{\partial\tau_{\ka\a p}}\Delta\tau_{\ka\a p}\psi(\br)
        =\varepsilon_{\rm p}\psi(\br)\,,\\
        &\Delta\tau_{\ka \a p}=-\sum_{\ka'\a'p'}(C^{0})_{\ka\a p,\ka'\a' p'}^{-1}\int d^{3}\br\frac{\partial V_{\text{KS}}^{0}}{\partial\tau_{\ka'\a' p'}}|\psi(\br)|^{2}\,,
    \end{aligned}\end{equation}
where $\varepsilon_{\rm p}$ is the Lagrange multiplier associated with the normalization of the wavefunction, and can be interpreted as the quasi-particle excitation energy of the polaron \cite{PhysRevB.106.075119,JLB-2022d}.

To solve the above equations, one expands the polaron wavefunction $\psi(\br)$ and the atomic displacements $\Delta\tau_{\ka\a p}$ in a basis of Kohn-Sham states $\psi_{n\bk}$ and vibrational eigenmodes $e_{\ka\a,\nu}(\bq)$ of the unperturbed crystal without excess electron or hole:
\begin{eqnarray}
        \psi(\br)&=&\frac{1}{\sqrt{N_{p}}}\sum_{n\bk}A_{n\bk}\psi_{n\bk}\,, \label{eq.elpol}\\
        \Delta\tau_{\ka\a p}&=&-\frac{2}{N_{p}}\sum_{\bq\nu}
        B_{\bq\nu}^{*}\sqrt{\frac{\hbar}{2M_{\ka}\omega_{\bq\nu}}}e_{\ka\a,\nu}(\bq)e^{i\bq\cdot\bR_{p}}\,,
        \label{eq.disp1}
    \end{eqnarray}
where $N_p$ is the number of unit cells in the Born-von K\'arm\'an supercell, $M_\kappa$ denote atomic masses, $\omega_{\bq\nu}$ are phonon frequencies, and $\bR_{p}$ are lattice vectors. 
With this change of coordinates, the system in \eqref{eq:polaron-eq1} can be recast into an equivalent system of equations for the amplitudes $A_{n\bk}$ and $B_{\bq\nu}$, which have been called the ``\textit{ab initio} polaron equations'' in Ref.~\citenum{Sio2019b}:
    \begin{equation}\begin{aligned}\label{eq:polaron-eq}
    &\varepsilon_{n{\bf k}}A_{n{\bf k}}
    -\frac{2}{N_p} \sum_{\bq m\nu} B_{\bq\nu}
    \, g_{mn\nu}^{*}({\bf k},\bq) \, A_{m{\bf k+q}}
    =\varepsilon_{\rm p}A_{n{\bf k}}\,,\\
    &B_{\bq\nu} 
    = \frac{1}{N_p}
    \sum_{mn{\bf k}} A^{*}_{m{\bf k+q}}
    \frac{g_{mn\nu}({\bf k},\bq)}{\hbar\omega_{\bq\nu}} A_{n{\bf k}}\,.
    \end{aligned}\end{equation}
In this expression, $g_{mn\nu}({\bf k},\bq )=\langle\psi_{m\bk +\bq }|\Delta_{\bq \nu} V_{\text{KS}}|\psi_{n\bk }\rangle$ represents the electron-phonon coupling matrix element.
These equations are implemented in the EPW code \cite{lee2023}.

\subsection*{Visualization of polarons}

To visualize polarons, the wavefunction is rewritten in a basis of Wannier functions as follows:
    \begin{equation}\begin{aligned}
        \psi(\br)=&\sum_{mp}A_{m}(\bR_{p})w_{m}(\br-\bR_{p})\,,
    \end{aligned}\end{equation}
where the functions $w_{m}(\br)$ are the maximally-localized Wannier functions used for electron-phonon interpolation. The coefficients $A_{m}(\bR_{p})$ in this expression are related to their counterparts in the Bloch representation by the unitary transformation: 
    \begin{equation}\begin{aligned}
        A_{m}(\bR_{p})=&\frac{1}{N_{p}}\sum_{n\bk}e^{i\bk\cdot\bR_{p}}U_{mn\bk}^{\dagger}A_{n\bk}\,.
    \end{aligned}\end{equation}
When plotted as a function of $\bR_{p}$, the quantity $|A_{m}(\bR_{p})|^{2}$ provides the envelope function of the polaron. This quantity is shown as the orange isosurfaces in Fig.~2(e)-(h) of the main text; the complete polaron wavefunctions are shown in Supplemental Fig.~6.

\subsection*{Convergence of polaron formation energy with supercell size}

To correctly describe polarons in the dilute limit (one polaron in an infinitely extended crystal), we solve the \textit{ab initio} polaron equations with increasingly denser ${\bf k}$- and $\bq$-meshes, and take the limit using the Makov-Payne scheme \cite{Makov-1995i}. These calculations are shown in Supplemental Fig.~\ref{figS:extrapolation}. In the Makov-Payne method,
the energy of a charged defect and its periodic replicas in a finite supercell is written as:
    \begin{equation}\label{eq:polaron-extra}
        \Delta E(L) = \Delta E_{\text{f}} + c_{1}L^{-1} + c_{3}L^{-3}+\mathcal{O}(L^{-5})\,,
    \end{equation} 
where $L$ is the characteristic size of the supercell, and
$\Delta E_{\text{f}}$ is the formation energy in the dilute limit $L\rightarrow\infty$ \cite{freysoldt_defect_2014}. The quantities $\Delta E_{\text{f}}$, $c_1$, and $c_3$ are obtained from least-square fitting to the calculated energies.

For future reproducibility we emphasize that, in the case of large polarons it is important to correctly sample the valley structure by making sure that high-symmetry points are included in the fine Brillouin zone grid. For example, in the case of 2D h-BN it is important to employ grids that are multiples of 3, e.g.  45$\times $45 and 90$\times $90, as shown in Supplemental Fig.~S3(c).
In this work, we perform calculations with grids reaching up to $80^3$, $48^3$, $210^2$, and $50^3$ for zb-BeO, $\gamma$-\ce{LiAlO2}, 2D h-BN and \ce{PbTiO3}, respectively. The corresponding number of atoms in the equivalent real-space supercells are 1024000 atoms, 1769472 atoms, 88200 atoms, and 625,000 atoms, respectively.

\subsection*{Calculation of topological invariants}

Analytical evaluations of topological invariants are discussed in Supplemental Note~3. For numerical evaluations, we proceed as follows.
For computational convenience, instead of a sphere centered on the polaron center, we consider a cubic surface, as illustrated in Supplemental Fig.~\ref{figS:topo}(p).
Given this cube, we evaluate the integral needed for the topological charge $Q$ [see definition in main text or in \eqref{eq.charge.def} of Supplemental Note~3] by adding up the integrals over its faces:
\begin{equation}
    Q=\frac{1}{4\pi}\left[
    \int d S_{\pm x}\bu\cdot(\partial_{y}\bu\times\partial_{z}\bu)
    +\int d S_{\pm y}\bu\cdot(\partial_{z}\bu\times\partial_{x}\bu)
    +\int d S_{\pm z}\bu\cdot(\partial_{x}\bu\times\partial_{y}\bu)
    \right]\,,
\end{equation}
where $dS_{\pm x}$, $dS_{\pm y}$, and $dS_{\pm z}$ correspond to the area elements of the faces with normal vectors along the $\pm x$, $\pm y$ and $\pm z$ axes, respectively, and $\bu(\br)$ is the normalized displacement field of the polaron. This displacement is calculated as an average over the atom of each unit cell, as described in Supplemental Note~1. 
As an example, the right panel of Supplemental Fig.~\ref{figS:topo}(p) shows the displacements on the $S_{+z}$ surface. 

The side length of the cube is taken to be $L=6\sigma_{\rm p}$, with $\sigma_{\rm p}$ being the standard deviation of the electron charge density of the polaron obtained by solving the \textit{ab initio} polaron equations. All atoms falling within the ranges $|x|,|y|<L/2$ and $|z-\delta|<L/2$ are counted as belonging to this surface; we set the ``thickness'' $\delta$ to half the lattice constant $a$.
For each face, we interpolate the function $\bu$ from discrete atomic positions to a dense $100\times100$ uniform grid via a curvature-minimizing, piecewise cubic interpolation method implemented in the ``griddata'' function of the SciPy package \cite{virtanen2020scipy}. 
The spatial derivatives of $\bu$ are computed via second-order central finite differences (``gradient'' function in the NumPy package \cite{harris2020array}). Numerical integration is performed via standard trapezoid quadrature in 2D (``dblquad'' function in SciPy). 
We checked that, by varying the side length between $5\sigma_{\rm p}$ and $7\sigma_{\rm p}$, the results do not change. 

To evaluate the vorticity $v$ and helicity $\gamma$ as a function of the polar angle $\theta$ (which is defined between 0 and $\pi$), we consider a sphere of radius $6\sigma_{\rm p}$ around the polaron center, as shown in Supplemental Fig.~\ref{figS:topo}(q). All atoms within the spherical shell $|r-\delta|< 6\sigma_{\rm p}$ are considered, and $\delta=a/2$ as for the cube surfaces described above.
The vectors $\bu$ belonging to the circle formed by intersecting this sphere with the plane $z=6\sigma_{\rm p}\cos\theta$ are used to calculate $v(\theta)$ and $\gamma(\theta)$. 

The azimuthal angle $\varphi$ of the vector field is defined as:
    $\varphi(\phi;\theta)=\arctan (u_{y}/u_{x})$, where $\theta$ is the polar angle and $\phi$ is the azimuthal angle.
When calculating $\varphi$ from this expression, one needs to pay attention to the fact that, for each solution $\varphi_0$, $\tan(\varphi)=u_{y}/u_{x}$ has additional solutions $\varphi_{0}+n\pi$, where $n$ is a integer. This can pose problems when calculating $\partial_{\phi}\varphi$ and $\varphi-\phi$ defined in \eqref{eq.vorticity.def0} and ~\eqref{eq.helicity.def0}; to overcome this issue, we add a piecewise constant $n\pi$ to $\varphi(\phi;\theta)$ so that all discontinuities are removed for $\phi$ between 0 and $2\pi$.
Once obtained the quantity $\varphi$ at discrete atomic sites, we interpolate it to a denser uniform $\phi$-grid with 5000 samples between 0 and $2\pi$, using third-order spines (``interp1d'' function in SciPy). 
The gradient $\partial_{\phi}\varphi$ is evaluated via second-order central finite differences. 
Then integrals needed for the vorticity and for the averaged helicity,
\begin{equation}
    v(\theta)=\frac{1}{2\pi}\int_{0}^{2\pi}\partial_{\phi}\varphi d\phi\,,~
    \gamma(\theta)= \frac{1}{2\pi}\int_{0}^{2\pi}(\varphi-\phi)d\phi\,, 
\end{equation}
are calculated via standard rectangular integration. 
For illustration purposes, the indices $(v,\gamma)$ of zb-\ce{BeO} and $\gamma$-\ce{LiAlO2} are shown as a function of the polar angle in Supplemental Fig.~\ref{figS:comparison}(i,j) and (k,l), respectively.

\subsection*{Calculation of Huang diffuse scattering intensity}

In crystalline solids, the amplitude for X-ray or electron scattering is given by the Fourier transform of the atomic density:
    \begin{equation}
        F^{0}({\bf Q})=\sum_{\kappa p}f_{\kappa}({\bf Q})e^{-i{\bf Q}\cdot({\bf R}_{p}+\bm{\tau}_{\kappa p}^{0}+\Delta\bm{\tau}_{\kappa p}^{\text{th}})}\,,
    \end{equation}
where $\hbar{\bf Q}$ is the momentum transferred in the scattering process, and the atomic coordinates are written as  ${\bf R}_{p}+\bm{\tau}_{\kappa p}^{0}+\Delta\bm{\tau}_{\kappa p}^{\rm th}$. $f_{\kappa}({\bf Q})$ is the atomic form factor.
The $\Delta\bm{\tau}_{\kappa p}^{\rm th}$'s represent thermal fluctuations, which we here describe using the special displacement method as implemented in the ZG code of EPW \cite{Zacharias-PhysRevLett-2015x,Zacharias-PhysRevB-2016r,zacharias_theory_2020,lee2023}; this method was shown to provide accurate Huang diffuse scattering patterns in agreement with experiments \cite{zacharias_multiphonon_2021,zacharias_efficient_2021}. 
Here, we consider thermal displacements at zero temperature to be consistent with the polaron equations which provide the ground-state polaron energy and structure. 

In the presence of a polaronic distortion given by $\Delta\bm{\tau}_{\kappa p}^{\rm P}$, the scattering amplitude is modified as:
    \begin{equation}
        F({\bf Q})=\sum_{\kappa p}f_{\kappa}({\bf Q})e^{-i{\bf Q}\cdot({\bf R}_{p}+\bm{\tau}_{\kappa p}^{0}+\Delta\tau_{\kappa p}^{\rm th}+\Delta\bm{\tau}_{\kappa p}^{\rm P})}\,.
    \end{equation}
This distortion causes the scattering amplitude to deviate from the multivariate Gaussian profile which leads to the usual thermal ellipsoids, and introduces distinctive anisotropic features and nodal structure, as is well-known for point defects \cite{trinkaus1972determination,defect1973,ehrhart_configuration_1978,you_diffuse_1997,Barabash-1999n,Stoller-2007m,Ma-2019a}.
We obtain the polaron signature in Huang diffuse scattering by taking the relative difference between the last two equations:
    \begin{equation}
        \frac{\Delta I}{I}
        =\frac{|F({\bf Q})|^{2}-|F^{0}({\bf Q})|^{2}}{|F^{0}({\bf Q})|^{2}}\,.
    \end{equation}
This expression is used to compute the patterns in Fig.~4 of the main text, as well as Supplemental Fig.~\ref{figS:Huang}. The atomic form factors $f_{\kappa}({\bf Q})$ depend on the experiment \cite{Chantler-JPhysChemRefData-2000f}; for simplicity, here we employ the data tabulated in Ref.~\citenum{peng_electron_1999} for X-ray scattering.  

We use the following calculation parameters for the figures reporting Huang scattering patterns.
Fig.~\ref{figS:Huang} of the main text: $80\times80\times80$ supercell of a simple cubic lattice for the antivortex, vortex, and vertical flow polarons; $100\times100$ supercell of a hexagonal lattice for the double antivortex polaron.
Supplemental Fig.~\ref{figS:Huang}: $74\times74\times74$, $40\times40\times40$, $210\times210\times1$, and $40\times40\times40$ supercells for zb-BeO, $\gamma$-\ce{LiAlO2}, 2D h-BN and \ce{PbTiO3}, respectively. 
The special displacements are computed within the same supercells as the polarons.
In each case, the polaron is initialized at the the center of the Wigner-Seitz supercell. 
In order to facilitate visualization, the computed scattering intensity is broadened by a normalized Gaussian function with standard deviation $3\cdot10^{-3}$\,\AA$^{-1}$.

\newpage
\section*{SUPPLEMENTAL NOTE 1: Transformation of polaron displacements under symmetry operations}

Large polarons primarily consist of coherent superpositions of low-energy electrons and holes near the valley extrema, and of long-wavelength phonons, as seen in Supplemental Fig.~4.
As a result, the symmetry of the polaron Hamiltonian and of the dynamical matrix are dictated by the little group of the valleys (for electrons and holes) and of the zone center (for phonons). The latter coincides with the crystal point group. 
These symmetries are reflected into the electron-phonon matrix elements, as shown in detail in Supplemental Fig.~\ref{figS:EPI} for all the materials considered in this work.

In the following, we analyze how crystal symmetry dictates the symmetry of the atomic displacement patterns of a large polaron.
The relation between displacements $\Delta\bm{\tau}_{\kappa p}$ and Fourier amplitudes $B_{\bq\nu}$ is given by \eqref{eq.disp1}, which we reproduce here for convenience \cite{Sio2019b}:
\begin{equation}\begin{aligned}\label{eq:disp-1}
    \Delta\bm{\tau}_{\kappa p}
    =&-\frac{2}{N_{p}}\sum_{\bq \nu}B_{\bq \nu}^{*}\sqrt{\frac{\hbar}{2M_{\kappa}\omega_{\bq \nu}}}{\bf e}_{\kappa\nu}(\bq )e^{i\bq \cdot\bR_{p}}\,.
\end{aligned}\end{equation}
To analyze the symmetry of the displacements, we investigate how $B_{\bq\nu}$ transforms under symmetry by considering the electron-phonon matrix element $g_{mn\nu}(\bk,\bq)$ in the long-wavelength limit $|\bq|\rightarrow 0$ \cite{giustino_electron-phonon_2007}. In this limit, the variation of the Kohn-Sham potential is dominated by dipole and quadrupole effects:
\begin{equation}\label{eq.tmp.1}
    \Delta_{\bq \nu} V^0_{\text{KS}}
    =\frac{e^{2}}{\Omega\eps_{0}}
    \sqrt{\frac{\hbar}{2\omega_{\bq \nu}}}
    \sum_{\kappa}
    \sum_{{\bf G}\neq-\bq }
    \frac{(\bq +{\bf G})_{\alpha}
    [iZ_{\kappa,\alpha\beta}^{*}+
    \frac{1}{2}Q_{\kappa,\alpha\beta\gamma}^{*}(\bq +{\bf G})_{\gamma}]e_{\kappa\beta,\nu}(\bq )}
    {\sqrt{M_{\kappa}}(\bq +{\bf G})_{\alpha'}\epsilon_{\alpha'\beta'}^{\infty}(\bq +{\bf G})_{\beta'}}
    e^{i(\bq +{\bf G})\cdot(\br -{\bf\tau}_{\kappa})}\,,
\end{equation}
where $Z_{\kappa,\alpha\beta}^{*}$ are the Born effective charges \cite{Vogl1976,sjakste2015wannier,Verdi2015,Sohier2016}, 
and $Q_{\kappa,\alpha\beta\gamma}^{*}$ are the dynamical quadrupoles \cite{Royo2019,Brunin2020,Park2020,ponce2021,Ponce2023,Macheda-PhysRevB-2024h}. 
${\bf G}$ denotes a reciprocal lattice vector, and $\epsilon_{\alpha\beta}^{\infty}$ is the high-frequency electronic dielectric tensor. Einstein's summation convention is implied.
The quadrupole term also contains two additional contributions arising from the Berry connection and from the variation of the Hartree and exchange-correlation potentials, but these terms have been shown to be negligible in most cases \cite{Brunin2020} and are neglected here.
To analyze the long-wavelength limit, we neglect Umklapp processes and the phase factors for each sublattice, i.e., we set ${\bf G}=0$, $e^{-i\bq\cdot\tau_{\kappa}} = 1$. Furthermore, we expand the vibrational eigenmodes near $\bq=0$ to first order in the wavevector as $e_{\kappa\alpha,\nu}(\bq)=e_{\kappa\alpha,\nu}^{(0)}(\bq)+e_{\kappa\alpha\beta,\nu}^{(1)}(\bq)q_{\beta}+\mathcal{O}(q^{2})$, where $q=|\bq|$. By making these replacements in \eqref{eq.tmp.1} we obtain the electron-phonon matrix element correct to second-order in $q$:
\begin{equation}\label{eq.g.lr}
    g_{mn\nu}(\bk ,\bq ) = g_{mn\nu}^{\rm F}(\bk ,\bq ) 
        + g_{mn\nu}^{\rm PZ}(\bk ,\bq )\,,
\end{equation}
where $g_{mn\nu}^{\rm F}(\bk ,\bq )$ and $g_{mn\nu}^{\rm PZ}(\bk ,\bq )$ describe Fr\"ohlich coupling
and piezoelectric coupling, respectively. The Fr\"ohlich term is given by [Eqs. (3.15), (3.16) in Ref.~\citenum{Vogl1976}]:
\begin{equation}\label{eq.g.frohlich}
    g_{mn\nu}^{\rm F}(\bk ,\bq )
    =i\langle u_{m\bk +\bq }|u_{n\bk }\rangle_{\rm uc}\frac{e^{2}}{\Omega\eps_{0}}
    \sqrt{\frac{\hbar}{2\omega_{\bq \nu}}}
    \frac{   
    \hat q_\alpha}{\epsilon^{\infty}(\hat \bq)\, q} \sum_{\kappa}\frac{   
     Z_{\kappa,\alpha\beta}^{*}
    e_{\kappa\beta,\nu}(\bq)
    }
    {\sqrt{M_\kappa}}\,,
\end{equation}
where $u_{n\bk }$ is the cell-periodic part of the Kohn-Sham state, $\psi_{n\bk} = N_p^{-1/2}u_{n\bk }\exp(i\bk\cdot \br)$, the overlap integral is evaluated over the unit cell, and $\epsilon^{\infty}(\hat \bq)$ is the projection of the dielectric tensor along the direction $\hat \bq = \bq/q$:
\begin{equation}
   \epsilon^{\infty}(\hat \bq) = \frac{q_{\alpha}\epsilon_{\alpha\beta}^{\infty}q_{\beta}}{q^2}\,.
\end{equation}
The piezoelectric term is given by:
\begin{equation}\label{eq.pz.tmp.1}
    g_{mn\nu}^{\rm PZ}(\bk ,\bq )
    =\langle u_{m\bk +\bq }|u_{n\bk }\rangle_{\rm uc}
    \frac{e^{2}}{\Omega\eps_{0}}
    \sqrt{\frac{\hbar}{2\omega_{\bq \nu}}}
    \sum_{\kappa}\frac{
    \hat q_{\alpha}
    \left[iZ_{\kappa,\alpha\beta}^{*} e_{\kappa\beta\gamma,\nu}^{(1)}(\bq)
    +
    \displaystyle\frac{1}{2}{Q_{\kappa,\alpha\beta\gamma}^{*}} 
    e_{\kappa\beta,\nu}^{(0)}(\bq)\right]\hat q_\gamma
    }
    {\sqrt{M_{\kappa}} \epsilon^{\infty}(\hat \bq)}\,.
\end{equation}
Since the Fr\"ohlich term diverges as $1/q$ at long wavelength while the piezoelectric term 
remains finite, for optical modes $g_{mn\nu}^{\rm F}$ dominates and we can ignore $g_{mn\nu}^{\rm PZ}$
in our symmetry analysis. Conversely, for acoustic modes $g_{mn\nu}^{\rm F}$ vanishes owing to the acoustic sum rule $\sum_\kappa Z_{\kappa,\alpha\beta}^{*} = 0$, therefore when analyzing the symmetry of acoustic modes we need only consider $g_{mn\nu}^{\rm PZ}$. 

The linear variation $e_{\kappa\beta\gamma,\nu}^{(1)}$ in \eqref{eq.pz.tmp.1} can expressed in terms of $e_{\kappa\beta,\nu}^{(0)}$ via perturbation theory \cite{born1954dynamical,Vogl1976}, so that the square bracket can be recast into $-\Omega\, {\rm e}^{\rm PZ}_{\alpha,\beta\gamma}e_{\kappa\beta,\nu}$ by using  the piezoelectric tensor ${\rm e}^{\rm PZ}_{\alpha,\beta\gamma}$ \cite{Martin1972,Vanderbilt-2000g,Wu2005}. With this replacement, \eqref{eq.pz.tmp.1} takes the compact form:
\begin{equation}\label{eq.g.pz3}
    g_{mn\nu}^{\rm PZ}(\bk ,\bq )
    =-\langle u_{m\bk +\bq }|u_{n\bk }\rangle_{\rm uc}
    \frac{e^{2}}{\epsilon_0}\sqrt{\frac{\hbar}{2\omega_{\bq \nu}}}
    \frac{\hat q_{\alpha}{\rm e}^{\rm PZ}_{\alpha,\beta\gamma}  
          e_{\kappa\beta,\nu}(\bq)
          \hat q_{\gamma}}
    {\sqrt{M_{\kappa}}\epsilon^{\infty}(\hat \bq)}~,
\end{equation}
which coincides with Mahan's phenomenological model [see Eq.~(4.14) of Ref.~\citenum{Vogl1976} or Eq.~(3.30) of Ref.~\citenum{Yu-2010t}].  

By inserting the long-range electron-phonon matrix elements from \eqref{eq.g.lr}, \eqref{eq.g.frohlich}, and \eqref{eq.g.pz3} into the second line of \textit{ab initio} polaron equations, \eqref{eq:polaron-eq}, the displacements of \eqref{eq:disp-1} can be expressed in terms of macroscopic tensors, phonon frequencies and eigenmodes, and the electronic charge density of the polaron:
\begin{equation}\label{eq.disp.tmp.1}
  \Delta\bm{\tau}_{\kappa}(\bR_{p}) =
     -\frac{2}{N_{p}}\sum_{\bq\nu}
     \sqrt{\frac{\hbar}{2M_{\kappa}\omega_{\bq \nu}}}
     {\bf e}_{\kappa\nu}(\bq )e^{i\bq \cdot\bR_{p}} \big(\alpha^{\rm F}_{\bq\nu}+\alpha^{\rm PZ}_{\bq\nu}\big)^*
     \frac{1}{N_{p}} \sum_{mn\bk} A_{m\bk+\bq}A_{n\bk}^*
          \langle u_{n\bk }|u_{m\bk +\bq }\rangle_{\rm uc}~,
\end{equation}
where the dimensionless coupling constants are defined as $\alpha^{\rm F,PZ}_{\bq\nu} = g_{mn\nu}^{\rm F,PZ}(\bk ,\bq ) / (\hbar\omega_{\bq\nu} \langle u_{m\bk +\bq }|u_{n\bk }\rangle_{\rm uc})$ and are independent of the electron indices. Using \eqref{eq.elpol}, the sum on the right hand side can be recast into:
\begin{equation}
     \frac{1}{N_{p}} \sum_{mn\bk} A_{m\bk+\bq}A_{n\bk}^*
          \langle u_{n\bk }|u_{m\bk +\bq }\rangle_{\rm uc} =
     \int_{\rm sc}d\br\, e^{-i\bq\cdot \br} 
         |\psi(\br)|^2~, 
\end{equation}
where the integral is over the Born-von K\'arm\'an supercell. This quantity corresponds to the Fourier transform of the electron density of the polaron, which we call $n(\bq)$ for brevity. Using this more compact expression, \eqref{eq.disp.tmp.1} becomes:
\begin{equation}\label{eq.disp.final}
  \Delta\bm{\tau}_{\kappa}(\bR_{p}) =
     -\frac{2}{N_{p}}\sum_{\bq\nu}
     \sqrt{\frac{\hbar}{2M_{\kappa}\omega_{\bq \nu}}}
     {\bf e}_{\kappa\nu}(\bq )e^{i\bq \cdot\bR_{p}} \big(\alpha^{\rm F}_{\bq\nu}+\alpha^{\rm PZ}_{\bq\nu}\big)^*
     \,n(\bq)~.
\end{equation}
In order to recover Eq.~(1) of the main text, we need to transform these atomic displacements into cell-averaged displacements. It is convenient to distinguish two cases, the position-dependent center-of-mass displacement and the position-dependent macroscopic polarization.

\subsection*{Position-dependent center-of-mass displacement} 

The position-dependent center-of-mass displacement is the mass-weighted average of the atomic displacements in the unit cell: $\Delta{\bm{\tau}}(\bR_{p}) = {\sum}_{\kappa}M_{\kappa}\Delta\bm{\tau}_{\kappa}(\bR_{p}) / M_{\rm uc}$, where $M_{\rm uc} = {\sum}_{\kappa}M_{\kappa}$. This displacement is relevant for polarons driven primarily by piezoelectric coupling. 

By considering only $\alpha^{\rm PZ}_{\bq\nu}$ in \eqref{eq.disp.final}, and using \eqref{eq.g.pz3}, the average displacement can be written as:
\begin{equation}
   \Delta{\tau}_\alpha(\bR_{p}) = \frac{e^{2}}{\epsilon_0 M_{\rm uc}}
     \frac{1}{N_{p}}\sum_{\bq} \frac{n(\bq)}{\epsilon^{\infty}(\hat \bq)} e^{i\bq \cdot\bR_{p}} 
    \hat q_{\beta}\,{\rm e}^{\rm PZ}_{\beta,\gamma\delta}\, \hat q_{\delta} 
     \sum_{\nu} \frac{1}{\omega_{\bq \nu}^2}\sum_{\kappa} 
     e_{\kappa\alpha,\nu}(\bq ) e_{\kappa\gamma,\nu}(\bq\!\rightarrow \!0)~.
\end{equation}
This expression is convenient to inspect the symmetry of $\Delta{\tau}_\alpha(\bR_{p})$.
To this end, we evaluate the displacements at $\hat S \bR_p$ and we multiply both sides by
$S^{-1}_{\a'\a}$. After some algebra we find: 
\begin{equation}\label{eq.transf.tmp.1}
   S^{-1}_{\a'\a}\Delta{\tau}_\alpha(\hat S \bR_{p}) = \frac{e^{2}}{\epsilon_0 M_{\rm uc}}
     \frac{1}{N_{p}}\sum_{\bq} \frac{n(\hat S \bq)}{\epsilon^{\infty}(\hat S\hat \bq)} e^{i\bq \cdot\bR_{p}} 
    S_{\beta\beta'} \hat q_{\beta'}\, {\rm e}^{\rm PZ}_{\beta,\gamma\delta} \,S_{\delta\delta'} \hat q_{\delta'} 
    \sum_{\nu} \frac{1}{\omega_{\hat S\bq \nu}^2}\sum_{\kappa}
     S^{-1}_{\a'\a} e_{\kappa\alpha,\nu}(\hat S \bq ) e_{\kappa\gamma,\nu}(\hat S\bq\rightarrow 0)~.
\end{equation}
Since the dielectric tensor has the full point group symmetry of the crystal, we have
$S^{-1}_{\alpha\alpha'}\epsilon^\infty_{\alpha'\beta'} S_{\beta'\beta} = 
\epsilon^\infty_{\alpha\beta}$, which implies $\epsilon^{\infty}(\hat S\hat \bq) = \epsilon^{\infty}(\hat \bq)$. Similarly, the piezoelectric tensor transforms as $S_{\alpha'\alpha}S_{\beta'\beta}S_{\gamma'\gamma}{\rm e}^{\rm PZ}_{\alpha',\beta'\gamma'} 
={\rm e}^{\rm PZ}_{\alpha,\beta\gamma}$. Furthermore, we have $\omega_{\hat S\bq \nu} = \omega_{\bq \nu}$ \cite{maradudin_symmetry_1968}. Finally, in the small $\bq$ limit, which is relevant for the long-wavelength couplings considered here, the transformation law of the vibrational eigenmodes is 
$e_{\kappa\alpha,\nu}(\hat S \bq ) = 
  S_{\alpha\beta}  \delta_{\kappa,\hat S(\kappa')}  
  e_{\kappa'\beta,\nu}(\bq )
$; 
$\hat S(\kappa')$ represents the atom that $\kappa'$ is brought into by the symmetry operation, see Eq.~(3.9) of Ref.~\citenum{maradudin_symmetry_1968} or Eq.~(40) of Ref.~\citenum{giustino_2007}. Using these rules in \eqref{eq.transf.tmp.1}, we obtain:
\begin{equation}\label{eq.transf.tmp.2}
   S^{-1}_{\a'\a}\Delta{\tau}_\alpha(\hat S \bR_{p}) = 
   \Delta{\tau}_\alpha(\bR_{p})
   \quad \mbox{if and only if} \quad 
     |\psi(\hat S \br)|^2=|\psi(\br)|^2~,
\end{equation}
having noted that the transformation law of $n(\bq)$ is the same as that of the electronic charge density of the polaron.
Therefore, if the polaron wavefunction has the full point group symmetry of the crystal, the displacements transform as in Eq.~(1) of the main text. This is the case when the electronic Fourier components are concentrated around a single valley at the zone center, or are equally distributed among symmetry-related valleys away from the zone center. In other cases, Eq.~(1) remains valid but the relevant point group is the little group of the valleys, see for example Supplemental Fig.~\ref{figS:SSB-BeO}.

\subsection*{Position-dependent macroscopic polarization} 

When the polaron is primarily driven by Fr\"ohlich coupling, the position-dependent center-of-mass displacement vanishes identically. This can be shown by combining \eqref{eq.disp.final} and \eqref{eq.g.frohlich}, the sum rule $\sum_\kappa Z_{\kappa,\alpha\beta}^{*} = 0$, and the fact that for longitudinal optical (LO) modes we can replace $\omega_{\bq\nu}=\omega_{\rm LO} = \mbox{const}$.

In these cases, we characterize the polaronic distortion via a related quantity, the position-dependent macroscopic polarization. This is defined as the sum of the microscopic electric dipoles arising in the unit cell. Following Ref.~\citenum{Verdi2015}, we write the position-dependent macroscopic polarization as ${\bf p}(\bR_{p}) = \Omega^{-1}e{\sum}_{\kappa}\,{\bf Z}_{\kappa}^{*}\cdot\Delta\bm{\tau}_{\kappa}(\bR_{p})$. Using \eqref{eq.g.frohlich}, and \eqref{eq.disp.final}, this quantity becomes:
\begin{equation}\label{eq.pol.cell}
  p_\alpha(\bR_{p}) = \frac{e}{\Omega}\big(\epsilon^{0}_{\alpha\beta}-\epsilon^\infty_{\alpha\beta}\big)\frac{1}{N_{p}}\sum_{\bq} 
  \frac{i q_\beta n(\bq)}{ \epsilon^{\infty}(\hat \bq)\, q^2}e^{i\bq \cdot\bR_{p}}~,
\end{equation}
having used the relation between Born effective charges and dielectric tensors, see Eq.~(52) of Ref.~\citenum{Gonze1997} or Eq.~(6) of Ref.~\citenum{Sio2022} for S.I. units.
This expression carries the intuitive meaning of the dipole potential generated by the polarization
cloud induced by $|\psi|^2$. 
Using \eqref{eq.pol.cell}, we can readily check the symmetry of the position-dependent macroscopic polarization. To this end, we evaluate $p_\alpha$ at $\hat S \bR_p$ and we multiply both sides by
$S^{-1}_{\delta\a}$. After a change of summation variable, we find: 
\begin{equation}\label{eq.pol.cell2}
  S^{-1}_{\delta\a} p_\alpha(\hat S\bR_{p}) = \frac{e}{\Omega}S^{-1}_{\delta\a}\big(\epsilon^{0}_{\alpha\beta}-\epsilon^\infty_{\alpha\beta}\big) S_{\beta\gamma}\frac{1}{N_{p}}\sum_{\bq} 
  \frac{i q_\gamma n(\hat S\bq)}{ \epsilon^{\infty}(\hat S\hat \bq)\, q^2}e^{i\bq \cdot\bR_{p}}~.
\end{equation}
Since the dielectric tensors have the full point group symmetry of the crystal, we have
$S^{-1}_{\delta\a}\big(\epsilon^{0}_{\alpha\beta}-\epsilon^\infty_{\alpha\beta}\big) S_{\beta\gamma} = 
\epsilon^{0}_{\delta\gamma}-\epsilon^\infty_{\delta\gamma}$, and
$\epsilon^{\infty}(\hat S\hat \bq) = \epsilon^{\infty}(\hat \bq)$. Therefore, as in \eqref{eq.transf.tmp.2}, we find the transformation law:
\begin{equation}
  {\bf p}(\hat S \bR_{p}) =
  \hat S {\bf p} \quad \mbox{if and only if} \quad 
     |\psi(\hat S \br)|^2=|\psi(\br)|^2~,
\end{equation}
and the same considerations as above about the valley structure apply.
In the discussion of the hedgehog-type polaron in the main text, the average displacement is defined as $\bu(\br)={\bf p}/e$. 

\newpage
\section*{SUPPLEMENTAL NOTE 2: Symmetry-allowed polaron displacements}

We provide a detailed derivation of the displacements fields that one obtains from Eq.~(1) of the main text. In particular, we derive the transformation laws for the displacement fields under the symmetry operations of the point groups that support symmetry-protected topological polarons. 

For conciseness, here we focus on the antivortex and double antivortex polaron textures. The other textures are obtained following the same reasoning, and are reported in Supplemental Table~\ref{table:vector-fields} for completeness.

\begin{figure*}
    \centering
    \subfigure{
        \begin{overpic}[width=0.70\textwidth,page=1]{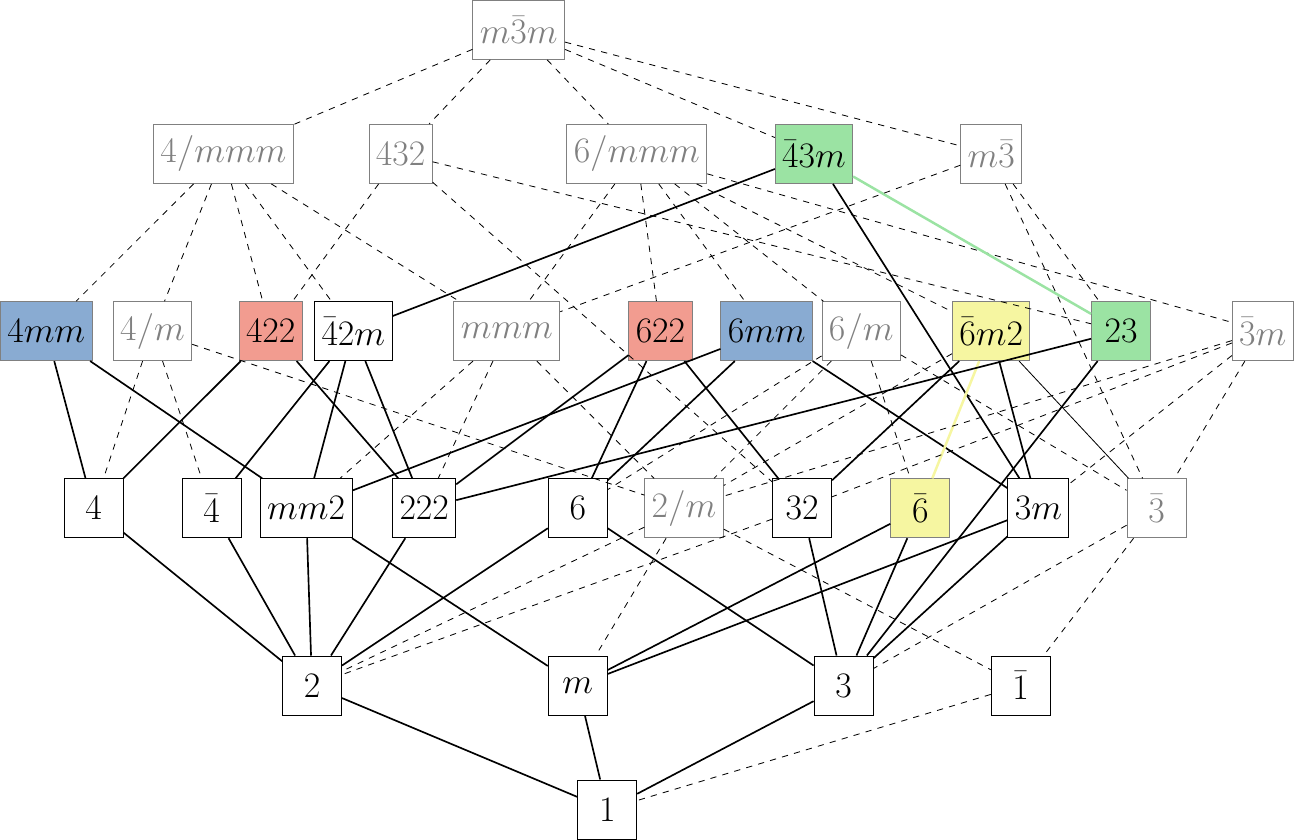}
        \put(0cm,7cm){\fontfig a}
        \end{overpic}}
    \hspace{1cm}
    \subfigure{
        \begin{overpic}[width=0.22\textwidth,page=2]{figS1.pdf}
        \put(0cm,7cm){\fontfig b}
        \end{overpic}}
    \caption{\justifying\textbf{Group-subgroup relations}.
    Group–subgroup hierarchies used to identify symmetry-protected topological polarons. 
Groups in black boxes connected by black lines are piezoelectric and are included in Fig.~1 of the 
main text; groups in gray boxes connected by
gray dashed lines are non-piezoelectric, and are excluded from Fig.~1 of the main text.
\textbf{a}, Three-dimensional point groups. The colors indicate symmetry groups that
protect distinct classes of polaron displacement fields: antivortex (green), vortex (red), double antivortex (yellow),
and vertical flow (blue). The root group is the cubic $m\bar{3}m$ group.
\textbf{b}, Two-dimensional rosette subgroups of the $6mm$ and $4mm$ groups. Only the groups $3m1/31m$ and $3$ support symmetry-protected displacements in 2D, and correspond to the double antivortex field (yellow).}
    \label{fig:group}
\end{figure*}

In the main text, we state that the antivortex polaron is symmetry-protected in the point groups $23$ and $\bar{4}3m$, see Fig.~\ref{fig:group}. To prove this statement, we first consider the point group $23$, and then proceed to its supergroup $\bar{4}3m$. The point group $23$ is generated by three proper rotations: $\hat{R}_{2z}$ and $\hat{R}_{2y}$, which are twofold (180$^\circ$) rotations around the $z$ and $y$ axis, respectively; and $\hat{R}_{3,111}$, which is a threefold (120$^\circ$) rotation around the [111] direction. 
Under these symmetry operations, a slowly-varying vector field ${\bf u}(\br )$ that vanishes at the origin transforms as follows:
    \begin{equation}\label{eq:sym-eq1}
        \bu(\hat{R}\,\br)=\hat{R}\,{\bf u}(\br), \quad \mbox{for}\quad \hat R = \hat R_{2z}, R_{2y}, \hat{R}_{3,111}\,.
    \end{equation}
Given the representation of the above symmetry operations in Cartesian coordinates:
    \begin{equation}
        \hat{R}_{2z}=\begin{pmatrix}
                    -1 &  0 & 0\\
                     0 & -1 & 0\\
                     0 &  0 & 1\\
                    \end{pmatrix}\,, \quad
        \hat{R}_{2y}=\begin{pmatrix}
                    -1 & 0 & 0\\
                     0 & 1 & 0\\
                     0 & 0 & -1\\
                    \end{pmatrix}\,, \quad
        \hat{R}_{3,111}=\begin{pmatrix}
                     0 & 1 & 0\\
                     0 & 0 & 1\\
                     1 & 0 & 0\\
                    \end{pmatrix}\,,
    \end{equation}
the transformation laws in \eqref{eq:sym-eq1} translate into:
    \begin{equation}\begin{aligned}
        &[u_{x}(-x,-y,z),u_{y}(-x,-y,z),u_{z}(-x,-y,z)]=[-u_{x}(x,y,z),-u_{y}(x,y,z),u_{z}(x,y,z)]\,,\\
        &[u_{x}(-x,y,-z),u_{y}(-x,y,-z),u_{z}(-x,y,-z)]=[-u_{x}(x,y,z),u_{y}(x,y,z),-u_{z}(x,y,z)]\,,\\
        &[u_{x}(y,z,x),u_{y}(y,z,x),u_{z}(y,z,x)]=[u_{y}(x,y,z),u_{z}(x,y,z),u_{x}(x,y,z)]\,.
    \end{aligned}\end{equation}
Near the origin, the displacement field is a polynomial. The only polynomials that simultaneously satisfy the above rules to second order in the coordinates are of the form:
    \begin{equation}\begin{aligned}\label{eq:antivortex}
    {\bf u}(\br) = a (x,y,z) + b(yz,zx,xy)\,,
    \end{aligned}\end{equation}
which corresponds to a superposition of the hedgehog pattern and the antivortex pattern. Moving to $\bar{4}3m$, this supergroup can be obtained from $23$ by acting on all its 12 symmetries with the mirror operation $\hat{M}_{110}$, which is a reflection around the plane $x+y=0$:
    \begin{equation}
        \hat{M}_{110}=\begin{pmatrix}
                     0 &  -1 & 0\\
                     -1 &  0 & 0\\
                     0 &  0 & 1\\
                    \end{pmatrix}\,, 
    \end{equation}
Using this expression in \eqref{eq:sym-eq1}, we obtain:
    \begin{equation}\begin{aligned}
        &[u_{x}(-y,-x,z),u_{y}(-y,-x,z),u_{z}(-y,-x,z)]=[-u_{y}(x,y,z),-u_{x}(x,y,z),u_{z}(x,y,z)]\,.
    \end{aligned}\end{equation}
It is immediate to verify that the field specified by \eqref{eq:antivortex} fulfills this relation, therefore the antivortex polaron is also allowed in the supergroup $\bar{4}3m$.

By repeating the same reasoning for the other point groups shown in Fig.~\ref{fig:group}(a), we obtain all possible symmetry-protected polaron displacement fields discussed in the main text and summarized in Supplemental Table~\ref{table:vector-fields}. The 12 groups shown in light gray in Fig.~\ref{fig:group} do not admit quadratic displacements since they contain inversion symmetry, $\hat P$, and therefore do not host nontrivial polaron patterns. Of the remaining 20 groups, 6 groups admit a unique type of vector field and therefore offer symmetry protection: $\bar 43m$ and $23$ (green), leading to the antivortex pattern; 
$422$ and $622$ (red), leading to the vortex pattern;
$\bar 6m2$ and $\bar 6$ (yellow) leading to the double antivortex pattern; and
$4mm$ and $6mm$ (blue), leading to the vertical flow pattern.

In the case of 2D materials, such as 2D h-BN considered in the main text, we can perform a classification of polarons along the same lines. The relevant 2D point groups, the 10 rosette groups, are shown in Fig.~\ref{fig:group}(b). 
In this case, the twofold rotation around the $z$ axis, $\hat{R}_{2z}$, plays the same role of the inversion operation $\hat P$ in 3D crystals: 6 rosette groups contain this operation and hence do not admit quadratic displacements. Of the 4 remaining groups, only two ($3m1/31m$ and $3$) admit unique symmetry-protected displacements.

For completeness, we derive the symmetry-allowed vector fields in the $31m$ rosette group of 2D h-BN. This group is generated by a threefold rotation $\hat{R}_{3z}$ and a mirror $\hat{M}_{y}$, as follows:
    \begin{equation}\begin{aligned}
        \hat{R}_{3z}=\begin{bmatrix}
                    \cos (2\pi/3) & -\sin (2\pi/3) \\
                    \sin (2\pi/3) & \phantom{-}\cos (2\pi/3)
                    \end{bmatrix} = -\frac{1}{2}\begin{pmatrix}
                    \phantom{-}1 & \sqrt{3} \\
                    -\sqrt{3} & 1
                    \end{pmatrix}\,, \qquad
        \hat{M}_{x}=\begin{pmatrix}
                    -1 & \phantom{-}0 \\
                     0 & 1
                    \end{pmatrix}\,.
    \end{aligned}\end{equation}
By using these operations in \eqref{eq:sym-eq1}, we obtain the following transformation laws:
    \begin{equation}\begin{aligned}
        &[u_x(-x/2-\sqrt{3}y/2,\sqrt{3}x/2-y/2),u_y(-x/2-\sqrt{3}y/2,\sqrt{3}x/2-y/2)]\\
        & \hspace{20pt}=
        [-u_x(x,y)/2-\sqrt{3}u_y(x,y)/2,\sqrt{3}u_x(x,y)/2-u_y(x,y)/2],
        \\
        &[u_x(-x,y),u_y(-x,y)]=[-u_x(x,y),u_y(x,y)]\,,
    \end{aligned}\end{equation}
which lead to the following symmetry-allowed vector field (up to quadratic order in the coordinates):    
    \begin{equation}\begin{aligned}
    {\bf u}(\br )
    =& a (x,y) + b (2xy, x^{2}-y^{2}).
    \end{aligned}\end{equation}
The linear term is the hedgehog-type displacement pattern, and the quadratic term represents
the double antivortex, as shown for 2D h-BN in Fig.~2(g) of the main text.

The above considerations apply to those cases where the electronic part of the polaron draw weight primarily from the $\Gamma$ point. When the band structure exhibits multiple symmetry-related valleys away from $\Gamma$, it is possible to find solutions to the \textit{ab initio} polaron equations where only a subset of these valleys contributes to the polaron wavefunction. In these cases, the above symmetry analysis continues to hold, but one needs to replace the crystal point group by the little group of the valley.

To illustrate this point, we consider zb-BeO. This system exhibits three inequivalent $X$-point valleys; when the polaron wavefunction resides on all three valleys, we have a fully symmetric polaron that follows the symmetry rules outlined above, see Supplemental Fig.~\ref{figS:SSB-BeO}(b).
However, a small amount of symmetry breaking, as arising for example from numerical noise, can lead to three polaron solutions that do not possess the $R_{3,111}$ symmetry, as shown in Supplemental Fig.~\ref{figS:SSB-BeO}(c)-(e). These solutions belong to the group $\bar{4}m2$, which is precisely the little group of the $X$ valleys and a subgroup of $\bar{4}m3$. These symmetry-breaking solutions can be found by initializing the trial polaron wavefunctions to have electronic weights concentrated in selected valleys.

\newpage
\section*{SUPPLEMENTAL NOTE 3: Topological invariants and classification of polarons}

We derive the topological invariants of the five classes of polaron vector fields discussed
in the main text. The following analysis is summarized in Supplemental Table~\ref{table:topo}.

Given a smooth and normalized vector field $\hat{\bf u}(\br) = {\bf u}({\bf r})/|{\bf u}({\bf r})|$, we consider its topological
charge $Q$, vorticity $v$, and helicity $\gamma$. For completeness, we reproduce the
definitions of these quantities given in the main text. The topological charge $Q$ is the 
flux of the topological density $\bm\Omega$ through a closed surface $S$ enclosing a topological
singularity, e.g., the center of the polaron where the normalized displacement is ill-defined \cite{Volovik-1987f,nagaosa2013topological,Everschor-Sitte-JApplPhys-2014g}:
    \begin{equation}\label{eq.omega.def}
        \bm{\Omega}_{\alpha}
        =\frac{1}{2}\epsilon_{\alpha\beta\gamma}\,
        \hat{\bu}\cdot(\partial_{\beta}\hbu\times\partial_{\gamma}\hbu)\,,
    \end{equation}
    \begin{equation}\label{eq.charge.def}
        Q=\frac{1}{4\pi}\int_{S}\bm{\Omega}\cdot d{\bf S}\,,
    \end{equation}
with $d{\bf S}$ being the surface element. The density $\bm\Omega$ quantifies the local winding of the vector field, and $Q$ counts
how many times this field wraps around the singularity, i.e. the total solid angle spanned by 
$\hat{\bf u}({\bf r})$ as $\bf r$ spans $S$.
The vorticity $v$ of this field is defined on a 2D plane cut, which we take to be a plane of constant $z$ for
definiteness:
\begin{equation}\label{eq.vorticity.def0}
    v = \frac{1}{2\pi}\oint_C \nabla \varphi \cdot d{\bf l}~, \qquad \varphi = \tan^{-1}[u_y({\bf r})/ u_x({\bf r})]~.
\end{equation}
Here, the integral is along a loop $C$ at constant $z$ enclosing the singularity, and $d{\bf l}$ is the line element. $v$ counts how many times the field rotates in the plane as ${\bf r}$ completes one loop over $C$.
The same path $C$ is used to determine the helicity $\gamma$, which is the average angle between the vector field and the radial direction over the loop:
\begin{equation}\label{eq.helicity.def0}
       \gamma = \frac{1}{2\pi}\oint_C (\varphi-\phi)~,
       \qquad \phi = \tan^{-1}(y/ x)\,.
\end{equation}

\vspace{3pt}
\textit{Hedgehog vector field ($m\bar{3}m$ point group)} 

This field is shown in Supplemental Fig.~\ref{figS:topo}(a) and is defined by the radial vector:
    \begin{equation}\label{eq:berry-monopole1}
        \hat{\bf u} =\sgn(a)\,\hat{\bf r}\,,
    \end{equation}
where $a$ is a material-specific constant as discussed in Supplemental Note~2, and $\hat{\bf r}={\bf r}/|{\bf r}|$.
We compute the topological density using the definition in \eqref{eq.omega.def}. Using the identity $(\partial_\beta \hat{\bf u})_\alpha =  (r^2 \delta_{\alpha\beta}-r_\alpha r_\beta)/r^3$, the expression for the triple product in terms of the Levi-Civita symbol, and the identities $\epsilon_{\alpha\beta\gamma}\epsilon_{\alpha'\beta\gamma}= 2\delta_{\alpha\alpha'}$, $\epsilon_{\alpha\beta\gamma}\delta_{\alpha\beta}=0$, and $\epsilon_{\alpha\beta\gamma}=-\epsilon_{\beta\alpha\gamma}$, we find:
    \begin{equation}\label{eq.hedgehog}
        \bm{\Omega}
        =\sgn(a)\,\frac{\hat{\bf r}}{r^2}\,.
    \end{equation}
Using this expression inside \eqref{eq.charge.def}, and choosing $S$ to be a sphere centered around ${\bf r}=0$, we obtain immediately $Q=\sgn(a)$. This charge is shown in Supplemental Fig.~\ref{figS:topo}(f) as a function of the parameter $a$. 

To determine the vorticity, we use \eqref{eq.vorticity.def0} and choose the loop $C$ to be a circle at constant $z$. Using spherical polar coordinates, we have $x=r \sin\theta \cos\phi$, $y=r\sin\theta \sin\phi$, and $d{\bf l} = r\sin\theta (-\sin \phi, \cos\phi) d\phi$. Furthermore, from \eqref{eq:berry-monopole1} we have $\varphi = \tan^{-1}(y/x)$, therefore
$\nabla \varphi = (-\sin\phi,\cos\phi)/r \sin\theta$. The line integral becomes: 
\begin{equation}
    v = \frac{1}{2\pi}\int_{0}^{2\pi} \frac{(-\sin\phi,\cos\phi)}{r \sin\theta} \cdot r\sin\theta (-\sin \phi, \cos\phi) d\phi = 1\,.
\end{equation}
For the hedgehog field, the angles $\varphi$ and $\phi$ coincide, therefore the helicity defined in \eqref{eq.helicity.def0} vanishes identically: $\gamma=0$. Supplemental Fig.~\ref{figS:topo}(k) shows the vorticity and helicity thus obtained, as a function of the polar angle $\theta$.

\vspace{3pt}
\textit{Antivortex vector field ($\bar{4}3m$ and $23$ point groups)} 

The antivortex field prior to normalization is given by (see Supplemental Table~\ref{table:vector-fields}):
    \begin{equation}\label{eq.Berry.anti}
        \bu
        =\frac{1}{r^2}\left[a(x,y,z)+b(yz,zx,xy)\right]\,,
    \end{equation}
with $a$ and $b$ being materials-specific constants. By performing similar steps as for the hedgehog field above, we obtain the topological density: 
    \begin{equation}\label{eq.Berry.anti2}
        \bm{\Omega}
        =\frac{1}{|r^2 \bu|^3}
        \left[
        (a x-b y z)(a^2 - b^2 x^2),~
        (a y-b z x)(a^2 - b^2 y^2),~
        (a z-b x y)(a^2 - b^2 z^2)\right]\,,
    \end{equation}
In the limit, $|b| \ll |a|/r$, this expression correctly reduces to \eqref{eq.hedgehog} obtained for the hedgehog field. Using the density, we evaluate the topological charge of \eqref{eq.charge.def} by performing explicit integration in spherical polar coordinates, with $d{\bf S} = \hat{\bf r}\, r^2 \sin\theta d\theta d\phi$:
   \begin{equation}
         Q =\begin{cases}-3\,\sgn(a), & r>r_{\rm c}\,, \\
         \sgn(a), & r<r_{\rm c}~, 
     \end{cases}\qquad \mbox{with} \qquad r_{\rm c}=\sqrt{3}|a/b|\,.
   \end{equation}  
The critical radius $r_{\rm c}$ separates the region where the vector field is dominated by the linear term (hedgehog pattern) from the region where the field is dominated by the quadratic term (antivortex pattern). Therefore the charge of the antivortex is $Q = -3\,\sgn(a)$, see Supplemental Fig.~\ref{figS:topo}(g).

To determine the vorticity, we focus on the case $r> r_{\rm c}$ where the antivortex contribution dominates. The topology does not change if we take $r\gg r_{\rm c}$, therefore we consider this
limit to simplify the analysis. The field reduces to 
$\bu=b(yz,zx,xy)/r^2$, and the in-plane phase becomes $\varphi = \tan^{-1}(bzx/bzy)$. Therefore we have
$\nabla \varphi = (\sin\phi,-\cos\phi)/r \sin\theta$ and the vorticity integral becomes:
\begin{equation}
    v = \frac{1}{2\pi}\int_{0}^{2\pi} \frac{(\sin\phi,-\cos\phi)}{r \sin\theta} \cdot r\sin\theta (-\sin \phi, \cos\phi) d\phi = -1\,.
\end{equation}
Since $\varphi = \sgn(bz)\tan^{-1}(x/y)$, we have $\varphi-\phi = \sgn(bz)\pi/2-[\sgn(bz)+1]\phi$, and the helicity integral in \eqref{eq.helicity.def0} yields $\gamma=\sgn(bz)\pi/2$. The vorticity and helicity for the antivortex field are shown in Supplemental Fig.~\ref{figS:topo}(l).
To correctly recover vanishing $\sgn(bz)$ in $\nabla\varphi$ and synchronize the relative phase between $\varphi$ and $\phi$, we only take $bz$ out of $\tan^{-1}$ when calculating $\varphi-\phi$. The same as in vortex case.

\vspace{3pt}
\textit{Vortex vector field ($422$ and $622$ point groups)} 

The vortex field is shown in Supplemental Fig.~\ref{figS:topo}(c) and is given by:
    \begin{equation}
    \bu=\frac{1}{r^2}[a(x,y,z)+b(yz,-xz,0)]~. 
    \end{equation}
In this case, the evaluation of the topological density yields:
    \begin{equation}
    \bm{\Omega}
    =\frac{1}{|r^2\bu|^3}
    \left[a^3(x,y,z)-a^2b(yz,-xz,0)+ab^2(0,0,z^3)\right]~,
    \end{equation}
and the topological charge is $Q=\sgn(a)$, see Supplemental Fig.~\ref{figS:topo}(h).
Since this charge is the same as for the hedgehog field, we need additional topological characteristics to distinguish these patterns.

To determine the vorticity, we focus on the large radius limit, where
the field reduces to $\bu=b(yz,-xz,0)/r^{2}$. In this case, the phase becomes $\varphi=\tan^{-1}(-bzx/bzy)$, therefore $\nabla\varphi=(\sin\phi,-\cos\phi)/r\sin\theta$ and $\varphi-\phi=-\sgn(bz)\pi/2+[\sgn(bz)-1]\phi$. 
Using these expressions, \eqref{eq.vorticity.def0} and \eqref{eq.helicity.def0} yield the vorticity $v=1$ and the helicity $\gamma=-\sgn(bz)\pi/2$, as shown in Supplemental Fig.~\ref{figS:topo}(m).

\vspace{3pt}
\textit{Double antivortex vector field ($\bar{6}m2$ and $\bar{6}$ point groups)} 

In the $\bar{6}$ point group we have the double-antivortex field given by:
    \begin{equation}
    \bu=\frac{1}{r^{2}}
    \left[a(x,y,z)+b_{1}(2xy,x^{2}-y^{2},0)+b_{2}(x^{2}-y^{2},-2xy,0)\right]\,.
    \end{equation}
The component with the $b_2$ prefactor is eliminated when moving to the parent group $\bar{6}m2$, and the resulting field is shown in Supplemental Fig.~\ref{figS:topo}(d). 
The topological density in this case is:
    \begin{equation}\begin{aligned}
    \bm{\Omega}=&\frac{1}{|r^{2}\bu|^{3}}
        \left[a^{3}(x,y,z)
             -a^{2}b_{1}(2xy,x^{2}-y^{2},0)
             -a^{2}b_{2}(x^{2}-y^{2},-2xy,0)
             -2a (b_{1}^{2}+b_{2}^{2})(x^{2}+y^{2})(x,y,2z)\right],
    \end{aligned}\end{equation}
which yields the topological charge:
   \begin{equation}
         Q =\begin{cases}-2\,\sgn(a)\,, & r>r_{\rm c}\,, \\
         \sgn(a)\,, & r<r_{\rm c}\,, 
     \end{cases}\qquad \mbox{with} \qquad r_{\rm c}=|a|/\sqrt{b_{1}^{2}+b_{2}^{2}}\,.
   \end{equation}  
This charge is shown in in Supplemental Fig.~\ref{figS:topo}(i), where we have taken $b_{2}=0$ for simplicity. 
In the limit $r\gg r_{c}$, the phase read: 
\begin{equation}
    \varphi=\tan^{-1}\left[\frac{b_{1}(x^{2}-y^{2})+b_{2}(-2xy)}{b_{1}(2xy)+b_{2}(x^{2}-y^{2})}\right]\,, 
\end{equation}
and its gradient is $\nabla\varphi=(2\sin\phi,-2\cos\phi)/r\sin\theta$. 
Using these expressions, \eqref{eq.vorticity.def0} and \eqref{eq.helicity.def0} yield
vorticity $v=-2$ and helicity $\gamma=\tan^{-1}(b_{1}/b_{2})$. 

For the specific case of the parent group $\bar{6}m2$, we can take the limit $b_{2}\rightarrow 0^{+}$ to  find $\gamma=\sgn(b_{1})\pi/2$, as shown in Supplemental Fig.~\ref{figS:topo}(n).
While the helicity of fields in the $\bar{6}m2$ and $\bar{6}$ groups can differ, they share the same topological charge and vorticity, and the pair of indices $(Q,v)$ distinguishes them from the other four classes of topological polarons. For this reason, we consider them as a single polaron species to simplify the classification.

\vspace{3pt}
\textit{Vertical flow vector field ($4mm$ and $6mm$ point groups)} 

The vertical flow field is shown in Supplemental Fig.~\ref{figS:topo}(e) and is given by:
    \begin{equation}
        \bu=\frac{1}{r^{2}}
        \left[a(x,y,z)+b_{1}(0,0,z^{2})+b_{2}(2xz,2yz,x^{2}+y^{2})\right]\,.
    \end{equation}
The topological density reads:
\begin{equation}\begin{aligned}
    \bm{\Omega}=&\frac{1}{|r^{2}\bu|^{3}}(a+2b_{1}z)(a+2b_{2}z)[ax,ay,z(a+2b_{2}z)]\,\\
     +&\frac{1}{|r^{2}\bu|^{3}}(a+2b_{2}z)[2b_{2}x(b_{1}z^{2}-b_{2}x^{2}-b_{2}y^{2}),2b_{2}y(b_{1}z^{2}-b_{2}x^{2}-b_{2}y^{2}),-(a+2b_{2}z)(b_{1}z^{2}+b_{2}x^{2}+b_{2}y^{2})]\,,
\end{aligned}\end{equation}
which yields the following topological charge, also shown in Supplemental Fig.~\ref{figS:topo}(j):
   \begin{equation}
         Q =\begin{cases}0\,, & r>r_{\rm c}\,, \\
         \sgn(a)\,, & r<r_{\rm c}\,, 
     \end{cases}\qquad \mbox{with} \qquad r_{\rm c}=|a/b_{1}|\,.
   \end{equation}  
In this expression, we have set $b_{2}=0$ for simplicity, since it does not alter the topological charge. The same choice is made in the main text and Supplemental Table~\ref{table:topo}. 
The vorticity and the helicity of this field, in the limit $r\gg r_{c}$, are ill-defined since the quadratic term with $b_{1}$ has no in-plane components. For this reason, we leave Supplemental Fig.~\ref{figS:topo}(o) empty. 

The above integrations were performed with the assistance of Mathematica \cite{Mathematica}.

\newpage
\section*{SUPPLEMENTAL NOTE 4: Analysis of ferroaxial polarons in inversion-symmetric groups}

In this note, we apply the symmetry-based framework developed in this work to analyze the large electron and hole ferroaxial polarons reported in Ref.~\citenum{JLB_topological_2024} for the halide double perovskite \ce{Cs2AgBiBr6}. 

As seen in the planar cuts of Fig.~1(K),(L) and Fig.~1(P),(Q) of Ref.~\citenum{JLB_topological_2024}, the displacement patterns correspond to a vortex for the electron polaron, and an antivortex for the hole polaron. We rationalize these findings in terms of the point group $4/m$ of \ce{Cs2AgBiBr6}.

The $4/m$ group is generated by three symmetry operations: inversion $\hat P$, mirror $\hat M_{z}$, and rotation $\hat R_{4z}$. The symmetry-allowed center-of-mass displacements (see Supplemental Note~1) near the polaron center can be obtained from Eq.~(1) of the main text and are given by:
    \begin{equation}\label{eq:perov-1}
        \bu(\br)=a_{1}(x,y,0)+a_{2}(-y,x,0)+a_{3}(0,0,z)\,,
    \end{equation}
where $a_{1,2,3}$ are independent material-specific parameters. All quadratic displacements are vanishing due to the inversion symmetry, and we neglect third and higher order terms as in the main text. The $z$-component of this field is $a_3 z$, and therefore it changes sign upon crossing the $z=0$ plane, in agreement with Ref.~\citenum{JLB_topological_2024}.
The topological charge and vorticity of the above field are $Q=\sgn(a_{3})=-1$ and $v=1$, in agreement with the values obtained from numerical integration in Ref.~\citenum{JLB_topological_2024} (we note that, in Ref.~\citenum{JLB_topological_2024}, the charge carries a sign error as confirmed by the authors).
Since, in this point group, in-plane displacements can be either of the hedgehog type, $(x,y,0)$, or vortex type, $(-y,x,0)$, the helicity of the field is not quantized as in the pure hedgehog pattern ($v=0,\pi$) and in the pure vortex pattern ($v=\pm \pi/2$). This observation is consistent with the value $\gamma=3\pi/4$ obtained numerically in Ref.~\citenum{JLB_topological_2024} for the electron polaron.

The emergence of an antivortex texture in the hole polaron of \ce{Cs2AgBiBr6}, as reported in Ref.~\citenum{JLB_topological_2024}, has a more complex origin. The antivortex pattern is not present
in \eqref{eq:perov-1}, because an in-plane antivortex field is not compatible with an $\hat R_{4z}$ rotation. 

However, close inspection of the band structure of \ce{Cs2AgBiBr6} [Supplemental Fig.~6(e) of Ref.~\citenum{JLB_topological_2024}] shows that the low-energy valance bands are composed of four $\hat R_{4z}$-connected $X$-valleys, while only two valleys participate in the hole polaron. Since these valleys preserve $\hat R_{2z}$ rotation but not $\hat R_{4z}$, the polaron symmetry is reduced from $4/m$ to $2/m$, as seen in the elongation of the wavefunction in Fig.~1(C) of Ref.~\citenum{JLB_topological_2024}. This symmetry breaking originates from electronic degeneracy, as we discuss in Supplemental Fig.~\ref{figS:SSB-BeO} for zb-BeO. As a result, the allowed displacement patterns acquire two additional components:
    \begin{equation}
        \bu'(\br)=
        a_{4}(-x,y,0)+a_{5}(y,x,0)\,,
    \end{equation}
which are both antivortex fields with helicity $\gamma=\pi/2$ or $\gamma=\pi$. 
If we ignore the terms with prefactors $a_{1}$ and $a_{2}$ which do not contribute to the antivortex pattern, we obtain topological charge $Q=-\sgn(a_{3})$ and vorticity $v=-1$, which are in agreement with Ref.~\citenum{JLB_topological_2024}. Also in this case, since the calculated field is an admixture of two antivortex fields with different helicity, the latter is not quantized. This observation is consistent with the calculations of Ref.~\citenum{JLB_topological_2024} yielding $\gamma=0.9\pi$.

In summary, both the vortex electron polaron and the antivortex hole polaron identified in \ce{Cs2AgBiBr6} can be rationalized in terms of symmetry principles. 

\newpage
\section*{SUPPLEMENTAL NOTE 5: Displacement fields from converse piezoelectric effect}

The electric field ${\bf E}$ generates a local strain field $\varepsilon_{\alpha\beta}$ via the converse piezoelectric strain coefficients $d_{\alpha\beta\gamma}$ \cite{nye1985physical}, 
    \begin{equation}\label{eq:strain-1}
        \varepsilon_{\alpha\beta} = d_{\alpha\beta\gamma} E_\gamma\,,
    \end{equation}
where the strain tensor is related to the displacement field via:
    \begin{equation}\label{eq.strain.tmp.1}
        \varepsilon_{\alpha\beta}
        =\frac{1}{2}\left(\frac{\partial u_\alpha}{\partial r_{\beta}} + \frac{\partial u_\beta}{\partial r_{\alpha}}\right)\,.
    \end{equation}
The electric field generated by a Gaussian charge density centered at $\br=0$ goes as $E_{\alpha}\propto r_{\alpha}$ near the origin. We take this field to be isotropic since topological properties are insensitive to continuous deformations. By combining this expression for the electric field with \eqref{eq:strain-1},\eqref{eq.strain.tmp.1}, and absorbing all coefficients inside the displacement, we obtain:
    \begin{equation}\label{eq:strain-2}
        \frac{\partial u_\alpha}{\partial r_{\beta}} + \frac{\partial u_\beta}{\partial r_{\alpha}}
        = d_{\alpha\beta\gamma} r_{\gamma}~.
    \end{equation}
The field satisfying this equation must be a second-order polynomial:
    \begin{equation}\label{eq.strain.tmp.2}
        u_\alpha = c_\alpha + a_{\alpha\beta} r_{\beta} + b_{\alpha\beta\gamma}r_{\beta} r_{\gamma}\,.
    \end{equation}
By requiring the displacement to vanish at $\br=0$, which we take to be the polaron center, we have $c_\alpha=0$.  Using this result and \eqref{eq.strain.tmp.2} inside \eqref{eq:strain-2}, we find: 
 \begin{equation}
     ( a_{\alpha\beta} + a_{\beta\alpha}) + (b_{\alpha\beta \gamma}  + b_{\alpha \gamma\beta}
     + b_{\beta\alpha \gamma} + b_{\beta \gamma\alpha} )r_{\gamma} =
          d_{\alpha\beta\gamma} r_{\gamma}\,.
 \end{equation}
Comparing equal powers $r_\alpha$, we find $a_{\alpha\beta}=-a_{\beta\alpha}$, hence $a_{\alpha\beta}$ must be antisymmetric. 
However, an antisymmetric tensor generates a displacement that corresponds to a pure rigid rotation, without any strain \cite{nye1985physical}; 
therefore we remove this component from the solution and we are left with:
 \begin{equation}
   b_{\alpha\beta \gamma}  + b_{\alpha \gamma\beta} + b_{\beta\alpha \gamma} + b_{\beta \gamma\alpha}
  = d_{\alpha\beta\gamma}\,.
 \end{equation}
These are 27 equations for the 27 unknown $b$'s. By removing again the rigid-body rotation of the displacement, we find that the tensor $b$ must be symmetric in the first two indices. Therefore the last expression becomes:
 \begin{equation}\label{eq:C-d-1}
     2b_{\alpha\beta \gamma} + b_{\gamma\alpha\beta} + b_{\gamma\beta\alpha}
    = d_{\alpha\beta\gamma}\,.
 \end{equation}
Now we have 18 unknowns and 18 equations (the $d$'s are also symmetric in the first two indices).
By exchanging $\alpha$ and $\gamma$, as well as $\beta$ and $\gamma$ in this expression, we obtain two additional relations:
    \begin{eqnarray}
    \label{eq:C-d-2}&2b_{\beta\gamma\alpha}  + b_{\alpha\beta\gamma} + b_{\gamma\alpha\beta}  
        = d_{\beta\gamma\alpha}\,,\\
    \label{eq:C-d-3}&2b_{\gamma\alpha\beta}  + b_{\beta\gamma\alpha} + b_{\alpha\beta\gamma}  
        = d_{\gamma\alpha\beta}\,.
    \end{eqnarray}
We can now obtain an explicit expression for the $b$ coefficients by multiplying \eqref{eq:C-d-1} by a factor 3 and subtracting both \eqref{eq:C-d-2} and \eqref{eq:C-d-3}:
    \begin{equation}
        4b_{\alpha\beta\gamma}=3 d_{\alpha\beta\gamma} -  d_{\beta\gamma\alpha} - d_{\gamma\alpha\beta}\,,
    \end{equation}
From these relations and the symmetry-allowed components of the converse piezoelectric tensor, as reported in the Bilbao Crystallographic Server \cite{Aroyo2006BilbaoCS,Aroyo:xo5013}, we obtain the following displacements for each point group:
\begin{enumerate}
\item $\bar{4}3m$ (zb-BeO) and $23$: the symmetry-allowed components are $d_{xyz}=d_{yzx}=d_{xzy}$, which implies $b_{xyz}=b_{yzx}=b_{xzy}$, and  $\bu=(yz,zx,xy)$.
\item $422$ ($\gamma$-\ce{LiAlO2}) and $622$: the symmetry-allowed components are  $d_{xzy}=-d_{yzx}$, which implies $b_{xzy}=-b_{yzx}$, and $\bu=(yz,-xz,0)$.
\item  $\bar{6}m2$ (2D h-BN): the symmetry-allowed components are  $d_{xxy}=d_{xyx}=-d_{yyy}$, which implies $b_{xxy}=b_{xyx}=-b_{yyy}$, and $\bu=(2xy,x^{2}-y^{2})$.
\item $\bar{6}$: the symmetry-allowed components are $d_{xxy}=d_{xyx}=-d_{yyy}$, $d_{xxx}=-d_{yyx}=-d_{xyy}$, which implies $b_{xxy}=b_{xyx}=-b_{yyy}$, $b_{xxx}=-b_{yyx}=-b_{xyy}$, and $\bu=(2xy,x^{2}-y^{2}),~(x^{2}-y^{2},-2xy)$.
\item $4mm$ (\ce{PbTiO3}) and $6mm$: the symmetry-allowed components are $d_{xzx}=d_{yzy}$, $d_{xxz}=d_{yyz}$, and $d_{zzz}$, which implies $b_{zzz}$, $b_{xzx}=b_{yzy}=b_{xxz}=b_{yyz}$, $b_{xxz}=b_{yyz}=-3b_{xzx}=-3b_{yzy}$. Therefore $\bu=(0,0,z^{2}),~(2xz,2yz,x^{2}+y^{2})$.
\end{enumerate}
These displacement fields are summarized in Supplemental Table~\ref{table:piezo}. These strain-induced textures match the patterns that we obtained via explicit symmetry analysis in the main text (Supplemental Table~\ref{table:vector-fields}) after removing rigid rotations.

For completeness we mention that, in the case of the hedgehog-type polaron, the relevant displacement is the position-dependent macroscopic polarization (see Supplenental Note~1). In this case, the relation between polarization and electric field is simply $P_{\alpha}=(\epsilon_{\alpha\beta}^{0}-\epsilon_{\alpha\beta}^{\infty})E_{\beta}$, therefore the prototypical displacement field is 
$\bu=(x,y,z)$ (barring anisotropic distortions that do not change the topology).

\newpage
\section*{SUPPLEMENTAL NOTE 6: Pseudo-magnetic field of topological polarons}

We show how the inhomogeneous strain associated with topological polarons induces a Berry curvature and a pseudo-magnetic field. For simplicity, we focus on 2D h-BN, since in this case we can draw from a large body of literature on strained graphene \cite{suzuura_phonons_2002,manes_symmetry-based_2007,de_juan_space_2012,castro_neto_electronic_2009,levy_strain-induced_2010,hsu_nanoscale_2020}.

Near the $K$-valley, the electronic Hamiltonian of 2D h-BN reads \cite{fang_electronic_2018}:
    \begin{equation}\label{eq.Ham.pseudo}
        H(\bk)
        =\hbar v_{F}(k_{x}\sigma_{x}+k_{y}\sigma_{y})+\Delta\sigma_{z}\,,
    \end{equation}
where $v_{F}$ is the counterpart of the Fermi velocity in graphene, $\Delta$ is half the band gap energy and represents the Semenoff mass, $\sigma_x$, $\sigma_y$, and $\sigma_z$ are the Pauli matrices. 
The Hamiltonian acts on a two-component pseudo-spinor corresponding to the amplitudes on the boron and nitrogen sublattices, and admits two bands with energy $\epsilon_\bk=\pm\sqrt{\Delta^{2}+v_{F}^{2}|\bk|^2}$. Note that we are using a reference frame with $\bk=0$ at the $K$ point.

Since \eqref{eq.Ham.pseudo} describes a (massive) Dirac fermion, the introduction of inhomogeneous strain leads to a modification of the band structure that can be described via the coupling to a pseudo-vector potential, see Eqs.~(63), (64) of Ref.~\citenum{manes_symmetry-based_2007}:
    \begin{equation}\begin{aligned}
        {\bf A}_K
        =\frac{\beta}{2a}(\varepsilon_{xx}-\varepsilon_{yy},-2\varepsilon_{xy},0)\,,
    \end{aligned}\end{equation}
where $a=2.504$~\AA\ is the lattice constant of 2D h-BN, and $\beta=1.7$ is a dimensionless electron-phonon coupling parameter \cite{fang_electronic_2018}.
From Supplemental Table~\ref{table:piezo}, we have $\varepsilon_{xy}=cx$, $\varepsilon_{xx}=-\varepsilon_{yy}=cy$, where $c$ is a constant depending on the magnitude of the polaronic distortion. By replacing these expressions in the last equation and taking the curl, we obtain the pseudo-magnetic field:
    \begin{equation}\begin{aligned}\label{eq.pseudomag}
        {\bf B}_K =(0,0,-2\beta c/a)\,.
    \end{aligned}\end{equation}
It can be shown that, through the same mechanism, the $K'$ valley experiences a pseudo-magnetic field $-{\bf B}_K$. These emergent gauge fields can lead to the formation of pseudo-Landau levels, and to modifications of the electronic Berry curvature, with implications for electron wavepacket dynamics \cite{sundaram_wave-packet_1999,niu2010,PhysRevLett.112.166601}.

To estimate the magnitude of these effects, we evaluate the parameters entering \eqref{eq.pseudomag} from our \textit{ab initio} calculations of the large hole polaron in 2D h-BN. To this end, we first convert discrete atomic displacements into a unique vector for each unit cell, as described in Supplemental Note~1:
\begin{equation}
u_{\alpha}(\bR_{p})={\sum}_{\kappa} M_{\kappa}u_{\kappa\alpha}(\bR_{p})/{\sum}_{\kappa'} M_{\kappa'}\,.
\end{equation}
Next, we use the finite-difference version of Eq~\eqref{eq.strain.tmp.1} to evaluate the strain on each lattice point. The central-difference formulas for a hexagonal lattice with points $m\,{\bf a}_1+n\,{\bf a}_2$, where ${\bf a}_1$ is along the $x$ axis and ${\bf a}_2$ is rotated with respect to ${\bf a}_1$ by 120$^\circ$, the central finite difference formulas are: 
\begin{eqnarray}
        \frac{\partial f}{\partial x}(m,n) &\simeq&
        \frac{f(m+1,n)-f(m-1,n)}{2a}\,, \\
        \frac{\partial f}{\partial y}(m,n) &\simeq&
        \frac{f(m+1,n+1)+f(m,n+1)-f(m+1,n-1)-f(m,n-1)}{2\sqrt{3}a}\,.
    \end{eqnarray}
Using $f(m,n) = u_{\alpha}(\bR_{p}+m{\bf a}_1+n{\bf a}_2)$ in these expressions, we obtain the strain field shown in Supplemental Fig.~\ref{figS:strain-hBN}. For visualization purposes, in the figure we plot the Frobenius norm of each field.
The maximum strain is achieved for $r_{\rm max}\simeq 30$~\AA\ and is $\varepsilon_{\rm max} = 0.019\%$, therefore the missing constant in \eqref{eq.pseudomag} is $c = \varepsilon_{\rm max}/r_{\rm max} = 6.33\cdot 10^{-6}$\,\AA$^{-1}$. The corresponding magnetic field from \eqref{eq.pseudomag} is $B = 1.3$\,T, which is moderately strong and could lead to measurable effects.

\clearpage


\begin{table*}[ht]
\begin{center}
\caption{\justifying\textbf{Symmetry-allowed displacement fields by point groups}. 
We report the point groups that support symmetry-protected topological polarons (first column), their symmetry operations (second column), and the linear (third column) and quadratic (fourth column) vector fields compatible with these symmetries.}\label{table:vector-fields}
\setlength{\extrarowheight}{0.2cm}
\begin{tabular}{l|l|l|l}
\hline\hline
Point group \qquad & Generators & Linear vector fields & Quadratic vector fields\\
 \hline
 $m\bar{3}m$ & $\{R_{2y},R_{2z},R_{2,110},R_{3,111},P\} \quad$ 
        & $(x,y,z)$ & n/a  \\
 $\bar{4}3m$ & $\{R_{2y},R_{2z},R_{3,111},M_{110}\}$ 
        & $(x,y,z)$ 
        & $(yz,zx,xy)$  \\ 
 $422$  & $\{R_{2y},R_{2z},R_{4z}\}$  
        & $(x,y,0)$, $(0,0,z)$ 
        & $(yz,-xz,0)$  \\
 $\bar{6}m2$ & $\{R_{3z},M_{z},M_{110}\}$
        & $(x,y,0)$, $(0,0,z)$ 
        & $(2xy,x^{2}-y^{2},0)$ \\
 $\bar{6}$  & $\{R_{3z},M_{z}\}$     
        & $(x,y,0)$, $(-y,x,0)$, $(0,0,z)$ 
        & $(2xy,x^{2}-y^{2},0)$, $(x^{2}-y^{2},-2xy,0)$ \\
 $4mm$  & $\{R_{2z},R_{4z},M_{y}\}$
        & $(x,y,0)$, $(0,0,z)$ 
        & $(xz,yz,0)$, $(0,0,x^{2}+y^{2})$, $(0,0,z^{2})$ \\
 \hline\hline
\end{tabular}
\end{center}
\end{table*}
\clearpage

\begin{table}[ht]
\begin{center}
\caption{{\justifying\textbf{Topological invariants of displacement fields by polaron type}}. Summary of topological invariants evaluated in Supplemental Note~3 for each type of symmetry-protected topological polaron: topological charge $Q$, vorticity $v$, and helicity $\gamma$. The pair of indices $(Q,\gamma)$ uniquely identifies the polaron type. The polaron class follows the classification of vectorlike physical quantities in Ref.~\citenum{hlinka_eight_2014}. The axial polarons absent from this table are discussed in Supplemental Note 4 for the halide double perovskite \ce{Cs2AgBiBr6}.
}\label{table:topo}
\setlength{\extrarowheight}{0.2cm}
\begin{tabular}{l|l|l|r|r|r}
 \hline\hline
  Polaron type & Polaron class & Vector field & Charge $Q$  &  Vorticity $v$ & Helicity $\gamma$ \\
 \hline
Hedgehog & Neutral & $a(x,y,z)$ & \sgn($a$) & \phantom{$-$}1 & 0  \\[0.1cm]
Antivortex & Polar & $a(x,y,z)+b(yz,zx,xy)$ & $-$3\,\sgn($a$) & $-1$ & 
$\phantom{-}(\pi/2)$\sgn($bz$) \\
Vortex & Chiral & $a(x,y,z)+b(yz,-xz,0)$ & \sgn($a$) & \phantom{$-$}1 & 
$-(\pi/2)$\sgn($bz$) \\
Double antivortex \quad & Polar & $a(x,y,z)+b(2xy,x^{2}-y^{2},0)$ & $-$2\,\sgn($a$) & $-2$ &  
$\phantom{-}(\pi/2)$\sgn($b$) \\
Vertical flow & Polar & $a(x,y,z)+b(0,0,z^{2})$& 0 & n/a & n/a \\
 \hline\hline
\end{tabular}
\end{center}
\end{table}
\clearpage

\begin{table*}[ht]
\begin{center}
\caption{
{\justifying\textbf{Dimensionality and topological stability of piezoelectric displacement fields by polaron type}.} 
Summary of topological stability for the four symmetry-protected polaron species as classified in Supplemental Table \ref{table:piezo}. The criterion follows the theory of topological defects in ordered media \cite{poenaru1977crossing,mermin1979topological,chaikin1995principles}: a defect corresponding to a $d_{s}$-dimensional singularity in a $d$-dimensional space is stable if and only if the codimension $d'=d-d_{s}$ is greater than or equal to the degrees of freedom $n$. All polarons discussed here are point defects with $d_{s}=0$.
}
\setlength{\extrarowheight}{0.2cm}
\begin{tabular}{l|l|l|l|l}
 \hline\hline
  Polaron type & Dimension of space $d$ & Codimension $d'$ &  Degrees of freedom $n$& Stable ? ($d'\geq n$ ?)\\
  \hline
  Antivortex & 3 & 3 & 3 & Yes\\
  Vortex     & 3 & 3 & 2 & Yes\\
  Double antivortex & 3 & 3 & 2 & Yes\\
  Double antivortex & 2 & 2 & 2 & Yes\\
  Vertical flow & 3 & 3 & 1 & Yes\\
  \hline\hline
\end{tabular}\label{tab:dimension}
\end{center}
\end{table*}

\clearpage

\begin{table*}[ht]
\begin{center}
\caption{\justifying\textbf{Displacements induced by the converse piezoelectric effect by polaron type}. 
Converse piezoelectric tensor, strain fields, and displacements of symmetry-protected topological polarons.
The first column reports the polaron type; the second column indicates the underlying point group; the third column summarizes the nonzero elements and symmetry of the converse piezoeletric tensor. The fourth and fifth columns give the resulting strain fields and associated displacements, respectively, as derived in Supplemental Note~5. The converse piezoelectric tensor is symmetric in the first two indices, $d_{\alpha\beta\gamma}=d_{\beta\alpha\gamma}$. 
For the vertical flow field, the components of $d_{\alpha\beta\gamma}$ shown in parentheses are not included in the evaluation of the strain and displacement, because these components do not contribute to the topological invariants.}
\label{table:piezo}
\setlength{\extrarowheight}{0.2cm}
\begin{tabular}{l|l|l|l|l}
 \hline\hline
Polaron type & Point group & Conv. piezoel. tensor $d_{\alpha\beta\gamma}$ & Strain $\varepsilon_{\alpha\beta}$ & Displacements $(u_x, u_y,u_z)$\\[3pt]
 \hline
 Antivortex & $\bar{4}3m$, $23$ 
        & $d_{xyz}=d_{yzx}=d_{zxy}$
        & $\varepsilon_{xy}\propto z$, 
          $\varepsilon_{xz}\propto y$,
          $\varepsilon_{yz}\propto x$
        & $(yz,zx,xy)$\\[6pt]
 Vortex & $422$, $622$  
        & $d_{xzy}=-d_{yzx}$
        & $\varepsilon_{xz}\propto y$, 
          $\varepsilon_{yz}\propto  -x$
        & $(yz,-xz,0)$\\[5pt]
 Double antivortex \quad & $\bar{6}m2$ 
        & $d_{xxy}=-d_{yyy}=d_{xyx}$
        & $\varepsilon_{xx}=-\varepsilon_{yy}\propto y$, 
          $\varepsilon_{xy}\propto x$
        & $(2xy,x^{2}-y^{2},0)$\\[10pt]
 Double antivortex \quad & $\bar{6}$ 
        & \shortstack[l]{$d_{xxy}=-d_{yyy}=d_{xyx}$ \\[2pt] 
          $d_{xxx}=-d_{yyx}=-d_{xyy}$}
        & \shortstack[l]{
          $\varepsilon_{xx}=-\varepsilon_{yy}\propto y$, $\varepsilon_{xy}\propto x$\\
          $\varepsilon_{xx}=-\varepsilon_{yy}\propto -x$, 
          $\varepsilon_{xy}\propto y$}
        & \shortstack[l]{
          $(2xy,x^{2}-y^{2},0)$\\
          $(x^{2}-y^{2},-2xy,0)$}\\[10pt]
 Vertical flow & $4mm$, $6mm$ \quad
        & \shortstack[l]{$d_{zzz}$\\($d_{xzx}=d_{yzy}$, $d_{xxz}=d_{yyz}$)}
        & $\varepsilon_{zz}\propto z$
        & $(0,0,z^{2})$\\
 \hline\hline
\end{tabular}
\end{center}
\end{table*}
\clearpage


\clearpage

\begin{figure}
    \centering
    \begin{overpic}[width=1.0\textwidth]{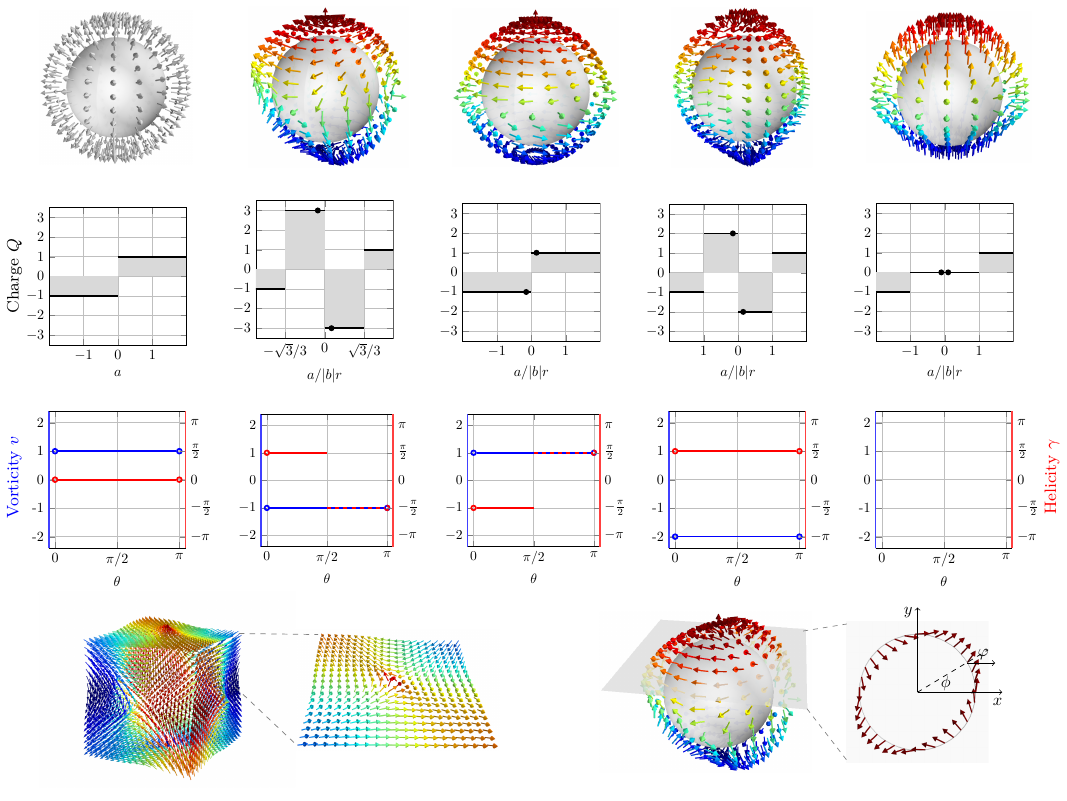}
        \put(0.2cm, 13cm){\fontfig a}
        \put(3.8cm, 13cm){\fontfig b}
        \put(7.3cm, 13cm){\fontfig c}
        \put(10.8cm,13cm){\fontfig d}
        \put(14.3cm,13cm){\fontfig e}
        \put(0.2cm, 9.8cm){\fontfig f}
        \put(3.8cm, 9.8cm){\fontfig g}
        \put(7.3cm, 9.8cm){\fontfig h}
        \put(10.7cm,9.8cm){\fontfig i}
        \put(14.3cm,9.8cm){\fontfig j}
        \put(0.2cm, 6.7cm){\fontfig k}
        \put(3.8cm, 6.7cm){\fontfig l}
        \put(7.3cm, 6.7cm){\fontfig m}
        \put(10.7cm,6.7cm){\fontfig n}
        \put(14.3cm,6.7cm){\fontfig o}
        \put(1.2cm,3.0cm){\fontfig p}
        \put(10cm,3.0cm){\fontfig q}
    \end{overpic}
    \caption{\justifying\textbf{Topological invariants}. 
    Panels \textbf{a}-\textbf{e} in the first row show the idealized displacement fields of topological polarons on a sphere enclosing the polaron center, as derived in Supplemental Notes~2 and 3 and summarized in Supplemental Table~\ref{table:topo}. The crystal lattices in \textbf{a-c} and \textbf{e} are simple cubic, the lattice in \textbf{d} is hexagonal. In all cases, we set the parameters $a=b=1$ (dimensionless) and consider a sphere of radius $r=10$ (dimensionless); these parameters correspond to the case $r\gg r_{\rm c}$ discussed in Supplemental Note~3 and shown as dots in the second row. All vectors are normalized. 
    Panels \textbf{f}-\textbf{j} in the second row show the topological charge $Q$ for each vector field in the same column, as a function of the parameters $a$, $b$, and $r$.
    Panels \textbf{k}-\textbf{o} in the third row show the corresponding vorticity $v$ (blue, left axis) and helicity $\gamma$ (red, right axis) in each case. Open circles denote cases where these indices are ill-defined.
    \textbf{p} The cube employed to calculate numerically the topological charge from \textit{ab initio} polaron displacements (left), with topological density evaluated on each face, for example the upper face $S_{+z}$ (right). The color codeing of the vectors is a visual aid. 
    \textbf{q} Left panel: Spherical cut of the antivortex field from \textbf{b}, used to evaluate vorticity and helicity. Right panel: Definition of the angle $\varphi=\tan^{-1}(u_{y}/u_{x})$ of the vector for a given polar angle in spherical coordinates; $\phi$ is the azimuthal angle.
    }
    \label{figS:topo}
\end{figure}
\clearpage

\begin{figure}
    \centering
    \begin{overpic}[width=0.75\textwidth]{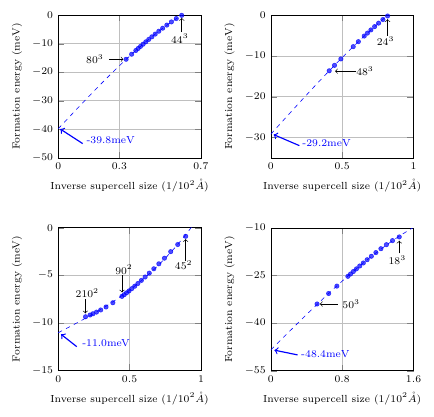}
        \put(0.4cm,12.5cm){\fontfig a}
        \put(7cm,12.5cm){\fontfig b}
        \put(0.4cm,6cm){\fontfig c}
        \put(7cm,6cm){\fontfig d}
    \end{overpic}
    \caption{\justifying\textbf{Convergence tests of polaron formation energies
    }.
    Blue disks are polaron formation energies calculated via the \textit{ab initio} polaron equations, as a function of supercell size. In this approach, the size of the supercell is set by the corresponding Brillouin zone sampling; for example, the label $80^3$ means 80$\times$80$\times$80 $\bk$- and $\bq$-points in \eqref{eq:polaron-eq}, which corresponds to an equivalent supercell with  80$\times$80$\times$80 primitive unit cells.
    The zero of the energy axis is set to the energy of the undistorted structure. 
    The dashed blue lines represent the fit to the formation energy via \eqref{eq:polaron-extra}, and the intercept with the vertical axis gives the formation energy in the dilute limit. 
    {\bf a}, electron polaron in zb-BeO; 
    {\bf b}, electron polaron in $\gamma$-\ce{LiAlO2}; 
    {\bf c}, hole polaron in 2D h-BN; 
    {\bf d}, electron polaron in \ce{PbTiO3}.
    }
    \label{figS:extrapolation}
\end{figure}
\clearpage

\begin{figure}
    \centering
    \begin{overpic}[width=1.0\textwidth,page=1]{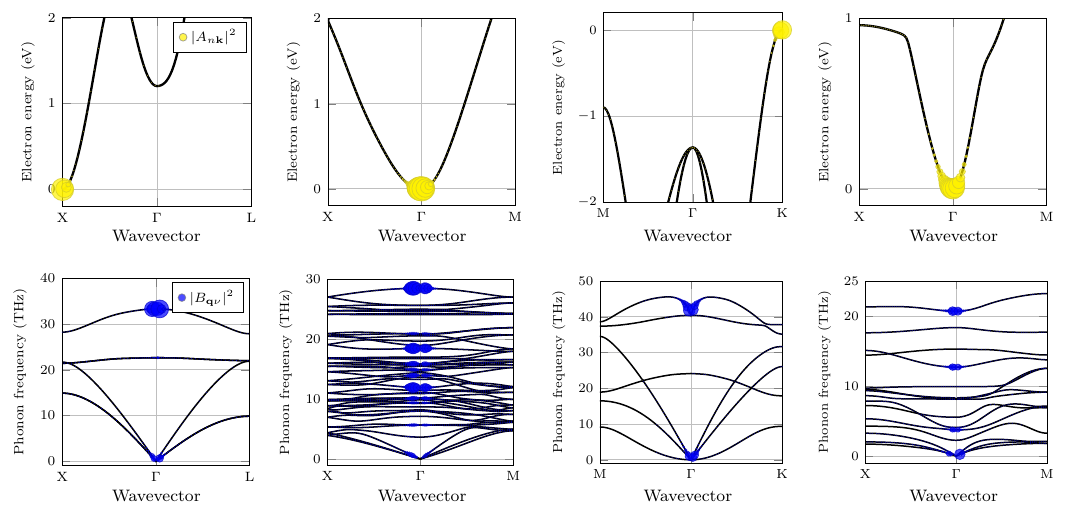}
        \put(0.35cm, 8.1cm){\fontfig a} \put(0.35cm,3.8cm){\fontfig e}
        \put(4.80cm, 8.1cm){\fontfig b} \put(4.80cm,3.8cm){\fontfig f}
        \put(9.27cm, 8.1cm){\fontfig c} \put(9.27cm,3.8cm){\fontfig g}
        \put(13.75cm,8.1cm){\fontfig d} \put(13.75cm,3.8cm){\fontfig h}
    \end{overpic}
    \caption{\justifying\textbf{Band and mode decomposition of \textit{ab initio} polarons}.
    Panels {\bf a}-{\bf d} in the top row report the band structures of zb-BeO (conduction), $\gamma$-\ce{LiAlO2} (conduction), 2D h-BN (valence), and \ce{PbTiO3} (conduction), respectively. 
    The Fourier amplitudes $A_{n{\bk}}$ appearing in \eqref{eq.elpol} are superimposed to the bands as yellow circles, with radii proportional $|A_{n{\bk}}|^2$. 
    Panels {\bf e}-{\bf g} in the bottom row report the phonon dispersions 
    of of zb-BeO, $\gamma$-\ce{LiAlO2}, 2D h-BN, and \ce{PbTiO3}, respectively.
    The Fourier amplitudes $B_{\bq\nu}$ appearing in \eqref{eq.disp1} are superimposed to the bands as blue circles, with radii proportional $|B_{\bq\nu}|^2$.}
    \label{figS:proj-AB}
\end{figure}
\clearpage

\begin{figure}
    \centering
    \begin{overpic}[width=1.0\textwidth,page=1]{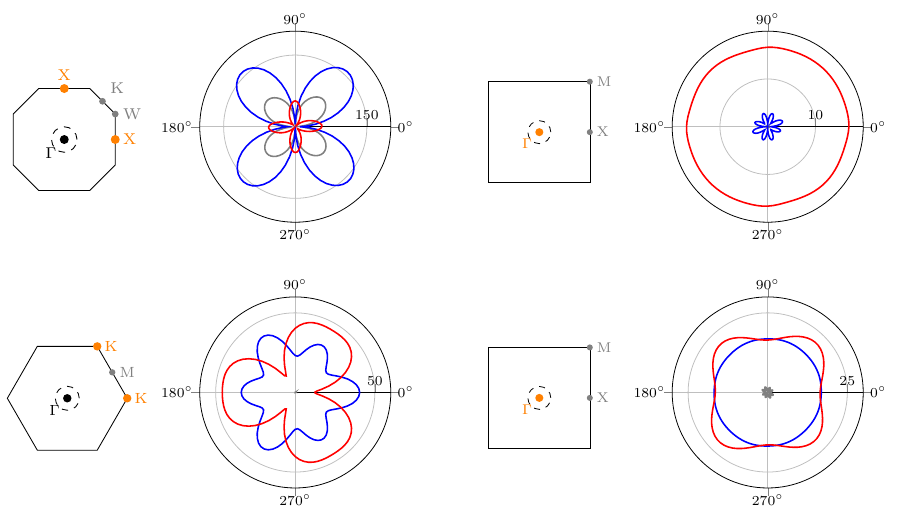}
        \put(0.2cm,9.5cm){\fontfig a}
        \put(9.7cm,9.5cm){\fontfig b}
        \put(0.2cm,4.3cm){\fontfig c}
        \put(9.7cm,4.3cm){\fontfig d}
        \put(2.5cm,9.5cm){\fonttext zb-\ce{BeO}}
        \put(11.8cm,9.5cm){\fonttext $\gamma$-\ce{LiAlO2}}
        \put(2.5cm,4.3cm){\fonttext 2D h-BN}
        \put(11.8cm,4.3cm){\fonttext \ce{PbTiO3}}
        \put(7.5cm,6.1cm){$|g_{mn\nu}(\bk,\bq)|$~(meV)}
        \linethickness{0.5mm}
        \put(7.5cm,5.8cm){\color{gray}\line(1,0){0.7cm}}
        \put(8.3cm,5.7cm){$\nu=1$}
        \put(7.5cm,5.5cm){\color{blue}\line(1,0){0.7cm}}
        \put(8.3cm,5.4cm){$\nu=2$}
        \put(7.5cm,5.2cm){\color{red}\line(1,0){0.7cm}}
        \put(8.3cm,5.1cm){$\nu=3$}
    \end{overpic}
    \caption{\justifying\textbf{Symmetry of \textit{ab initio} electron-phonon matrix elements}. 
    {\bf a} Left: 2D plane cut of the Brillouin zone of zb-BeO for $q_z = 0$.
    Gray dots indicate the locations of high-symmetry points.
    Orange dots indicate the electron valleys that contribute the most to the polaron wavefunction.
    The dashed circle indicates the path of $\bq$-points that we use to produce the polar plot on the right; the radius of this circle is $0.1\,(2\pi/a)$, with $a$ being the lattice constant.
    Right: Electron-phonon coupling matrix elements $|g_{mn\nu}(\bk,\bq)|$, for $m=n=1$ and $\bk$ at one of the valleys. The matrix elements for the acoustic branches $\nu=1,2,3$ are shown as gray, blue, and red lines, respectively. Modes $\nu=1,2$ are transverse acoustic, mode $\nu=3$ is longitudinal acoustic.
    These curves exhibit all the symmetries expected from our analysis in Supplemental Note~2 and Supplemental Table~\ref{table:vector-fields}.
    {\bf b}, {\bf c}, and {\bf d}: Same information as in {\bf a}, this time for     
    $\gamma$-\ce{LiAlO2},  2D h-BN, and \ce{PbTiO3}, respectively.}
    \label{figS:EPI}
\end{figure}
\clearpage

\begin{figure}
    \centering
    \begin{overpic}[width=1.0\textwidth]{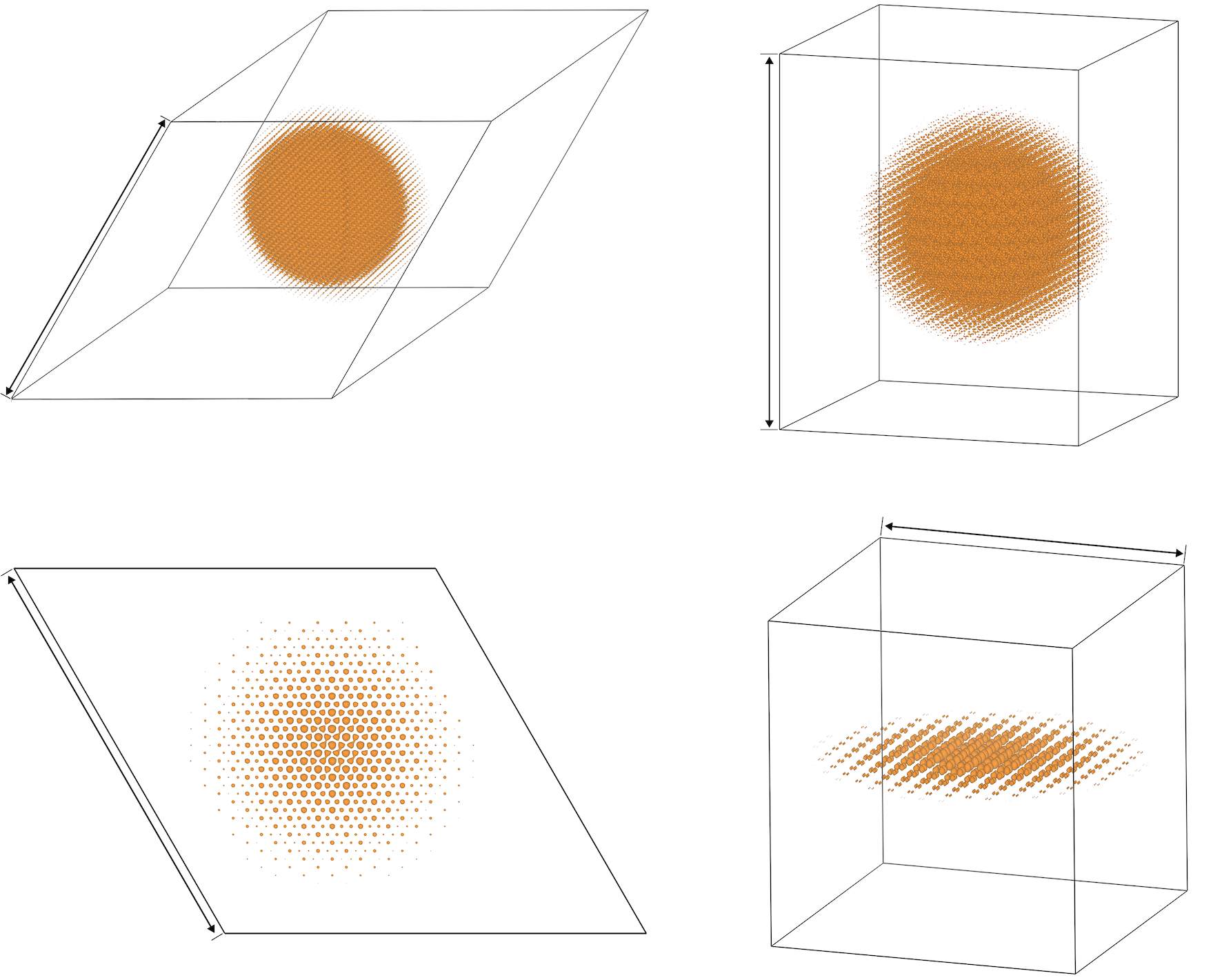}
        \put(0cm,14.2cm){\fontfig a} \put(0.7cm,10.3cm){\rotatebox{60}{12.2 nm}}
        \put(11cm,14.2cm){\fontfig b}\put(1cm,3.6cm){\rotatebox{-60}{11.2 nm}}
        \put(0cm,6.6cm){\fontfig c}  \put(10.9cm,10.4cm){\rotatebox{90}{15.4 nm}}
        \put(11cm,6.6cm){\fontfig d} \put(14.6cm,6.7cm){\rotatebox{-6}{7.7 nm}}
    \end{overpic}
    \caption{\justifying
    \textbf{Electron and hole wavefunctions of \textit{ab initio} polarons}. 
    \textbf{a} Isosurface of the charge density of the electron polaron in a $46^{3}$ supercell of zb-BeO. 
    \textbf{b} Isosurface of the charge density of the electron polaron in a $25^{3}$ supercell of $\gamma$-\ce{LiAlO2}. 
    \textbf{c} Isosurface of the charge density of the hole polaron in a $45^{2}$ supercell of 2D h-BN.
    \textbf{d} Isosurface of the charge density of the electron polaron in a $20^{3}$ supercell of \ce{PbTiO3}.
    The standard deviations of these distributions are $1\sim2$\,nm, and the isosurfaces contain $95\%$ of the charge. 
    For clarity, atoms and their displacements are not rendered. 
    }
    \label{fig:psir}
\end{figure}
\clearpage

\begin{figure*}[tbh]
	\centering
    \begin{overpic}[width=\textwidth]{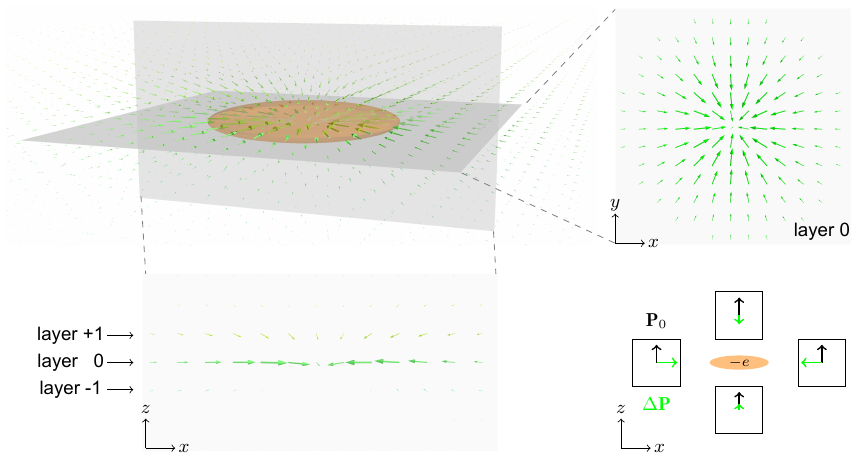}
    \put(0.8cm, 8.8cm){\fontfig a}
    \put(11.6cm, 8.8cm){\fontfig b}
    \put(0.8cm, 3.5cm){\fontfig c}
    \put(11.6cm, 3.5cm){\fontfig d}
    \end{overpic}
	\caption{
    \justifying\textbf{Position-dependent macroscopic polarization induced by an electron polaron in \ce{PbTiO3}.} 
    \textbf{a} The cell-averaged polarization changes $\Delta{\bf P}$ (arrows) computed from first principles. The color code is a visual aid. The orange ellipsoid represents the envelope function of the electron wavefunction. 
    The induced polarizations along $\alpha$-direction is $\Delta P_{\alpha}=\sum_{\kappa\alpha'}Z_{\kappa,\alpha\alpha'}^{*}\Delta\tau_{\kappa\alpha' p}$ with both Born-effective charges $Z_{\kappa,\alpha\alpha'}^{*}$ and atomic displacements $\Delta\tau_{\kappa\alpha p}$ of atom $\kappa$ in each unit cell at ${\bf R}_{p}$.  
    Two planar cuts in the $xy$ and $xz$ planes crossing the polaron center are shown in \textbf{b} and \textbf{c}, respectively. 
    \textbf{b} In-plane ($xy$) polarization changes are isotropic and reach a maximum of $1.4~\%$ of the ferroelectric polarization ${\bf P}_{0}=(0,0,0.77)~\text{C}/\text{m}^2$ calculated using the modern theory of polarization \cite{king1993theory,resta1992theory,resta1994macroscopic};
    \textbf{c} The out-of-plane ($z$) polarization reaches a maximum magnitude of $0.34~\% P_0$. 
    \textbf{d} Schematic illustration of polaron-induced polarization $\Delta{\bf P}$ (green arrows) relative to the ferroelectric polarization (black arrows in each unit cell). For clarity, $\Delta{\bf P}$ is magnified by a factor of 80. 
    The pronounced polarization anisotropy reflects the anisotropy of the static dielectric tensor, which we calculate to be $\epsilon_{xx}^0 = \epsilon_{yy}^0 = 211.2$ and $\epsilon_{zz}^0 =42.9$: for the same electric field strength, the anisotropy ratio of the induced macroscopic polarization is $\Delta P_{z}/\Delta P_{x}=0.20$, which is similar to our \textit{ab initio} result $0.34/1.4 = 0.24$.
    }
	\label{figS:FE}
\end{figure*}
\clearpage

\begin{figure}
    \centering
    \begin{overpic}[width=\textwidth]{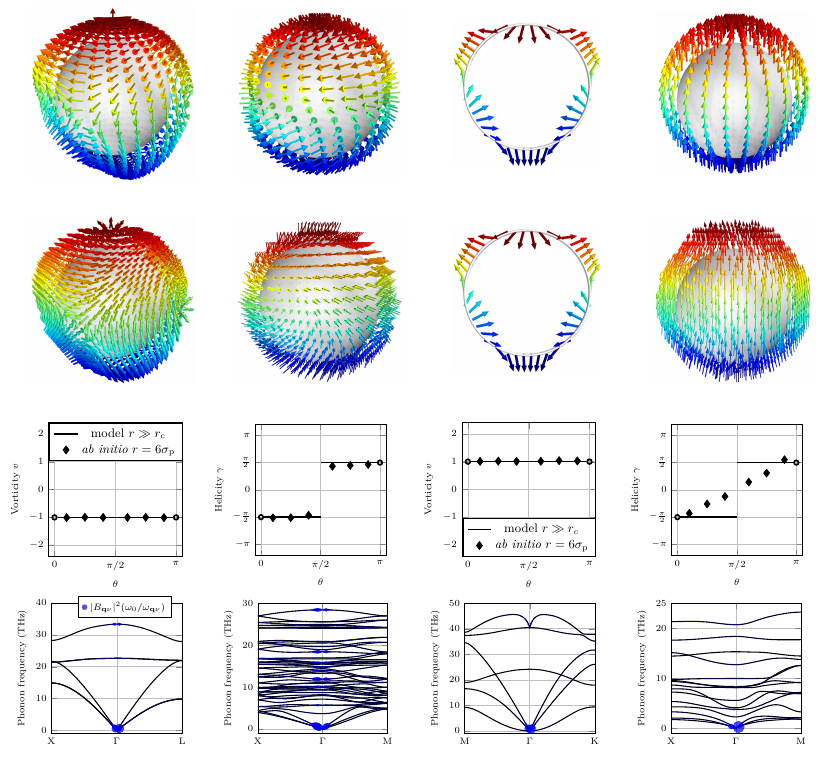}
        \put(0.2cm,15.9cm){\fontfig a} 
        \put(0.2cm,11.3cm){\fontfig e} 
        \put(0.2cm,7.2cm){\fontfig i}
        \put(1.9cm,7.4cm){\fonttext zb-\ce{BeO}}
        \put(0.2cm,3.3cm){\fontfig m}
        \put(4.7cm,15.9cm){\fontfig b} 
        \put(4.7cm,11.3cm){\fontfig f} 
        \put(4.7cm,7.2cm){\fontfig j}
        \put(6.4cm,7.4cm){\fonttext zb-\ce{BeO}}
        \put(4.7cm,3.3cm){\fontfig n}
        \put(9.3cm,15.9cm){\fontfig c} 
        \put(9.3cm,11.3cm){\fontfig g}
        \put(9.3cm,7.2cm){\fontfig k}
        \put(10.8cm,7.4cm){\fonttext $\gamma$-\ce{LiAlO2}}
        \put(9.3cm,3.3cm){\fontfig o}
        \put(13.7cm,15.9cm){\fontfig d} 
        \put(13.7cm,11.3cm){\fontfig h} 
        \put(13.7cm,7.2cm){\fontfig l}
        \put(15.4cm,7.4cm){\fonttext $\gamma$-\ce{LiAlO2}}
        \put(13.7cm,3.3cm){\fontfig p}
        \put(1.8cm,12.1cm){$Q=-3$}
        \put(6.3cm,12.1cm){$Q=-1$}
        \put(10.2cm,12.1cm){$(v,\gamma)=(-2,\pi/2)$}
        \put(15.6cm,12.1cm){$Q=0$}
    \end{overpic}
    \caption{\justifying\textbf{Comparison between model polaron textures and \textit{ab initio} polarons}.
    Panels \textbf{a}-\textbf{d} in the first row: Idealized vector fields from Supplemental Table~\ref{table:topo}
    for the antivortex, vortex, double antivortex, and vertical flow fields, respectively.
    The parameters $a$,$b$, and $r$ are chosen as follows 
    $(a,b) = (1,-2)$, $(-3,1)$, $(1/3,1)$, $(0,1)$, respectively, and we set $r=10$ (arbitrary units). 
    These values are chosen to achieve close resemblances with the corresponding \textit{ab initio} polaron displacements in the second row. 
    The topological charge of these fields is insensitive to this choice, as shown in Supplemental Note~3 and Supplemental Fig.~\ref{figS:topo}(f-j).
    Panels \textbf{e}-\textbf{h} in the second row: Polaron displacements computed from first principles for zb-\ce{BeO}, $\gamma$-\ce{LiAlO2}, 2D h-BN, and \ce{PbTiO3}, respectively. 
    The radius of each sphere is taken to be $r=2\sigma_{\text{p}}$, with $\sigma_{\text{p}}$ being the standard deviation of the electron charge distributions: $r=21$~\AA, 29~\AA, 21~\AA, and 16~\AA, respectively.
    For clarity, we render only a single atomic species in each case: O for zb-BeO, Li for $\gamma$-\ce{LiAlO2}, B for 2D h-BN, and O for \ce{PbTiO3}. 
    We note that, in the case of 2D h-BN in panel \textbf{g}, the topological charge is ill-defined, and the key topological invariant is the vorticity. 
    We confirmed that the topological charge and vorticity $(Q', v')$ obtained numerically from the \textit{ab initio} polaron displacements are identical to those of the idealized vector fields $(Q,v)$ for each class, within a relative error $<1\%$.
    Panels \textbf{i}-\textbf{l} in the third row:    
    Vorticity $v$ and helicity $\gamma$ of zb-\ce{BeO} and $\gamma$-\ce{LiAlO2}, plotted as a function of the polar angle $\theta$. Six polar angles, $\theta/\pi=0.1,0.25,0.4,0.6,0.75,0.9$, are used for numerical calculations. The deviation between the helicity of the model and the \textit{ab initio} polaron in panel \textbf{l} is attributed to the weak piezoelectricity of $\gamma$-\ce{LiAlO2} and size of the simulation cell.
    Panels \textbf{m}-\textbf{p} in the fourth row:
    Decomposition of polaron displacements in phonon modes. The quantity $|B_{\bq\nu}|^{2}(\omega_{0}/\omega_{\bq\nu})$ shown in the plots is proportional to the squared polaron displacements, and $\omega_{0}$ is a reference frequency set to the highest phonon frequency. We see that long-wavelength acoustic modes are pronounced in all cases. 
    }
    \label{figS:comparison}
\end{figure}
\clearpage

\begin{figure}
    \centering
    \begin{overpic}[width=1.0\textwidth]{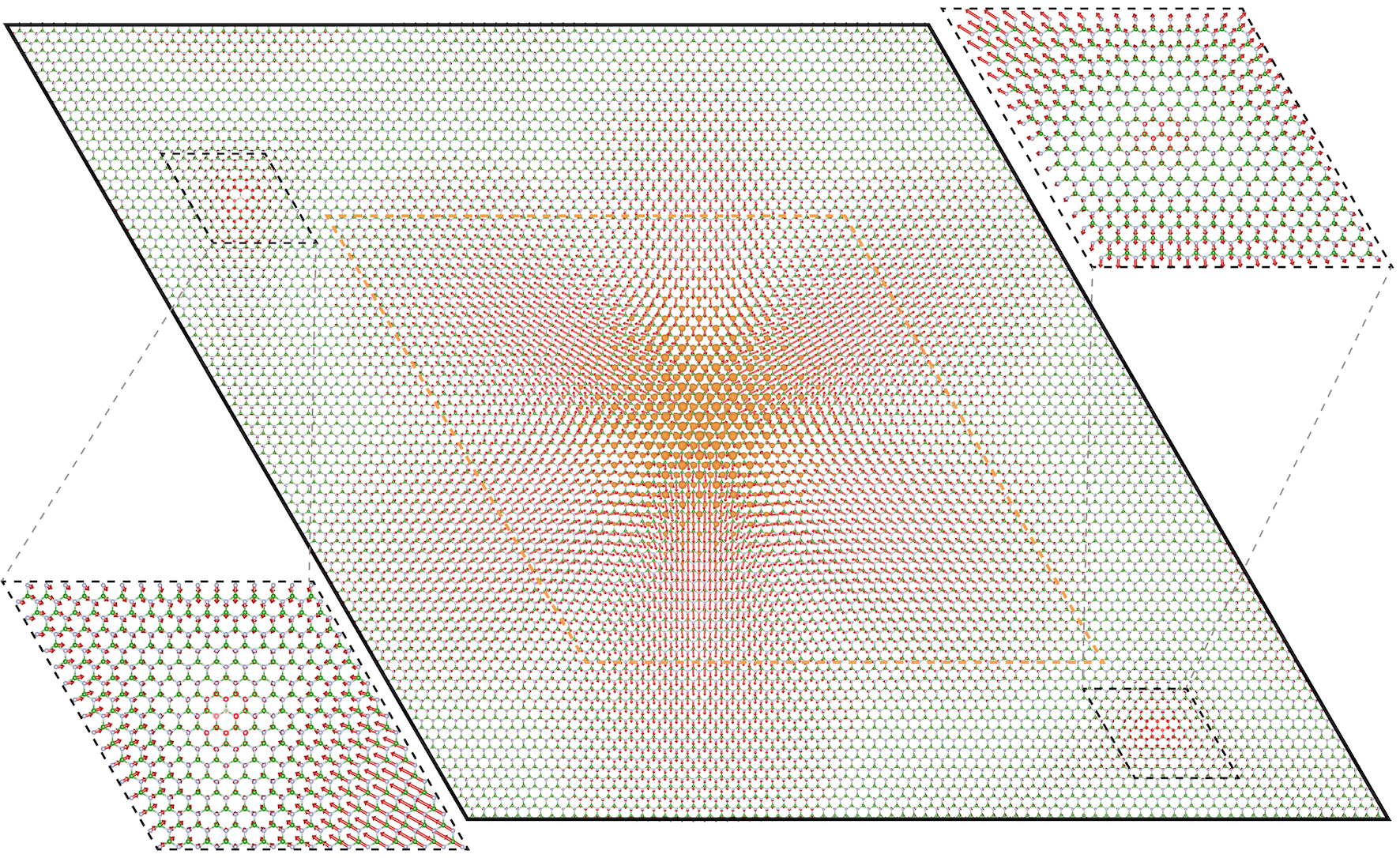}
        \put(6.3cm,8.5cm){\colorbox{white!60}{\large\color{orange}$v=-2$}}
        \put(1.6cm,3.8cm){\large $v=1$}
        \put(15.4cm,7.1cm){\large $v=1$}
    \end{overpic}
    \caption{\justifying\textbf{Numerical test of Poincar\'e-Hopf theorem}.
    The central panel shows a $81\times81$ Born-von K\'arm\'an supercell of 2D h-BN. Periodic boundary conditions make this supercell topologically equivalent to a torus with Euler characteristic $\chi=0$. 
    The orange isosurface in the middle represents the hole polaron density $|\psi(\br)|^{2}$, and the red arrows are the accompanying atomic displacements, magnified 1200-fold for clarity. 
    The vorticity of the displacement field along the closed loop defined by the dashed orange rhombus is $v=-2$, which corresponds to an antivortex. 
    Outside of this rhombus, we recognize two hedgehog-type fields within the dashed black rhombi (magnified 4-fold in the two side panels); for these fields we compute the vorticity $v=1$. 
    The existence of these two additional fields and their vorticity are dictated by the Poincar\'e-Hopf theorem \cite{Milnor1988g,crowley2017poincare,PT2021}, which requires the sum of all vorticities (i.e., winding numbers) of a continuous vector field to equal the Euler characteristic of the manifold. 
    As a result, integration along the boundary of the Born-von K\'arm\'an supercell produces a vorticity $v_{\rm tot}=0$.
    }\label{figS:poincare-hBN}
\end{figure}
\clearpage

\begin{figure}
    \centering    
    \begin{overpic}[width=1.0\textwidth]{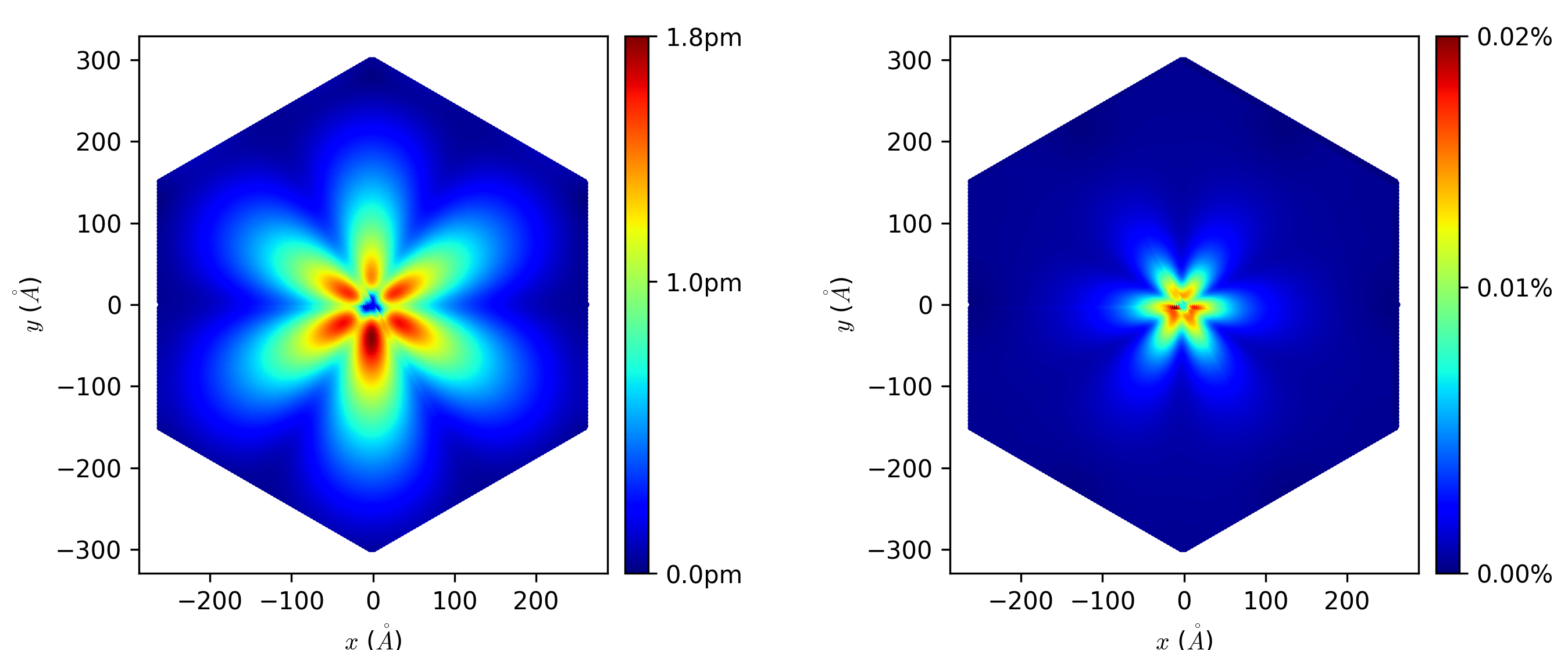}
    \put(0.2cm,7cm){\fontfig a}
    \put(9.4cm,7cm){\fontfig b}
    \end{overpic}
    \caption{\justifying
    \textbf{Comparison between polaron displacements and strain field in 2D h-BN}.
    Polaronic displacements and strain in a Wigner-Seitz supercell of 2D h-BN containing $88,200$ atoms. 
    \textbf{a} Magnitude of atomic displacements accompanying the hole polaron. The maximum displacement is $1.84$~pm. 
    \textbf{b} Strength of local strain induced by the electric field of the hole polaron. The maximum local strain is $0.019\%$. The spread of the displacement field is $\sim$200~\AA, while that of the strain field is $\sim$60~\AA. 
    }\label{figS:strain-hBN}
\end{figure}
\clearpage

\begin{figure}
    \centering
    \begin{overpic}[width=1.0\textwidth]{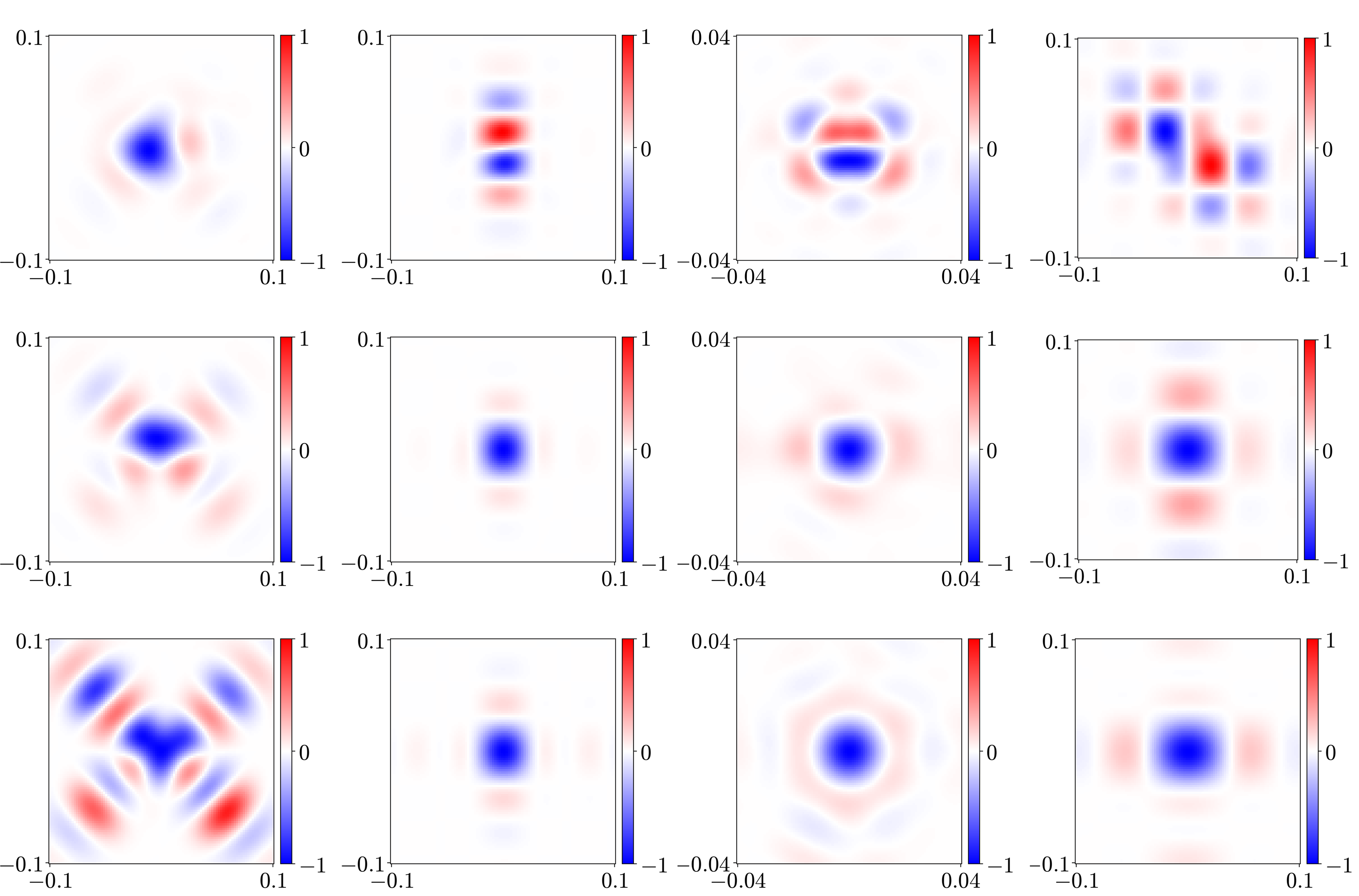}
    \put(1.5cm,11.5cm){\fonttext zb-BeO}
    \put(6cm,11.5cm){\fonttext $\gamma$-\ce{LiAlO2}}
    \put(10.6cm,11.5cm){\fonttext 2D h-BN}
    \put(15.2cm,11.5cm){\fonttext \ce{PbTiO3}}
    \put(1.5cm,7.5cm){\fonttext acoustic}
    \put(6.0cm,7.5cm){\fonttext acoustic}
    \put(10.5cm,7.5cm){\fonttext acoustic}
    \put(15.0cm,7.5cm){\fonttext acoustic}
    \put(1.5cm,3.5cm){\fonttext antivortex}
    \put(6.1cm,3.5cm){\fonttext vortex}
    \put(9.9cm,3.5cm){\fonttext double antivortex}
    \put(14.8cm,3.5cm){\fonttext vertical flow}
    \put(0.8cm,  11cm){\fontfig a}
    \put(5.3cm,  11cm){\fontfig b}
    \put(9.85cm, 11cm){\fontfig c}
    \put(14.35cm,11cm){\fontfig d}
    \put(0.8cm,  7cm){\fontfig e}
    \put(5.3cm,  7cm){\fontfig f}
    \put(9.85cm, 7cm){\fontfig g}
    \put(14.35cm,7cm){\fontfig h}
    \put(0.8cm,  3cm){\fontfig i}
    \put(5.3cm,  3cm){\fontfig j}
    \put(9.85cm, 3cm){\fontfig k}
    \put(14.35cm,3cm){\fontfig l}
    \put(2.7cm,  11cm){\fonttext [001]}
    \put(7.2cm,  11cm){\fonttext [001]}
    \put(11.8cm, 11cm){\fonttext [010]}
    \put(16.2cm, 11cm){\fonttext [001]}
    \put(1.5cm,0cm){\normalsize $q_{x}$~(\AA$^{-1}$)}
    \put(6cm,0cm){\normalsize $q_{x}$~(\AA$^{-1}$)}
    \put(10.5cm,0cm){\normalsize $q_{x}$~(\AA$^{-1}$)}
    \put(15cm,0cm){\normalsize $q_{x}$~(\AA$^{-1}$)}
    \put(0cm,1.3cm){\rotatebox{90}{\normalsize $q_{y}$~(\AA$^{-1}$)}}
    \put(0cm,5.3cm){\rotatebox{90}{\normalsize $q_{y}$~(\AA$^{-1}$)}}
    \put(0cm,9.3cm){\rotatebox{90}{\normalsize $q_{y}$~(\AA$^{-1}$)}}
    \put(13.5cm,1.3cm){\rotatebox{90}{\normalsize $q_{z}$~(\AA$^{-1}$)}}
    \put(13.5cm,5.3cm){\rotatebox{90}{\normalsize $q_{z}$~(\AA$^{-1}$)}}
    \put(13.5cm,9.3cm){\rotatebox{90}{\normalsize $q_{z}$~(\AA$^{-1}$)}}
    \end{overpic}
    \caption{\justifying\textbf{\textit{Ab initio} calculations of Huang diffuse scattering including thermal disorder}.
    Panels \textbf{a}-\textbf{d} in the first row: Normalized Huang scattering intensity including thermal disorder at room temperature, arising from the polaron displacements in zb-BeO, $\gamma$-\ce{LiAlO2}, 2D h-BN, and \ce{PbTiO3}, respectively. Only the heaviest element of each compound is included in the calculations, since the X-rays are less sensitive to light elements.
    Panels \textbf{e}-\textbf{h} in the second row:
    Same as in the first row, but this time only long-wavelength acoustic modes are included, by retaining contributions from modes with $\omega_{\bq\nu}\le 1$~THz.
    These patterns are expected to appear after a time delay $\Delta t\ge(2\pi/\max\{\omega_{\bq\nu}\})\sim10$\,ps from the excitation.
    To eliminate residual contributions from the hedgehog-type field, in \textbf{e}-\textbf{h} we show the average $\Delta\bm{\tau}(\bm{\tau}^{0}_{\kappa p})=[\Delta\bm{\tau}(\bm{\tau}^{0}_{\kappa p})+\Delta\bm{\tau}(-\bm{\tau}^{0}_{\kappa p})]/2$.
    Panels \textbf{i}-\textbf{l} in the third row: Normalized Huang scattering intensity for the idealized vector fields $\Delta\bm{\tau}(\bm{\tau}^{0}_{\kappa p})\propto \bu(\bm{\tau}_{\kappa p}^{0})\exp[-|\bm{\tau}_{\kappa p}^{0}|/2(2\sigma_{\text{p}})^2]$, where $\sigma_{\text{p}}$ is taken from the electron density distributions in the \textit{ab initio} calculations, and we choose to use $2\sigma_{\rm p}$ because the displacements are less localized than the wavefunctions.
    The magnitude of maximum displacement in each case is set to be the same as in the corresponding \textit{ab initio} calculations.
    The functions $\bu(\br)$ used in these panels are are $(yz,zx,xy)$, $(yz,-xz,0)$, $(2xy,x^{2}-y^{2},0)$, and $(0,0,z^{2})$, respectively. 
    We see that, even in the presence of thermal disorder, the model Huang diffraction patterns in \textbf{e}-\textbf{h} and the \textit{ab initio} calculatios in \textbf{i}-\textbf{l} are in excellent agreement.
    The slight difference between the idealized intensities shown in 
    \textbf{i}, \textbf{j}, and \textbf{l} and those shown in Fig.~4 of the main text are due to the fact that here we consider the actual crystals, while in Fig.~4 we used simple cubic lattices for clarity. In all cases shown in this figure (first, second, and third row), the Huang intensities of each polaron class are uniquely identifiable and constitute a characteristic fingerprint of the polaron topology.
     }\label{figS:Huang}
\end{figure}
\clearpage

\begin{figure*}[tbh!]
	\centering
    \begin{overpic}[width=\textwidth]{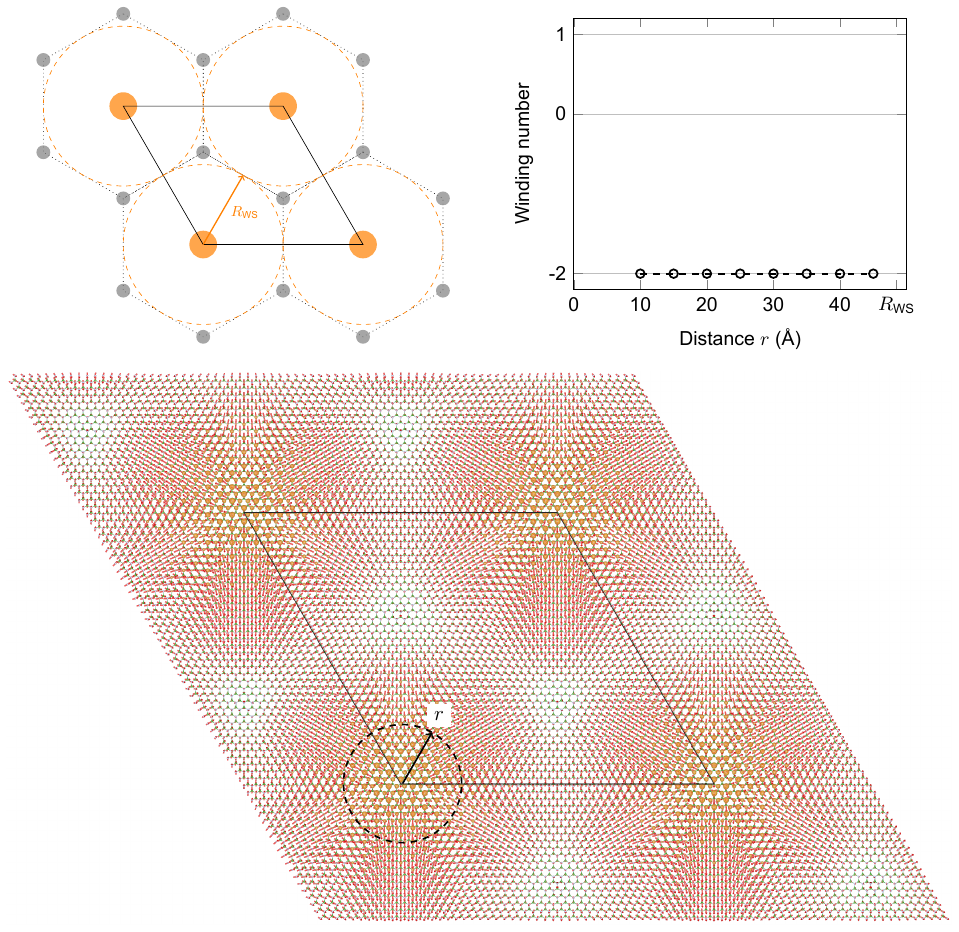}
        \put(0cm, 16.6cm){\fontfig a}
        \put(9.6cm, 16.6cm){\fontfig c}
        \put(0cm, 9.6cm){\fontfig b}
    \end{overpic}
    \caption{
    \justifying\textbf{Polaron Wigner crystal in 2D h-BN.} 
    \textbf{a} Schematic of a polaron Wigner crystal \cite{wigner1934interaction} on a triangular lattice, with orange regions showing polaron charge density. Gray regions (monopoles), enforced by periodic boundary conditions (Poincaré–Hopf theorem), form a honeycomb lattice matching the Wigner–Seitz cells (dotted hexagons). Half the distance between adjacent polarons is $R_{\text{WS}}$, which is also the radius of the largest inscribed orange dashed circle. 
    \textbf{b} \textit{Ab initio} simulation of a single polaron in 2D h-BN (lattice constant $a=4.7~a_{0}$ with $a_{0}$ the Bohr radius) with periodic boundary conditions, with one polaron per $N_{p}$ unit cells, corresponding to a Wigner radius $r_{s}/a_{0}=(\pi n a_{0}^{2})^{-1/2}$, where $n$ is the carrier density. Quantum Monte Carlo simulations \cite{tanatar1989ground} indicate that the electron crystallization occurs when $r_{s}>37$, translating to $N_{p}\geq15^{2}$. Here we plot a polaron array with $N_{p}=39^{2}$, for which $R_{\text{WS}}=48.5$~\AA.  
    \textbf{c} Circular integration paths centered at any polaron site in \textbf{b} with radii $r$ from 10 to 45~\AA\ ($r < R_{\text{WS}}$) yield a quantized winding number of $-2$. We note that such a lattice of double antivortices and monopoles has also been reported in twisted bilayer h-BN \cite{bennett2023polar}.
    }
	\label{figS:multipolaron}
\end{figure*}
\clearpage

\begin{figure}
    \centering
    \begin{overpic}[width=1.0\textwidth]{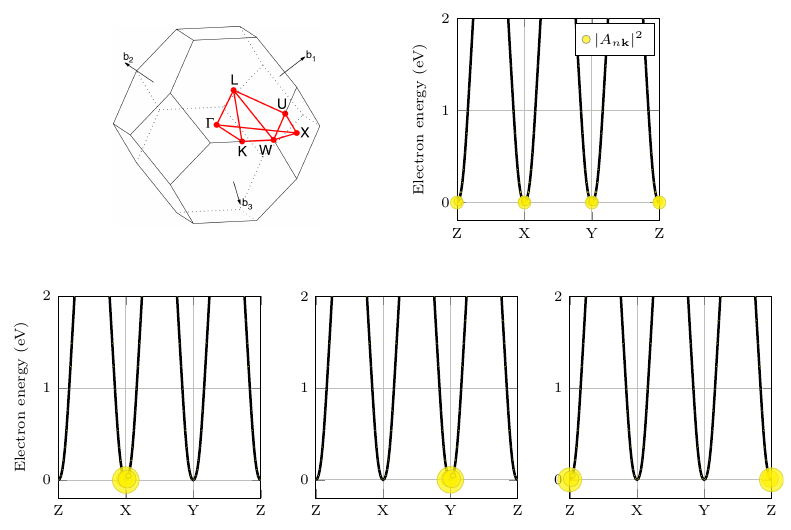}
    \put(2.5cm,11.5cm){\fontfig a}
    \put(9.4cm,11.5cm){\fontfig b}
    \put(0.3cm,5.3cm){\fontfig c}
    \put(6.5cm,5.3cm){\fontfig d}
    \put(12.2cm,5.3cm){\fontfig e}
    \end{overpic}
    \caption{\justifying
    \textbf{Electronic degeneracy and symmetry breaking}.
    \textbf{a} First Brillouin zone of face-centered-cubic zb-BeO. 
    There are three equivalent $X$ points connected by a 120$^\circ$ rotation around the [111] direction, which are labeled by $X$, $Y$, and $Z$ in panels \textbf{b}-\textbf{e}. 
    \textbf{b} Spectral decomposition of the electron polaron in terms of the Bloch states $\psi_{n\bk}$. The radius of the yellow circles is proportional to $|A_{n\bk}|^{2}$. In this panel, the three degenerate valleys contribute equally to the polaron wavefunction; as a result, the polaron displacements possess the full point-group symmetry of the crystal. This polaron corresponds to Supplemental Fig.~6(a). 
    \textbf{c}-\textbf{e} By initializing the Fourier amplitudes $A_{n\bk}$ to reside in only one of the valleys $X$, $Y$, or $Z$, the self-consistent polaron solution remains concentrated around that valley.
    These three solutions individually break the rotation symmetry around [111], but transform into each other via this symmetry. The symmetry group of these solutions is the little group of the valley, namely $\bar{4}2m$. 
    The formation energy difference between the solution in \textbf{b} and those in \textbf{c}-\textbf{e} is $E_{\rm c,d,e} - E_{\rm b} = 0.16$\,meV.
    }\label{figS:SSB-BeO}
\end{figure}
\clearpage


\bibliography{refs}